\documentclass[12pt]{article}

\usepackage{color,graphicx}
\usepackage{amsmath,amssymb}

\usepackage[numbers,compress]{natbib}

  \renewenvironment{thebibliography}[1]{%
    \begin{oldthebibliography}{#1}%
      \setlength{\parskip}{0ex}%
      \setlength{\itemsep}{0ex}%
      \small
  }%
  {%
    \end{oldthebibliography}%
  }

\newcommand{\Bx}{x_{\rm B}}
\newcommand{\GeV}{{\rm GeV}}
\newcommand{\ms}{$\overline{\rm MS}$}
\newcommand{\cs}{$\overline{\rm CS}$}
\font\cmss=cmss12 
\def\1{\hbox{{1}\kern-.25em\hbox{l}}}
\def\bfZ{\relax{\hbox{\cmss Z\kern-.4em Z}}}

\begin{document}

\begin{titlepage}

\vspace*{20mm}

\centerline{\Large \bf Deeply virtual Compton scattering at small
$\boldsymbol \Bx$}
\vspace*{0.5ex}
\centerline{\Large \bf  and the access to the GPD $\boldsymbol H$}

\vspace{20mm}

\centerline{\bf   Kre\v{s}imir Kumeri{\v c}ki$^{a}$ and Dieter M\"uller$^{b}$}

\vspace{8mm} \centerline{\it $^a$Department of Physics, Faculty of
Science, University of Zagreb} \centerline{\it P.O.B. 331,
HR-10002 Zagreb, Croatia}

\vspace{5mm} \centerline{\it $^b$Institut f\"ur Theoretische
Physik II, Ruhr-Universit\"at Bochum} \centerline{\it D-44780
Bochum, Germany}

\vspace{15mm}

\centerline{\bf Abstract}

\vspace{0.5cm}

\noindent We give a partonic interpretation for the deeply virtual
Compton scattering (DVCS) measurements of the H1 and ZEUS
Collaborations in the small-$\Bx$ region in terms of generalized
parton distributions. Thereby we have a closer look at the
skewness effect, parameterization of the $t$-dependence, revealing
the chromomagnetic pomeron, and at a model-dependent access to the
anomalous gravitomagnetic moment of nucleon. We also quantify the
reparameterization of generalized parton distributions resulting
from the inclusion of radiative corrections up to
next-to-next-to-leading order. Beyond the leading order
approximation, our findings are compatible with a `holographic'
principle that would  arise from  a (broken) SO(2,1) symmetry.
Utilizing our leading-order findings, we also perform a first
model-dependent ``dispersion relation'' fit of HERMES and JLAB  DVCS
measurements.  From that we extract the generalized parton
distribution  $H$ on its cross-over line and predict the beam
charge-spin asymmetry, measurable at COMPASS.

\vfill \noindent {\it Keywords:} deeply virtual Compton
scattering,
generalized parton distributions\\


\end{titlepage}

\newpage

\tableofcontents

\newpage

\section{Introduction}
\label{Sec-Int}

The electron/positron-proton collider experiments H1 and ZEUS at
the HERA ring in DESY improved not only the quantitative
understanding of inclusive processes, e.g., by pinning down the
small-$x$ behavior of parton distribution functions (PDFs)
\cite{Nadetal08,MarStiThoWat09}, but also led to new insights into
the proton structure. Two decades ago, it was mostly unforeseen
that at high energies the deep inelastic scattering (DIS) cross
section steeply rises, that the proton remains intact in almost
one third of all scattering events, and that even exclusive
processes become measurable, contributing considerably to the
total cross section. The reader may find comprehensive reviews
in Refs.~\cite{AbrFraStr95,AbrCal98,KleYos08}.

These exclusive processes, e.g., the electro- or photo-production
of a vector meson or a photon, were extensively studied in the
small-$\Bx$ kinematics by the H1 and ZEUS Collaborations
\cite{
Adletal01,Chekanov:2003ya,Aktas:2005ty,Aaretal07,Sch07,
Breetal97,Breetal98,Breetal99,Adletal99,Adletal02,Cheetal07,
Bre99a,Adletal00a,Cheetal05,
Deretal96,Breetal00,
Breetal98a,Adletal00,Cheetal02,Cheetal04,Aktetal05
}.
The amplitude of the subprocesses,
\begin{eqnarray}
\label{Excl-Proc}
\gamma^{(\ast)}(q_1)\, p(P_1) \to V(q_2)\,  p(P_2),
\qquad V= \gamma, \rho, \omega, \phi, J/\Psi, \Upsilon,
\end{eqnarray}
is  necessarily dominated by $t$-channel exchanges that carry the
quantum numbers of the vacuum. The large amount of HERA data calls
for a phenomenological description and it challenges the
theoretical understanding of the nucleon in terms of its partonic
substructure. Needless to say, a quantitative understanding of the
parton dynamics will be crucial at the frontier of exploration of
the structure of matter at LHC \cite{FraStrWei05,BooCheRoySch09}.
In this context, it is worth noting that exclusive Higgs
production via gluon fusion is a rather clean channel
\cite{KhoMarRys00,DeRKhoMarOraRys02}; however, for cross section
estimates the gluonic content of the proton must be quantified.

One might have hoped to master the phenomenology of such exclusive
processes in the framework of Regge theory. Unfortunately, for an
`incoming' virtual photon $S$-matrix theory is not applicable.
Thus, Regge theory loses the theoretical foundation and might possibly
be replaced by a pragmatic Regge phenomenology
\cite{DonLan86}. Consequently, firm conclusions valid for {\em
on-shell} scattering, like the one that unitarity requires that
the rightmost singularity in the complex angular momentum plane
belongs to a $J=1$ exchange, might not be appropriate for {\em
off-shell} processes. In fact, one of the lessons of H1/ZEUS
experiments is that cross sections, \emph{effectively} parameterized as
\begin{eqnarray}
\label{Cro-Sec-Off}
\frac{d\sigma^{\gamma^{\ast} p \to V p}}{dt}
\propto  \left(\frac{W^2}{W_0^2}\right)^{2 (\alpha(t)-1)}\,, \quad W^2 =
(P_1+q_1)^2\,,
\end{eqnarray}
rise steeply, contrarily to what is implied by the pomeron ($J=1$)
trajectory
\begin{eqnarray}
\label{Def-PomTra}
\alpha_{\mathbb{P}}(t) = 1 + 0.25\, t/\GeV^2\,.
\end{eqnarray}
In addition, the \emph{effective trajectory} $\alpha(t)$ varies
with the photon virtuality ${\cal Q}^2=-q_1^2$. The corresponding
\emph{effective Regge pole} in the complex angular momentum plane
might also be understood as a convenient implementation of cuts.

Inspired by QCD, a  phenomenological description of cross sections
at small $\Bx$ has been achieved in terms of the color dipole
model \cite{Muea94,MuePat94}, making direct contact to the high
energy approximation (BFKL) of scattering amplitudes
\cite{BalLip78,KurLipFad77}.  The physical picture might be set up
in the rest frame of the proton in which the highly energetic
virtual photon fluctuates into a quark-antiquark pair. The
quark-antiquark pair forms a small color dipole, spatially
distributed in transverse direction, that interacts with the
proton by a gluonic exchange. Finally, the quark-antiquark pair
forms a meson or annihilates into a photon. The physical
amplitude is given  as convolution (with respect to longitudinal
momentum fraction and transverse separation) of the color
dipole spectral function (cross section) with the corresponding
wave functions, describing the transition of the initial photon
into a quark-antiquark pair or the quark-antiquark pair into the
final state.

A perturbative QCD framework, applicable for longitudinally
polarized photons, is founded on factorization theorems
\cite{ColFraStr96}.  Here, in setting up the partonic space-time
picture, one may prefer a `brick wall' frame. The proton is viewed
as a bunch of partons travelling along the $z$-direction close to
the light-cone,  while the virtual photon goes in the opposite
direction and may have zero energy.  During the scattering
process, e.g., a quark is knocked out by the photon, it picks up
an antiquark and they form a meson, which travels in the direction
of the photon. Another subprocess is when the photon knocks out a
quark-antiquark pair arising from the fluctuation of gluonic
components in the proton wave function. In this picture the
quark-antiquark pair moves almost collinearly and its transition
into the final state is described by a collinear distribution
amplitude. The partonic content of the proton is encoded in
generalized parton distributions (GPDs)
\cite{MueRobGeyDitHor94,Rad96,Ji96a}, which can be interpreted as
a probability amplitude to emit and absorb a parton that moves
along the light cone or, equivalently, as a $t$-channel exchange
\cite{MueSch05}. For comprehensive reviews on GPDs see
\cite{Die03a,BelRad05}.

Obviously, the two space-time pictures are related by a boost
along the $z$-axis and in an {\em exact} QCD treatment both
approaches must yield the same results. The rise of the cross
section (\ref{Cro-Sec-Off}) at large $W$ requires a pomeron-like
$t$-channel exchange, encoded either in the color dipole spectral
function or in GPDs. The difference between the two approaches,
e.g., for the case of longitudinally polarized photons in hard
exclusive meson production, is mainly in the treatment of
transverse degrees of freedom and in the view on partons,
exchanged in the $t$-channel.  This partonic exchange is in the
color dipole model approach commonly viewed as entirely gluon
dominated, while the collinear perturbative framework provides an
implicit separation, defined by the factorization scheme, of the
quark and gluonic content of the proton.  The correspondence of
the objects in the two frameworks is obvious, and the color dipole
spectral function is often expressed in terms of so-called
$k_\perp$-unintegrated GPDs or, equivalently, so-called quantum
phase space distributions \cite{BelJiFen03}. We add that in the
two-gluon exchange color dipole model one may integrate out the
$k_\perp$-dependence in the target related part, which provides
then `hybrid' models that contain preasymptotic corrections due
to the transverse quark-antiquark pair fluctuation, elaborated,
e.g., in Refs.~\cite{FraKoeStr95,GolKro05,GolKro07}.

In this article we restrict ourselves to DVCS at small $\Bx$,
which has been measured at DESY on the electron/positron-proton
collider HERA experiments H1 and ZEUS
\cite{Aktas:2005ty,Aaretal07,Chekanov:2003ya,Sch07}. This has been
previously studied in the spirit of the aligned-jet model
\cite{FraFreStr98}, from high-energy/Regge perspective
\cite{BalKuc00,CapFazFioJenPac06,FazJen08}, in color dipole model
\cite{DonDos00,McDSanSha02,FavMac04,Mac07,KopSchSid08}, and in
collinear factorization approach at leading order (LO)
\cite{BelMueKir01,GuzTec06,GuzTec08}, next-leading order (NLO)
\cite{FreMcD01b,FreMcD01a,FreMcDStr02}, and next-to-next-leading
order (NNLO) \cite{KumMuePas07}. As said above, the view on the
partonic content of the proton varies with the approach. In DVCS
the different points of view are obvious. In the collinear DVCS
approach one starts with the hand-bag diagram, while in a color
dipole model one would draw a $t$-channel gluon ladder. In the
collinear factorization approach one usually stays with the
resolution conventions which are set in the standard PDF analysis
of deep inelastic scattering (DIS) at a given fixed order in the
coupling. Whether one expresses now the sea quark content in terms
of gluons, as in the color dipole approach, or stays with a large
amount of sea quarks, might be considered as a matter of taste.
However, one should bear in mind that what we shall call gluons is
not the same object that appears in a model approaches. Let us
also add that a model-dependent evaluation of power suppressed
corrections and their interpretation as higher-twist corrections
in the collinear factorization approach is likely an
oversimplified view on the dynamics of QCD.

We analyze the DVCS data of the H1 and ZEUS Collaborations within
the collinear factorization approach, going along the lines of
previous work \cite{KumMuePas07}. To set up non-perturbative GPD
models, we are to some extent motivated by the Regge phenomenology
and we employ its language. However, to avoid any
misunderstanding, we enclose in quotation marks Regge-theory
terms, appearing in our modelling. For instance, ``pomeron''
denotes a $t$-channel exchange with vacuum quantum numbers
responsible for a steep  rise of cross sections, associated with
an exchange of a colorless quark-antiquark or gluon pair. A ${\cal
Q}^2$-dependence of our ``pomeron trajectory'' will be solely
induced by evolution \cite{Mue06}. Using a least square fitting
procedure, we aim for a model-dependent access to both quark and
gluon GPDs at the LO of perturbation theory and
beyond.  There are two reasons to update our previous findings.
First, the H1 Collaboration provided new data from the HERA II run
including a significantly improved measurement of the
$t$-dependence of the cross section \cite{Aaretal07} and a
preliminary result for the beam charge asymmetry (BCA)
\cite{Sch07}. The second reason is that, in contrast to our
previous ad hoc model study, we utilize here for the first time
flexible GPD models, allowing us to describe DVCS also in the LO
(hand-bag) approximation. Such GPDs can then be used in a simple
GPD description of fixed target experiments, as pointed out in
Ref.~\cite{KumMuePas08}.

The outline of the paper is as follows. In Sect.~\ref{Sec-DVCS}
we present the observables, relevant for our DVCS analysis  in the
small-$\Bx$ kinematics. In Sect.~\ref{SubSect-Res} we give a short
overview of popular GPD models that have been employed for small-$\Bx$
phenomenology. 
Thereby, we adopt the terminology of quantum mechanics for both
CFF and GPD models, i.e., we distinguish between GPD model
and its representation.
A model refers to a specific GPD which can be given in different equivalent representations,
e.g., within double distribution representation, some version of conformal partial wave 
expansion, directly
in some SO(3) partial wave expansion or in the light-cone wave function overlap representation%
\footnote{In the literature specific GPD models are often associated with
representations. We have experienced that this indiscriminate use of language
can lead to confusion.}.
We shall also emphasize that the claim, that at
small $x$ GPDs should be rigidly tied to PDFs,
cannot be mathematically justified and, moreover,
its partonic/physical content is rather speculative. We
will then set up our model for integral (conformal) Mellin
moments.  Then we give a short insight into the technicalities of
the (conformal) Mellin-Barnes integral representation
of the DVCS amplitude.  In Sect.~\ref{Sec-Fits} we present then a detailed
analysis of H1 and ZEUS DVCS data and provide a GPD
interpretation. In particular, we shall discuss the failure of
ad hoc small-$x$ models at LO, and we shall give a detailed
analysis of the skewness effect, the parameterization of
$t$-dependence, the model-dependent access of the anomalous
gravitomagnetic  moment, and the reparameterization effect of
radiative corrections. In Sect.~\ref{Sec-DisRelFit} we include our
LO GPD findings in a global ``dispersion relation'' fit to  DVCS
data for unpolarized proton target, which includes the
measurements of HERMES, CLAS, and HALL A Collaborations. Thereby,
we aim for a first model-dependent extraction of the
dominant GPD $H$. This will allow us to predict the beam
charge-spin asymmetry for COMPASS kinematics. Finally, we
summarize and conclude.

\section{Deeply virtual Compton scattering at small $\boldsymbol \Bx$}
\label{Sec-DVCS}

In the deeply virtual electroproduction of photons both the
Bethe-Heitler (BH) bremsstrahlung and the DVCS process contribute
to the cross section. In the small-$\Bx$ region the DVCS
process dominates which allows one to extract
the DVCS cross section by a subtraction procedure.
There the integration over
the azimuthal angle $\phi$ projects on the transverse DVCS cross
section and guarantees that the contamination by the interference
term is negligibly small. In this way, both the H1
\cite{Aktas:2005ty,Aaretal07} and ZEUS \cite{Chekanov:2003ya}
Collaborations measured the cross section of DVCS on
an unpolarized proton in dependence of the photon virtuality
${\cal Q}^2$, the center-of-mass energy $W$, and the
momentum transfer squared $t$.

On the theoretical side, the unpolarized DVCS cross section in the small
$\Bx$-kinematics,
\begin{eqnarray}
\label{Def-CroSec}
\frac{d\sigma^{\rm DVCS}}{dt}(W,t,{\cal Q}^2) \approx
\frac{4   \pi \alpha^2 }{{\cal Q}^4} \frac{W^2 \xi^2}{W^2+{\cal Q}^2}
\left[\left| {\cal H} \right|^2  - \frac{t}{4 M^2_{p}}
\left| {\cal E} \right|^2
\right]
\left(\xi,t,{\cal Q}^2\right)\Big|_{\xi=\frac{{\cal Q}^2}{2 W^2+{\cal Q}^2}}\,,
\end{eqnarray}
is primarily given in terms of two Compton form factors (CFFs),
$ {\cal H}$ and ${\cal E}$.
Here $\alpha$ is the electromagnetic fine structure constant
and $M_{p}$ is the proton mass. Note that we consider CFFs as
functions of the symmetric scaling variable $\xi$, which can also be
expressed in terms of the Bjorken scaling variable $\Bx$:
$$
\xi=\frac{{\cal
Q}^2}{2 W^2+{\cal Q}^2} = \frac{\Bx}{2} + O(\Bx^2)\;.
$$
Both CFFs belong to the parity- and charge-even sector and, caused
by an effective ``pomeron exchange'', they might rise steeply at
small $\Bx$. In the collinear factorization approach their
dominant contribution arises from the twist-two GPDs $H$ and $E$.
In the cross section (\ref{Def-CroSec}) we neglected contributions
of parity-odd  CFFs $\widetilde {\cal H}$ and $\widetilde {\cal
E}$, which are supposed to be suppressed by $O(\Bx)$, for details%
\footnote{Let us add that similarly to polarized parton
densities the Regge phenomenology of the CFFs $\widetilde {\cal
H}$ and $\widetilde {\cal E}$ might not be so well understood. The
CFF $\xi \widetilde {\cal E}$ contains also a pion pole
contribution which yields a constant real part.  However, it
cannot compete with ${\cal H}$ in the DVCS cross section. Hence,
for small $\Bx$, we can safely neglect these CFFs in
(\ref{Def-CroSec}).} see, e.g.,
Refs.~\cite{BelMueKir01,KumMuePas07}. Owing to the integration
over the azimuthal angle $\phi$, there are no interferences of
twist-two with twist-three or gluon transversity contributions.
The squares of the both latter and possible twist-four and higher
contributions are all considered negligible.

Furthermore, the possibility to have both electrons and positrons as probes in
the HERA experiments allows direct access to the BH-DVCS
interference term. This serves as an experimental consistency
check of the aforementioned subtraction procedure and additionally
allows the measurement  of the beam charge asymmetry (BCA)
\begin{eqnarray}
\label{Def-ChaAsy}
A_{\rm BCA}(\phi)= \frac{d^+\sigma- d^-\sigma}{d^+\sigma+ d^-\sigma}\,,
\end{eqnarray}
reported in Ref.~\cite{Sch07}.
This asymmetry has a more intricate azimuthal angular $\phi$-dependence,
particularly when one integrates over a restricted phase space.
For unbinned data, it can be expressed in terms of the BH and DVCS
amplitudes and approximated to twist-two accuracy by the first and
third harmonics. Here and in the following we take the conventions
of Ref.~\cite{BelMueKir01} and find
\begin{eqnarray}
\label{Def-Asy}
A_{\rm BCA}(\phi)\! &\!\!\!=\!\!\! & -\frac{T_{\rm BH} T_{\rm DVCS}^\ast + T_{\rm BH}^\ast T_{\rm DVCS}}{|T_{\rm BH}|^2 + |T_{\rm DVCS}|^2}
\\
&\!\!\! \approx \!\!\! & \! \Bx \frac{F_1\, \Re {\rm e}{\cal H} -\frac{t}{4 M_{\rm p}^2} F_2\, \Re {\rm e} {\cal E}}{{\cal N}(\phi)} \cos(\phi)
\! + \! \Bx \frac{F_1\, \Re {\rm e}\! \left(\!{\cal E}_T + 2\widetilde {\cal H}_T\!\right)- F_2\, \Re {\rm e}\!\left(\! {\cal H}_T -
\frac{t}{2 M_{\rm p}^2} \widetilde{\cal H}_T\!\right)}{{\cal N}_T(\phi)}  \cos(3\phi)
\,,
\nonumber
\end{eqnarray}
where $F_1(t)$ and $F_2(t)$ are the Dirac and Pauli form factors
of the proton. We could again neglect the $\widetilde {\cal H}$
contribution in the first harmonic. The third harmonic is induced
by a photon helicity flip of two units. It is perturbatively tied
to gluon transversity  and might be contaminated by twist-four
(quark) contribution \cite{KivMan01}. The normalization factors,
\begin{eqnarray}
{\cal N}_T^{-1} &\!\!\! \approx \!\!\! & - \frac{t-t_{\rm min}}{4 M^2_{\rm p}}  \frac{1-y}{2-2y+y^2}\, {\cal N}^{-1}\,, \qquad
y =  \frac{1}{\Bx} \frac{{\cal Q}^2}{s}\,,
\end{eqnarray}
and
\begin{eqnarray}
{\cal N} &\!\!\! \approx \!\!\! &  \frac{y {\cal Q}}{8 \sqrt{1-y} (2-2 y+y^2)  \sqrt{t_{\rm min}-t}}
\left[\!\sum_{n=0}^3 c_n^{\rm BH} \cos(n \phi) +
 \frac{t}{{\cal Q}^2}  {\cal P}_1(\phi){\cal P}_2(\phi) \Bx^2 \sum_{n=0}^2 c_n^{\rm DVCS}\cos(n \phi)\!  \right],
 \nonumber\\
\end{eqnarray}
moderately induce a further azimuthal angle $\phi$-dependence.
Thereby, the DVCS Fourier coefficients $c_n^{\rm DVCS}$  depend of
course on the CFFs. Although a good approximation of the
normalization factors is obtained by restricting to the zeroth
harmonics in the BH and DVCS amplitude squared terms,
\begin{eqnarray}
c_0^{\rm DVCS} = 2 (2-2y+y^2) \left[\left| {\cal H} \right|^2  - \frac{t}{4 M^2_{p}}
\left| {\cal E} \right|^2
\right]\,,
\end{eqnarray}
a residual $\phi$-dependence is induced by the product ${\cal
P}_1(\phi){\cal P}_2(\phi)$ of BH propagators. At twist-three
level the zeroth and second harmonics of the interference term
complete the azimuthal angular dependencies in the BCA
(\ref{Def-Asy}). Note that this zeroth harmonic is dominated by
the same  twist-two CFF combination as the first harmonic,
displayed in Eq.~(\ref{Def-Asy}).

For the HERA kinematics we expect from the model studies in
Ref.~\cite{BelMueKir01}, see discussion in Sect.~7.1 there, a
moderate twist-two modulation ($\sim 10 - 15\%$) of the BCA with
a small twist-three admixture ($\sim 2-5 \%$), which is
mainly governed by the twist-two CFF $\cal H$. We
emphasize that the $\cos(3\phi)$ harmonic, related to gluon
transversity, is theoretically and
phenomenologically uncharted. To reveal the dominant twist-two
$\cos(\phi)$ harmonics of the interference term from a BCA
measurement, one might utilize the Fourier decomposition of the
charge asymmetry
\begin{eqnarray}
\label{BCA-FC}
A_{\rm BCA}(\phi) = p_0 +  p_1 \cos(\phi) + p_2 \cos(2\phi) + p_3 \cos(3\phi) + \cdots\,,
\end{eqnarray}
given as an infinite sum, and extract the dominant amplitude
$p_1$ from a fit. In this way both twist-three and
possible gluon transversity contributions diminish and we could on the theoretical side
safely employ the approximation for the amplitude
\begin{eqnarray}
\label{Def-Asy-p1}
p_1 \approx
\frac{8 \sqrt{1-y} (2-2 y+y^2)  \sqrt{t_{\rm min}-t}} {y {\cal Q}}
\frac{
\Bx \left[F_1\, \Re {\rm e}{\cal H} -\frac{t}{4 M_{\rm p}^2} F_2\, \Re {\rm e} {\cal E}\right]
}{
\! c_0^{\rm BH}  +
\frac{t}{{\cal Q}^2}  {\cal P}^{(0)}_{12}  2 (2-2y+y^2) \Bx^2
\left[\left| {\cal H} \right|^2  - \frac{t}{4 M^2_{p}}
 \left| {\cal E} \right|^2 \right]
}\,,
\end{eqnarray}
where ${\cal P}^{(0)}_{12}$ denotes the zeroth harmonic of the
product ${\cal P}_{1}(\phi){\cal P}_{2}(\phi)$. Obviously, the
(unbinned) observable $p_1$ is suppressed at large ${\cal Q}^2$
and small $-t$ and vanishes for $-t\to -t_{\min} \approx 0$. Note
also that twist-two contributions to other harmonics might arise
from the residual $\phi$-dependence in the normalization factor as
well.

{From} cross section measurements one extracts in the first place
the CFF $\cal H$, since $\cal E$ is kinematically suppressed by a
factor $-t/4M^2_{p}$, see Eq.~(\ref{Def-CroSec}). Both CFFs enter
in a different  combination in the first harmonic of the BCA
(\ref{Def-Asy-p1}), where $\Re{\rm e}{\cal E}$ is now accompanied
by $- F_2(t) t/4M^2_{p}$. A model study, in which ${\cal E}$ was
varied with fixed $\cal H$, has shown that it might be possible in a
BCA measurement, supplemented by the DVCS cross section
measurement, to access the CFF ${\cal E}$ \cite{BelMueKir01}.

To understand clearly which GPD degrees of freedom are accessible
in a DVCS measurement \cite{Ter05,KumMuePas08}, we remind that the
CFFs satisfy ``dispersion relations'', see, e.g.,
Refs.~\cite{FraFreGuzStr97,Che97,Ter05,KumMuePas07,DieIva07,AniTer07}.
Literally, we have to distinguish between the physical dispersion
relations that follow from general considerations and the
``partonic'' ones that are in {\em one-to-one} correspondence
\cite{Che97,Ter05,KumMuePas07,DieIva07,AniTer07} with
the leading approximation of CFFs in the perturbative
framework. For finite photon virtuality the physical and partonic
dispersion relations differ by higher twist contributions, see,
e.g., Refs.~\cite{KumMuePas07,AniTer07}. According to our goal, we
employ the set that is consistent with the perturbative framework
at leading twist and denote them with quotation marks. For
instance, ${\cal H}$ and ${\cal E}$ have even signature and their
subtracted ``dispersion relations'' might be written in the
following form \cite{KumMuePas07}:
\begin{eqnarray}
\label{Def-DisRel}
\Re {\rm e}
\left\{
{ {\cal H} \atop {\cal E} }
\right\}
(\xi,t,{\cal Q}^2) \approx
\frac{1}{\pi} {\rm PV} \int_0^1\! dx\, \frac{2x}{\xi^2-x^2} \Im {\rm m}
\left\{
{ {\cal H} \atop {\cal E} }
\right\}
(x,t,{\cal Q}^2)  \mp {\cal C}(t,{\cal Q}^2) \,,
\end{eqnarray}
where we set the threshold $\xi_{\rm th}=1 +O(1/{\cal Q}^2)$ equal to
one. Since the real part is determined by the imaginary part, one
might choose to extract it from the experiment indirectly, by
means of the ``dispersion relation'' \cite{KumMuePas08}. Then, when
fitting the data, one needs to model only the imaginary part and
the subtraction constant. It has been pointed out in
Refs.~\cite{Ter05,KumMuePas07,AniTer07} that the subtraction
constant is entirely related to a GPD term that completes
polynomiality.  In the representation of Ref.~\cite{PolWei99} it
is the so-called $D$-term, see also discussion in
Ref.~\cite{Ter01}.

The functional form for the CFFs in the small-$\Bx$ region might
be borrowed from the Regge theory. Thereby, we assume that the
``Regge trajectory'' is linear in $t$ and is same for both DIS and
DVCS. The imaginary part of CFF can then be written as:
\begin{eqnarray}
\label{Ans-CFF} \Im{\rm m}\left\{{\cal H} \atop {\cal E}
\right\}(\xi, t, {\cal Q}^2) = \pi \xi^{-\alpha(t,
{\cal Q}^2)}\, \left\{h_\alpha \atop e_\alpha \right\}(
t,  {\cal Q}^2) + \cdots\,.
\end{eqnarray}
For on-shell scattering, i.e., ${\cal Q}^2=0$, one would take for
$\alpha(t)\equiv\alpha(t, {\cal Q}^2=0)$ the soft pomeron
trajectory (\ref{Def-PomTra}) as the leading one. It appears in
the (target) helicity conserved CFF ${\cal H}$ and we shall assume
that it couples also to the helicity-flip CFF ${\cal E}$. This is
related to the so-called chromomagnetic pomeron, studied in
Ref.~\cite{Dia02} in the instanton approach. Interestingly, that
study suggests that the chromomagnetic pomeron should play a
dominant role in polarization phenomena at high-energy and so one
might expect that the CFF ${\cal E}$ is sizable. In principle,
the pomeron might also couple to the set of transversity CFFs,
which would be visible as a sizeable $\cos(3\phi)$ harmonics in
the BCA (\ref{Def-Asy}).

In accordance with phenomenology and perturbative QCD
the ``pomeron trajectory'' depends on the photon
virtuality ${\cal Q}^2$. Note that perturbation
theory states that in the double log approximation,
for large ${\cal Q}^2$ and small $\xi$,
the ``pomeron'' in the Regge-inspired ansatz (\ref{Ans-CFF}) is
replaced by a more intricate functional dependence
\cite{Mue06}. Also for this reason, we understand the
parameterization (\ref{Ans-CFF}) as an  {\em
effective} one. Next, we have non-dominant ``Reggeon
exchanges'', e.g., $\alpha(t) \approx 1/2 + t/{\rm
GeV}^2$, which we shall neglect in the small-$\Bx$
kinematics.

Plugging the Regge-like parameterization (\ref{Ans-CFF}) of the
imaginary part into the ``dispersion relation'' (\ref{Def-DisRel}), we
of course recover the  known Regge formula for the small-$\Bx$
region:
\begin{eqnarray}
\label{Ans-CFF-A}
\left\{{\cal H} \atop {\cal E} \right\}(\xi,t,{\cal Q}^2) = \pi \left[i - \cot\left(\frac{\pi
\alpha(t, {\cal Q}^2)}{2}\right)\right] \, \xi^{-\alpha(t, {\cal Q}^2)}
\left\{h_\alpha \atop e_\alpha \right\}(t,{\cal Q}^2) + \cdots\,.
\end{eqnarray}
Here the ellipsis stands for non-dominant contributions of
``Reggeon exchanges'' with typically
$$\alpha^{\rm Reg}(t \sim 0.2\,{\rm GeV}^2) \sim 0.3, $$ for a constant%
\footnote{From the viewpoint of the ``dispersion relation''
(\ref{Def-DisRel}) this constant, i.e., the so-called $J=0$ pole,
contains contributions from the subtraction constant and further
contributions induced by both ``pomeron'' and non-leading
``Regge exchanges''. \label{foonot-J=0}}, and for further terms
that vanish as $\xi\to 0$. In the considered kinematics the ``pomeron
trajectory'' has a typical value of
$$
\alpha(t \sim 0.2\,{\rm GeV}^2 , {\cal Q}^2 \sim 4\, {\rm GeV}^2 )
\sim 1.1 - 1.2.
$$
Hence, in contrast to on-shell forward Compton scattering at large
energies, where the classical pomeron intercept $\alpha(0)=1$ yields
a vanishing real part, we encounter in DVCS with $\alpha(0) > 1$ a
positive real part that should even dominate the negative ``Reggeon''
contributions and subtraction constant \cite{BelMueKir01}. This
already predicts that for H1/ZEUS kinematics the sign of the
leading $\cos\phi$ harmonic in the BCA (\ref{Def-Asy}) is positive and
even provides an estimate of its size, quoted above, which is consistent
with preliminary measurements of the H1 Collaboration \cite{Sch07}.

\section{Models for the flavor singlet GPDs at small $\boldsymbol x$}
\label{SubSect-Res}

In the collinear factorization approach the CFFs $\mathcal{F} = \{
\mathcal{H}, \mathcal{E} \}$ are given as convolution of a hard
scattering part with GPDs ${F}= \left\{ {H},{E}\right\} $, which
to LO accuracy reads
\begin{eqnarray}
\label{ConFor-LO} {\cal F}(\xi, t, {\cal Q}^2)
\stackrel{\rm LO}{=} \sum_{q=u,d,s,\cdots} e_q^2 \int_{-1}^1\! dx \left[ \frac{1}{\xi-x -i 0}
- \frac{1}{\xi+x- i 0} \right] F^q(x,\eta= \xi,t,{\cal
Q}^2)\,,
\end{eqnarray}
where $e_q$ is the fractional quark charge. Note that this is
nothing else but the so-called hand-bag approximation, where only
quarks%
\footnote{
Of course, this is a simplified partonic view, since an infinite number of longitudinal
gluons, radiated from the struck quark, are included in the GPD. The radiation of a
transverse gluon is taken into account in genuine twist-three contributions.
}
are resolved in a hard Compton scattering process.
Obviously, the imaginary part of the CFFs is given by the GPD on
the cross-over line ($\eta=x$)
\begin{eqnarray}
\label{GPD@LO2SpeFun}
{\Im}{\rm m}{\cal F}(\xi, t,
{\cal Q}^2) \stackrel{\rm LO}{=}
\pi \sum_{q=u,d,s,\cdots} e_q^2
\left[F^q(x=\xi,\eta=\xi,t,{\cal Q}^2)-F^q(x=-\xi,\eta=\xi,t,{\cal Q}^2)
\right]\,.
\end{eqnarray}
The GPD at negative momentum fraction $x=-\xi$ corresponds to the
negative antiquark contribution. The real part of the amplitude
can be calculated either by means of the ``dispersion relation''
(\ref{Def-DisRel}) or from the convolution formula
(\ref{ConFor-LO}). The equality of the two results is ensured by
construction \cite{Che97,KumMuePas07}, i.e., by the polynomiality
or support properties of the GPD \cite{Ter05,KumMuePas07}. For
fixed ${\cal Q}^2$ the LO perturbative approach is completely
equivalent to the (approximated) ``dispersion relation''
(\ref{Def-DisRel}) and the only information that can be accessed
at LO is the GPD on the cross-over line.

Perturbation theory also predicts the evolution of GPDs, e.g.,
their change on the cross-over line in the flavor singlet sector
is
\begin{eqnarray}
\label{EvoCroOvLin} {\cal Q}^2 \frac{d}{d{\cal Q}^2} \mbox{\boldmath $F$}(x,x,t,{\cal
Q}^2) = \int_x^1\frac{dy}{x}\, \mbox{\boldmath $V$}(1,y/x; \eta =x,\alpha_s({\cal Q}))
\mbox{\boldmath $F$}(y,x,t,{\cal Q}^2)\,,
\end{eqnarray}
where the flavor singlet quark ($\Sigma$) and gluon (G) GPDs are
collected in the column vector
\begin{eqnarray}
\label{Def-GPD-sin}
\mbox{\boldmath $F$}(x,\eta,t,{\cal Q}^2) =
\left(\!
\begin{array}{c}
F^{\Sigma}  \\
F^{\rm G}
\end{array}
\!\right)(x,\eta,t,{\cal Q}^2)\,,
\qquad
{F^{\Sigma}}(x,\cdots)= \sum_{q=u,d,s,\cdots}   \left[{F^q}(x,\cdots)-F^q(-x,\cdots) \right] \,,
\end{eqnarray}
and the evolution kernel is a two-dimensional matrix%
\footnote{It is common to adopt the convention in which
the gluon GPD reduces in the forward limit to $x G(x)$. In such a case the non-diagonal
entries in the evolution matrix are accompanied by $1/\eta$ or
$\eta$, which are set equal to $x$, see Ref.~\cite{BelFreMue99}.}
which leads to the mixing of the quark
and gluon GPDs. $F^{\Sigma}(x,\eta, \cdots)$ and $F^{\rm
G}(x,\eta,\cdots)$ are antisymmetric and symmetric in $x$,
respectively, and both are symmetric in $\eta$.  One realizes from
the evolution equation (\ref{EvoCroOvLin}) that the scale change
of the GPD on the cross-over line is governed by its value in the
outer region ($x\ge\eta$). Moreover, the evolution kernel in the
gluon-gluon channel has a ``pomeron''-like singularity. Hence, as
in the forward case, the evolution in the small-$x$ region is
driven by the gluons and can be solved in the double log
approximation. There one finds that the ``pomeron intercept''
$\alpha(0)$ {\em effectively} increases and the slope
$\alpha^\prime$ decreases with growing ${\cal Q}^2$ \cite{Mue06}.
We note that the change of the residue function in (\ref{Ans-CFF})
under evolution cannot be derived for the general case, since it
is determined by the GPD in the outer region.

Beyond LO accuracy, the imaginary part of CFFs is given by a
convolution of the hard-scattering part with the GPDs, analogous
to the evolution equation.  As in DIS, the momentum fraction
integration runs from $x=\xi$ to $x=1$. However, there is an
important difference. Namely, the GPD itself depends on the
scaling variable $\xi$. Hence, the deconvolution cannot be
performed and the outer GPD region can only be accessed through
evolution effects, see discussion in Ref.~\cite{Fre99}.

According to all this, a parameterization of the outer GPD region
is sufficient for a description of DVCS via imaginary part of
CFFs. Then, using simple Regge-motivated small-$\Bx$
parameterization of imaginary part of CFFs from (\ref{Ans-CFF}),
and in accordance with the LO relation (\ref{GPD@LO2SpeFun}), we
write a rather flexible GPD ansatz in the small-$x$ region at a
given input scale:
\begin{eqnarray}
\label{GPD-ans}
F(x\ge \eta,\eta,t) =
n_F\, x^{-\alpha(t)}\, r(\vartheta|F)\,
\mbox{\boldmath$\beta$}(t|F)  \,;
\qquad \vartheta \equiv \eta/x \;.
\end{eqnarray}
Here we factorized the GPD into the Regge part $x^{-\alpha(t)}$, the
skewness function $r(\eta/x|F)$, the residue function
$\mbox{\boldmath$\beta$}(t|F) $, and  the residue $n_F$ of the GPD
in the forward limit. Hence, in this limit, i.e., $\eta\to 0$ and
$t\to 0$, both $r(\eta/x|F)$ and $\mbox{\boldmath$\beta$}(t|F) $
are normalized to one. Note that skewness- and $t$-dependence
might not necessarily factorize. Since present data do not allow
to address their possible functional interplay, we stick here to the
most convenient model. Also, we would like to add that in this GPD
approach restoring of the GPD in the central region ($x \le \eta$)
is a mathematical problem, which is solved in the short distance operator
product expansion framework by construction \cite{Che97,
KumMuePas07,KumMuePas08}. Thereby, the Mellin moments of $F(x,
x\vartheta,t)$ are expanded in a Taylor series around the point
$\vartheta=0$, i.e., $r(\vartheta|F)$ possesses certain analytic
properties.

\subsection{Survey of GPD models at small ${\boldsymbol x}$}
\label{SurSec-GPD-mod}

GPD modelling can be done in different representations. It is
popular to employ either double distribution (DD) representation
\cite{MueRobGeyDitHor94,Rad97} or some version of conformal
partial wave (PW) expansion
\cite{Rad97,BelGeyMueSch97,Shu99,PolShu02,MueSch05,KirManSch05a},
adopted from the description of mesons
\cite{BroLep79,EfrRad80,EfrRad80a,BroLep80}. Mathematically, there
is a one-to-one mapping between GPDs in two different
representations. Since this mapping might be cumbersome, it has
been sometimes only partly worked out. In these circumstances,
one might hope to gain some new physical/partonic insight in the
small-$x$ region just by a change of the representation. However,
it is clear that all popular models finally lead in the small-$x$
region to the GPD of the form (\ref{GPD-ans}), where, however, the
skewness function might be rather restricted.
Surely, the small-$x$ GPD properties,  described in momentum
fraction representation, can be equivalently set up in any other
representation, too.

Our effective Regge-motivated ansatz (\ref{GPD-ans}) is supported
by a diagrammatic analysis from the $t$-channel view
\cite{FraFreGuzStr97}, where it was found that the ``Regge
trajectory'' is skewness-independent; however, the residue
function, i.e., the skewness function $r(\vartheta|F)$ in our
ansatz (\ref{GPD-ans}), depends on it. One arrives at the same
conclusion from an $s$-channel view, if one adopts the ideas of
Refs.~\cite{LanPol70,BroCloGun73} and convolutes a spectator quark
model, expressed in the DD representation,  with a constituent
quark mass spectral function. Of course, in Regge-inspired DVCS
models one starts with such a point of view, where the skewness
dependence of the impact form factors has to be modelled. We
already stated that Regge behavior is included in the popular GPD
models, too.

Such arguments and agreement with experiment, see also
Sect.~\ref{Sec-Int}, led to a broad acceptance of effective Regge
behavior for off-shell processes, even in the absence of a firm
theoretical foundation. Assuming this, the remaining task for GPD
phenomenology is the determination of the
\begin{itemize}
\item[i.]  skewness function $r(\vartheta|F)$ and
\item[ii.] residue function $\mbox{\boldmath$\beta$}(t|F)$.
\end{itemize}
Certainly, one can address these questions by fitting to the
experimental measurements using one's favorite GPD representation.
The problem remains always the same, namely, to have appropriate
functional GPD ansaetze.

The value of the skewness function at $\vartheta=1$, see
Eq.~(\ref{GPD-ans}), is an important characteristic of a GPD
model. For the  GPD $H$, with $r(\vartheta)\equiv r(\vartheta|H)$
and $r \equiv r(1)$, it can be expressed as the ratio of the GPD
at the cross-over line to the corresponding PDF, given as the GPD
at $\eta=0$ for $t=0$. Since we rely on the universality of the
``pomeron trajectory'', this \emph{skewness ratio} is for small
$x$ mostly independent of $x$. For the quark GPD it reads
\begin{eqnarray}
\label{Def-r-Rat-0}
r^\Sigma({\cal Q}^2) = \frac{H^\Sigma(x,\eta=x,t=0,{\cal Q}^2)}
{\Sigma(x,{\cal Q}^2)}\,,
\qquad
\Sigma(x,{\cal Q}^2) = H^\Sigma(x,\eta=0,t=0,{\cal Q}^2) \;,
\end{eqnarray}
and for gluons we use the convention:
\begin{eqnarray}
\label{Def-rG-Rat-0}
r^{\rm G}({\cal Q}^2) =
\frac{H^{\rm G}(x,\eta=x,t=0,{\cal Q}^2)}{x G(x,{\cal Q}^2)}\,,
\qquad
x G(x,{\cal Q}^2) =
H^{\rm G}(x,\eta=0,t=0,{\cal Q}^2)\,.
\end{eqnarray}
In an LO DVCS analysis the quark skewness (\ref{Def-r-Rat-0}) is
given as the ratio of $\Im{\rm m} {\cal H}(x,t=0,{\cal
Q}^2)/\pi$ and the transverse unpolarized DIS structure function
$F_T(x,{\cal Q}^2)/x$; therefore, it can be almost directly
measured, while the gluonic one (\ref{Def-rG-Rat-0}) can be
accessed only by utilizing a large lever arm in ${\cal Q}^2$.

The aligned-jet model considerations in Ref.~\cite{FraFreStr98}
provide estimates for both the DIS structure function, given as
imaginary part of the virtual forward Compton amplitude, and for
the imaginary part of DVCS amplitude. In
an LO approximation resulting quark skewness  ratio%
\footnote{The authors define quantity $R^{\rm FFS}$ as ratio of
imaginary DIS and DVCS amplitudes. In LO approximation this means
$r = 2^{-\alpha(0)}/R^{\rm FFS} \sim 2/R^{\rm FFS}$.} turns out to be
$r^\Sigma({\cal Q}^2\sim 2.5\, \GeV^2)\approx 1$. We note that
this model has been used to predict the size of the DVCS cross
section, which was afterwards experimentally confirmed, see Ref.~\cite{Chekanov:2003ya}.
The model was then generalized for the deeply virtual Compton
amplitude with two virtual photons, providing a prediction for the
quark skewness function \cite{FreMcDStr02}:
\begin{eqnarray}
\label{Def-SkeFunFSS}
r_{\rm FMcDS}^{\Sigma}(\vartheta)=
\frac{(1+\vartheta)^{1-\alpha(0)}}{2 \vartheta}
\left(\!
1+\frac{M_0^2}{{\cal Q}_0^2}\!\right)
\ln\!\left(\!
\frac{1+\vartheta
}{
1-\frac{{\cal Q}_0^2-M_0^2}{{\cal Q}_0^2+M_0^2}\, \vartheta}
\!\right)\,,\quad
M_0^2 \sim 0.5\,\GeV^2\,,\;\; {\cal Q}_0^2 \sim 2.5\, \GeV^2\,.
\end{eqnarray}

In the double distribution (DD) representation a GPD that has
support in the interval $-\eta \le x \le 1$ is expressed as
follows \cite{MueRobGeyDitHor94,Rad97}:
\begin{eqnarray}
\label{Def-DD}
F(x,\eta,t) =
\int_0^1\!dy\int_{-1+z}^{1-z}\!  \delta(x-y - \eta\, z)  f(y,z,t)\,.
\end{eqnarray}
Note that in this specific representation polynomiality is not
completed for $H$ and $E$. This artifact is not crucial for the
small-$x$ application and it can be cured in various ways, see,
e.g., Refs.~\cite{PolWei99,BelMueKirSch00,Ter01,HwaMue07}. It is
quite popular among phenomenologists to utilize Radyushkin's DD
ansatz (RDDA) \cite{Rad98a,Rad98}. Here the DD at $t=0$ is
factorized in a PDF and a profile function, namely,
\begin{eqnarray}
\label{RDDA}
h(y,z,t=0) =
\frac{\Gamma(3/2+b)}{\Gamma(1/2) \Gamma(1+b)}
\frac{q(y)}{1-y} \left(\!1-\frac{z^2}{(1-y)^2}\!\right)^b\,.
\end{eqnarray}
The profile function is chosen to be convex and its width is
controlled by the $b$ parameter. The  skewness function can be
easily evaluated in terms of hypergeometric functions:
\begin{eqnarray}
\label{Def-r-fun}
r^\Sigma_{\rm RDDA}(\vartheta) =
{_2F_1}\!\left(\!{\alpha/2,(1+\alpha)/2 \atop
3/2+b}\Big|\vartheta^2\!\right)\,, \qquad
r^{\rm G}_{\rm RDDA}(\vartheta) =
{_2F_1}\!\left(\!{(\alpha-1)/2,(\alpha)/2 \atop
3/2+b}\Big|\vartheta^2\!\right)\,.
\end{eqnarray}
Here and in the following we use a shorthand for the
intercept $\alpha \equiv \alpha(t=0)$. Setting $\vartheta=1$ in
the skewness functions (\ref{Def-r-fun}), the values of the
skewness ratios at the cross-over line $\eta=x$ follow:
\begin{eqnarray}
r^\Sigma_{\rm RDDA} = 2^{2b-\alpha}
\frac{\Gamma(3/2+b)\Gamma(1+b-\alpha)}{\Gamma(3/2)\Gamma(2+2
b-\alpha)}\,,\qquad
r^{\rm G}_{\rm RDDA} = 2^{1+2b-\alpha}
\frac{\Gamma(3/2+b)\Gamma(2+b-\alpha)}{\Gamma(3/2)\Gamma(3+2 b-\alpha)}\,.
\end{eqnarray}
Taking the originally proposed value $b=1$, one finds for $\alpha=1.2$:
\begin{eqnarray}
\label{r-ratio-RDDA}
r^\Sigma_{\rm RDDA} \approx 1.8\;,
\qquad r^{\rm G}_{\rm RDDA} \approx 1.04\,.
\end{eqnarray}
For growing $b$ the quark skewness ratio decreases and rapidly
approaches the value $r^{\Sigma}=1$, which corresponds to setting
GPD equal to a $t$-decorated PDF. For the gluon GPD we see that
the ratio $r^{\rm G}=1$ is almost realized for $b=1$. A skewness
ratio $r<1$ requires to give up the convexity of the profile
function.

It is also rather popular to expand  GPDs in tree-level conformal
partial waves. In some versions the conformal partial wave
amplitudes are then additionally mapped to forward-like functions.
In such expansions the evolution operator is diagonalized at LO.
Unfortunately, such representations involve certain mathematical
subtleties, which have been understood only in the last few years
\cite{Nor00,ManKirSch05,MueSch05,Pol07a}. Originally, based on an
oversimplified view on the inverse integral transformation
\cite{ShuBieMarRys99}, it has been erroneously stated that for
small $x$ the GPD transform reduces to a PDF.  Thus, the authors
of Ref.~\cite{ShuBieMarRys99} claim  that the GPDs in the
small-$x$ region are at $t=0$ tied to PDFs by the following ratio:
\begin{eqnarray}
\label{r-ratio-con} r_{\rm con}^\Sigma  = \frac{2^{\alpha}
\Gamma(3/2+\alpha)}{\Gamma(3/2)\Gamma(2+\alpha)}\,,
\qquad
r_{\rm con}^{\rm G}  = \frac{
2^{1+\alpha}
\Gamma(3/2+\alpha)
}{
\Gamma(3/2) \Gamma(3+\alpha)
}\,.
\end{eqnarray}
This ratio is in fact a Clebsch-Gordan coefficient, e.g.,
occurring in the conformal partial wave expansion of the product
of two currents (it can be seen, e.g., by taking a ``Regge pole''
value for complex conformal spin, $j=\alpha-1$, in
Eq.~(\ref{CPW-LO}) below). Therefore, we shall call it conformal
ratio. Although the small-$x$ claim, which corresponds to a specific double
distribution model \cite{MusRad99},  is not based on  general
grounds,  the arguments were repeated and
refined in a more recent version of the claim, where it was
assumed that the ``pomeron pole'' is the rightmost singularity in
the complex {\em conformal spin} plane \cite{MarNocRysShuTeu08}.
Below we shall spell out, what is already (implicitly) said in the
literature, e.g.,
Refs.~\cite{MusRad99,Nor00,ManKirSch05,MueSch05,Pol07a}, namely,
that the claim in Refs.~\cite{ShuBieMarRys99} and
\cite{MarNocRysShuTeu08} is based on certain mathematical
simplifications.  As long as the partonic content of the conformal
PW expansion is not understood, we consider this claim as not
necessary applicable for GPD phenomenology.

We would like to mention that GPDs can also be described by a
$t$-channel SO(3) partial wave expansion, formulated within
complex-valued angular momentum $J$. Of course, assuming effective
``Regge poles'', this description is fully equivalent to our
Regge-motivated ansatz (\ref{GPD-ans}) in momentum fraction space.
The GPD representations are somehow the analogue of a Legendre  and
power expansion of amplitudes at high energies. We recall that
both expansions are equivalent \cite{Khu63} and the latter became
more popular in Regge phenomenology in the sixties of the last
century.

A combination of
$t$-channel SO(3) and conformal partial wave expansion has been
proposed \cite{PolShu02}, where the difference of conformal spin
$j+2$ and $t$-channel angular momentum $J$, is used as a discrete
variable:
\begin{eqnarray}
\label{Def-rho}
\rho = j+1-J = \{0,2,4,\ldots \} \;,
\end{eqnarray}
which is always even. This GPD representation is set up with
forward-like functions $Q_{\rho}(z)$ of momentum fraction-like
variable $z$ and has been called ``dual''
parameterization%
\footnote{Usage of term ``dual'' was motivated by the fact that in
dual models \cite{Ven68} the $s$-channel amplitude is described by
the $t$-channel exchanges. We add that this feature is more
general and arises from crossing and the Sommerfeld-Watson
transform of the $t$-channel SO(3) partial wave expansion. In
Regge theory/phenomenology the resummation of $t$-channel
exchanges is encoded in the Regge trajectory.}. Below we employ
essentially the same model within a conformal partial wave
expansion in terms of a Mellin-Barnes integral. Taking in such a
model only the leading contribution, where $j+1=J$, one finds the
conformal skewness function and ratio (\ref{r-ratio-con}). Taking
into account that the singular behavior of the non-leading
contributions $Q_{\rho}(z)$ increases by $z^{-\rho}$, one can
obtain other values for the skewness ratio \cite{Pol07a}.  Hence,
$\rho$ cannot be a priori considered as an expansion parameter.
Note that in the minimal ``dual'' model of Ref.~\cite{GuzTec06}
and in its corrected version \cite{GuzTec08} the second forward-like
function $Q_{2}(z)$ has the same small-$z$ behavior as the leading
one $Q_{0}(z)$. Hence, this model possesses the conformal skewness
ratio (\ref{r-ratio-con}).

The inverse transform in the ``dual'' parameterization has not
been derived so far. Nevertheless, it has been exemplified within
model studies, where $\rho$ can be effectively employed as an
expansion parameter, that ``dual'' and DD representation can be
effectively mapped into each other \cite{Sem08,PolSem08}.

In the RDDA the conformal ratio (\ref{r-ratio-con}) arises when
choosing $b=\alpha$, and, moreover, the conformal skewness
function coincides then with the RDDA one, given in
Eq.~(\ref{Def-r-fun}). Since the value $\alpha \approx 1.2$ is rather
close to the originally proposed value for $b$, the conformal ratios
\begin{eqnarray}
\label{ConRat-num}
r^\Sigma_{\rm con} \approx 1.65\,, \qquad r^{\rm G}_{\rm con}
\approx 1.03 \;,
\end{eqnarray}
are rather close to the RDDA ones (\ref{r-ratio-RDDA}). In the
``dual'' model of Ref.~\cite{GuzTec06} both the leading $Q_0(z)$
and next-leading $Q_2(z)$ forward-like functions have been taken
into account; however, both of them have the same $z$ behavior.
Hence, in this model the skewness ratio also has the conformal value
(\ref{r-ratio-con}) and it is in the flavor singlet sector not
very different from the RDDA, see also
Ref.~\cite{GuzTec06} for numerical examples.
Therefore, both GPD models contradict the
aligned-jet model estimate and so also the experiment
\cite{FreMcDStr02,GuzTec08}.

To convert the predictions of the aligned-jet model into the GPD
language, the authors of Ref.~\cite{FreMcDStr02} set the GPD at
small $x$ and low input scale $\mu_0^2$ equal to the PDF:
\begin{eqnarray}
\label{Ans-GPD-wrong}
H(x,x,t=0,\mu_0^2) = q(x,\mu_0^2) \quad \mbox{for small}\; x\;
\mbox{and low}\; \mu_0^2\,.
\end{eqnarray}
This is nothing but the RDDA model in the limit $b\to \infty$;
practically, a large value $b \gg 1$ is sufficient.  This ansatz
implies that the skewness function is set $r(\eta/x) = 1$  for all
$x$. With such an initial condition, evolution, starting at a
rather low input scale,  will rapidly lead to an increase of the
$r$-ratio. Thus, this GPD model fails to describe data, too. We
will not go into details here, however, we would just like to
point out that the skewness function determines the evolution of
the skewness ratio. In RDDA ansatz its functional dependence
ensures that the conformal ratio will be approached with
increasing ${\cal Q}^2$, as numerically demonstrated in
Ref.~\cite{DieKug07a}.

Let us only shortly discuss the functional form of the residual
$t$-dependence. Historically, it is common in Regge phenomenology
to model it by an exponential ansatz, i.e,
\begin{eqnarray}
\label{Ans-bet-exp}
\mbox{\boldmath$\beta$}(t) = {\rm exp}\left \{B t\right\}\,.
\end{eqnarray}
In Regge framework it is clear that this is an effective
description for the small $-t$ region. An exponential
$t$-dependence of  GPDs arises in the overlap wave function
representation \cite{DieFelJakKro98,BroDieHwa00,DieFelJakKro00}
within wave functions that have an exponential fall-off in the
transverse momentum. It is questionable that in such GPD models
the GPD spectral properties, ensuring polynomiality, can be
implemented \cite{MukMusPauRad02}. On the other hand, power-like
wave functions allow us to represent GPDs having both properties.
With power-like wave functions one can easily evaluate DDs and
they meet the field theory inspired view. In this framework a
simple spectator quark GPD model was set up \cite{HwaMue07} and it
was found that at small $x$ the $t$-dependence indeed factorizes
and that a power-like behavior arises:
\begin{eqnarray}
\label{Ans-bet-dip}
\mbox{\boldmath$\beta$}(t)=\left(1-\frac{t}{M^2}\right)^{-p} \;.
\end{eqnarray}
Certainly, in such models Regge-like behavior is not implemented.
As stated above, this can be achieved by convolution of such model
with an appropriate constituent mass spectral function. However,
it remains unclear to us how should the $t$-dependence be treated,
so that common Regge behavior arises and, simultaneously, general
positivity constraints \cite{Pob02,Pob02a} are satisfied; see Ref.~\cite{Pob02b} 
for GPD integral representations. Hence, we have no
definite ansaetze for the residue function at smaller values of
$t$ at hand; however, a residual $t$-dependence as in
Eq.~(\ref{Ans-bet-dip}) at a (very) low input scale and smaller
values of $x$ looks to us rather plausible.

\subsection{Modelling of integral conformal GPD moments}
\label{subsec-Fits}

Our goal is to have a first empirical look at a more flexible GPD
modelling, where we rely on the physical Regge-inspired picture.
We shall introduce three different GPD models for the small-$x$
region, each having different skewness function. The models are
defined by their conformal Mellin moments. This is foremostly a
technical point, allowing us to employ existing stable numerical
code that includes radiative corrections at NNLO level.
Analogously to the singlet quark and gluon GPDs in
Eq.~(\ref{Def-GPD-sin}), we collect their conformal moments in a
two-dimensional vector
\begin{eqnarray}
\label{Ans-ConMomS} \mbox{\boldmath $H$}_j(\eta, t,\mu^2) = \left(
\begin{array}{c}
H_j^{\Sigma}  \\
H_j^{\rm G}
\end{array}
\right)(\eta, t,\mu^2)\,.
\end{eqnarray}
These moments are given by the expectation values of local operators
which are the lowest state in conformal towers; they are labelled by
integral conformal spin $j+2$, and they have spin projection $j+1$.
They can be defined by convolution of GPDs with Gegenbauer polynomials
\begin{eqnarray}
\label{Fj} \mbox{\boldmath $H$}_{j}(\eta,t,\mu^2) = \frac{
\Gamma(3/2)\Gamma(j+1)}{2^{j}  \Gamma(j+3/2)}
\frac{1}{2}\int_{-1}^1\! dx\; \eta^{j} \left(
\begin{array}{cc}
C_j^{3/2} & 0 \\
0 & \frac{3}{j} \frac{1}{\eta}\,  C_{j-1}^{5/2}
\end{array}
\right)\!\!\left(\frac{x}{\eta}\right)
 \mbox{\boldmath $H$}(x,\eta,t,\mu^2)\,,
\end{eqnarray}
where polynomial indices $\nu =\{3/2,5/2\}$ are
group-theoretically determined. Since $C^{\nu}_j(-x) = (-1)^j
C^{\nu}_j(x)$, our GPDs (\ref{Def-GPD-sin}) have only \emph{odd}
$j$ moments. For the GPD $\mbox{\boldmath $E$}$ we adopt the
analogous decomposition. If polynomiality is completed, the
integral conformal moments (\ref{Fj}) are polynomials in $\eta$ of
order $j+1$. Note that the highest order terms provide the
subtraction constant in the ``dispersion relation'' (\ref{Def-DisRel})
\cite{Ter05,KumMuePas07}.

The normalization in (\ref{Fj}) is chosen in such a way that the
basic property of GPDs $H$, namely, that they reduce in the
forward limit to the PDFs, translates into similarly simple
integer-$j$ moment-space relation
\begin{eqnarray}
\label{Def-GPD-for}
 {\boldsymbol H}_j(\eta,t,\mu^2)
 \xrightarrow[\eta\to 0]{} {\boldsymbol q}_j(t,\mu^2) \equiv
\left({ \Sigma_j \atop G_j}\right)(t,\mu^2)\,,
\qquad
\left({ \Sigma_j \atop G_j}\right)(t,\mu^2) \equiv
\int_0^1\!dx\, x^j \left({\Sigma \atop G}\right)(x,t,\mu^2)\,,
\end{eqnarray}
where  zero-skewness GPDs $\Sigma$ and  $G$  at $t=0$ are given by
the flavor singlet quark and gluon PDFs, respectively. For the
overall normalization we shall use the PDF momentum fraction
averages $N^{\Sigma, G}$, that are given by the first Mellin
moments and satisfy the following sum rule
\begin{eqnarray}
\label{MomSumRul}
N^{\Sigma}(\mu^2)+ N^{\rm G}(\mu^2) = 1
\quad \mbox{with}\quad
N^{\Sigma} = \Sigma_1(t=0,\mu^2)\;\;
\mbox{and}\;\;
N^{\rm G} = G_1(t=0,\mu^2)\,.
\end{eqnarray}

In our modelling we are going along the lines pointed out in the
momentum fraction space. The dominant small-$x$ behavior arises
from the ``pomeron exchange'' that is for zero-skewness GPD
related to sea quark contributions defined via anti-quarks
$\overline{q}$
\begin{eqnarray}
\label{Dec-PDFs}
\Sigma = q^{\rm sea} +  q^{\rm val}\,,
\qquad
q^{\rm sea} = 2 \overline{q}\,,
\qquad
q^{\rm val}=\sum_{q=u,d,s,\cdots}  \left[q - \overline{q}\right]\,.
\end{eqnarray}
The difference of quark and anti-quark PDFs we denote as
valence-like flavor singlet contributions. They are related to
``Reggeon exchanges'' with an intercept $\alpha\approx 1/2$. Thus,
they can be neglected for the small-$\Bx$ kinematics. The standard
DIS terminology (\ref{Dec-PDFs}) we adopt for GPDs, too. First we
set up the zero-skewness GPD, including the $t$-dependence, and
afterwards we skew it, i.e., model the skewness function.

We start with the standard ansatz for PDF Mellin moments at a given
input scale $\mu_0$
\begin{eqnarray}
\label{PDFx2j}
q(x) = N \frac{x^{-\alpha} (1-x)^\beta}{B(2-\alpha,\beta+1)}
\qquad
\stackrel{\rm Mellin\, transform}{\Leftarrow\! \Rightarrow}
\qquad  q_j =  N \frac{B(1-\alpha+j,\beta+1)}{B(2-\alpha,\beta+1)}\,,
\end{eqnarray}
for both quark and gluon PDFs. Here $\alpha \gtrsim 1$
is the intercept of an effective ``pomeron trajectory'', while $\beta$
parameterizes the large $x$ or $j$ behavior. The normalization is given
by the momentum fraction average $N$, see Eq.~(\ref{MomSumRul}).

Then, to obtain the model for the zero-skewness GPDs
(\ref{Def-GPD-for}), we decorate the PDF Mellin moments (\ref{PDFx2j})
with $t$-dependence by (i) extending the ``Regge intercept''
$\alpha$ to a  linear trajectory $\alpha(t)=\alpha +
\alpha^\prime t$, where we only introduce the leading pole, and
then (ii) by multiplying with a residue function $\mbox{\boldmath
$\beta$}(t)$, having either exponential (\ref{Ans-bet-exp}) or
power-like (\ref{Ans-bet-dip}) functional form. In accordance with our
$t$-factorized ansatz at small $x$, we  neglect the dependence of
the slope $B$ or cut-off mass $M$ on $j$.%
\footnote{This is also sufficient from the pragmatical point of
view because such a dependence anyway cannot be constrained by
fitting small-$\Bx$ DVCS data \cite{KumMuePas07}.} All parameters
can be separately adjusted for sea quarks and gluons, even the
effective ``pomeron trajectories'' might slightly differ. Thus,
our ansatz for the Mellin moments of the zero-skewness GPDs reads
\begin{eqnarray}
\label{Ans-MomSigm}
\Sigma_j(t)  &\!\!\! = \!\!\!& N_{\rm sea}
\frac{
B(1-\alpha^{\rm sea}+j,\beta^{\rm sea} +1)
}{
B(2-\alpha^{\rm sea},\beta^{\rm sea}+1)}\,
\frac{\mbox{\boldmath $\beta$}^{\rm sea}(t)
}{
1-\frac{t}{(m^{\rm sea}_j)^2}}+\cdots,
\qquad
(m^{\rm sea}_j)^2=
\frac{1+j-\alpha^{\rm sea}}{\alpha_{\rm sea}^\prime}\,,
\\
\label{Ans-MomG}
G_j(t) &\!\!\! =\!\!\! & N_{\rm G}
\frac{
B(1-\alpha^{\rm G}+j,\beta^{\rm G}+1)
}{
B(2-\alpha^{\rm G},\beta^{\rm G}+1)}\,
\frac{\mbox{\boldmath $\beta$}^{\rm G}(t)
}{
1-\frac{t}{(m^{\rm G}_j)^2}}\,,
\phantom{+\cdots e}\,\,
\qquad
(m^{\rm G}_j)^2=
\frac{1+j-\alpha^{\rm G}}{\alpha_{\rm G}^\prime}\,.
\end{eqnarray}
Here the ellipsis in the quark singlet sector indicates
valence-like contributions, cf.~(\ref{Dec-PDFs}). The
``pomeron pole'' is written as a monopole form factor,
where the cut-off mass squared $m^2_j$ depends on $j$.
The pole is located at
$$
m_j^2 =t  \qquad \Rightarrow \qquad  j=\alpha(t)-1 \;.
$$
Note that at $t=0$ these pole contributions are included in the
beta functions, appearing in
Eqs.~(\ref{Ans-MomSigm}, \ref{Ans-MomG}).

The skewing is achieved by relating the conformal GPD moments for
given conformal spin to the forward Mellin moments by a
linear transformation that depends on $\eta$, e.g., by
\begin{eqnarray}
\label{Ske-in-conSpa}
\mbox{\boldmath $H$}_j(\eta, t) =
\mbox{\boldmath $r$}_{j}(\eta)
\left(
{ q^{\rm sea}_j \atop G_j}
\right)(t) +\cdots\,,
\quad
\mbox{\boldmath $r$}_{j}(\eta)=
\left(
\begin{array}{cc}
r^{\rm sea}_{j}(\eta)  & 0 \\
0 & r^{\rm G}_{j}(\eta)
\end{array}
\right) \,.
\end{eqnarray}
It is required that for integral conformal spin $r^{\text{sea,G}}_{j}(\eta)$ are even
polynomials in $\eta$ of order $j+1$, which are at $\eta=0$
normalized to one. They are moment-space analogue of the
skewness functions $r(\eta/x)$.

Here, we should already point out a crucial property of
$r_{j}(\eta)$, given as a function of two variables $j$ and $\eta$.
On the first naive glance one might expect that normalization
$r_{j}(0)=1$ implies skewness ratio $r(\theta=1)=1$; however, this
is not the case. The inversion of Mellin transform, such as
(\ref{Fj}), involves an integration over complex-valued $j$ with
certain weight, e.g., the GPD on its cross-over line is given
by \cite{MueSch05}
\begin{equation}
{\boldsymbol H}(x,\eta=x,t)=\frac{1}{2\pi i}
 \int_{c-i\infty}^{c+i\infty} dj\, \left(\frac{x}{2}\right)^{-1-j}\,
 \frac{\Gamma(5/2+j)}{\Gamma(3/2)\Gamma(3+j)}
  \begin{pmatrix}
   1 & 0 \\ 0  & 2x/(3+j)
 \end{pmatrix}
 {\boldsymbol H}_{j}(\eta = x, t) \;.
\label{eq:GPDtrajMB}
\end{equation}
If one extends $j$ to complex values and assumes a smooth forward
limit of $r_j(\eta)$ and uses the pole value $j=\alpha-1$ in
Eq.~(\ref{eq:GPDtrajMB}), one is immediately led to the conclusion
that the skewness ratio is equal to its conformal value
(\ref{r-ratio-con}), e.g., for sea quarks,
\begin{eqnarray}
\label{Wro-Sta}
\lim_{\eta\to 0}{r}_{j}^{\rm sea}(\eta) = 1
\qquad \Rightarrow \qquad
r_{\rm con}^{\rm sea}(\vartheta=1)  = \frac{2^{\alpha}
\Gamma(3/2+\alpha)}{\Gamma(3/2)\Gamma(2+\alpha)}\,.
\end{eqnarray}
This is a naive thinking, too, because assuming that the function
$r_{j}(\eta)$ for complex-$j$ has the same $\eta\to0$ limit as its
integer-$j$ analogue is unwarranted in the case of a singularity
at $\eta=0$.  We will show in Sect.~\ref{SecSub-NumEva} that indeed
a branch point can develop there. Precisely this feature ensures
the equivalence of representations and will enable us to construct
flexible GPD models, not constrained by the conformal skewness
ratio (\ref{r-ratio-con}).

We proceed with our modelling for integer $j$. To find the
convenient functional form for $r_j(\eta)$, it is helpful to start
by evaluating it from a known GPD at $t=0$. A simple example has
been given in Ref.~\cite{MueSch05}, where the following
quark-antiquark unsymmetrized GPD has been evaluated from a toy DD
$f^{\rm toy}(y, z)=y^{-\alpha}$, taken to be equal to its ``Regge'' piece:
\begin{eqnarray}
\label{GPD-toy}
H^{\rm toy}(x,\eta) = \frac{1}{1-\alpha}\theta(-\eta \le x)\,
\frac{1}{\eta} \left(\frac{x+\eta}{1+\eta}\right)^{1-\alpha}
+ {\eta\to -\eta}.
\end{eqnarray}
At small $x$  this toy GPD provides in the forward limit the
realistic PDF behavior $x^{-\alpha}$ and its skewness
function at small $\eta$ is
\begin{eqnarray}
\label{toy-skefun}
r^{\rm toy}(\vartheta)=
\frac{(1+\vartheta)^{1-\alpha}}{2 \vartheta}
\frac{1}{\alpha-1}
\left[
\left(\frac{1-\vartheta}{1+\vartheta}\right)^{1-\alpha} -1
\right]\,,
\end{eqnarray}
which is reminiscent of the aligned-jet model one
(\ref{Def-SkeFunFSS}). Since this expression arises from a DD that
does not vanish on the support boundaries, it is ill-defined at
$\vartheta=1$ for $\alpha \ge 1$ and needs a regularization, e.g.,
by using a profile function that vanishes at the support
boundaries. Note that for RDDA (\ref{RDDA}) the  GPD behavior at
the end-point $x=-\eta$ and cross-over point $x=\eta$ is  governed
by both the Regge behavior of the PDF and the end-point behavior
of the profile function.

Nevertheless,  we can use this model as a guide for modelling
conformal GPD moments. The conformal moments of the GPD
(\ref{GPD-toy}) can be straightforwardly calculated by means of a
so-called beta transform of Gegenbauer polynomials into the
hypergeometric function ${_3F_2}$.   We write the resulting
conformal moments in the form of a transformation
(\ref{Ske-in-conSpa}) and read off:
\begin{eqnarray}
\label{Mod-GPD-Reg-Mom-1}
r_j^{\rm toy}(\eta) &\!\!\!=\!\!\! &
\frac{(1+j-\alpha)(2+j-\alpha)}{2(1-\alpha)(2-\alpha)}
\frac{\Gamma(3/2)  \Gamma(3 + j)}{\Gamma(3/2 + j)}
\\
&& \times
\left[
\frac{\eta-1}{2\eta} \left(\frac{\eta}{2}\right)^j
{_3F_2}\!\left(\!{-j, 3+j, 2-\alpha \atop 2, 3 - \alpha}\bigg|
\frac{\eta-1}{2 \eta }\!\right) + \left\{\eta \to -\eta\right\}
\right]\,.
\nonumber
\end{eqnarray}
Note that for odd $j$ these moments exist even for $\alpha =1$.
The divergent behavior of the model (\ref{GPD-toy}) at
$\vartheta=1$ shows up in this space as convergency problem for
large $j$. Hence, we can employ the skewness moments
(\ref{Mod-GPD-Reg-Mom-1}) only if the zero-skewness GPD decreases
faster than $1/j^2$ at $j\to \infty$, which is satisfied for
realistic values of $\beta$ in (\ref{Ans-MomSigm}) and
(\ref{Ans-MomG}). As already said, below we shall continue the odd
integer $j$ conformal moments to complex-valued $j$. Here we only
emphasize that $r_j^{\rm toy}(\eta)$ has a zero at $j=\alpha-1$
which will cancel the ``pomeron pole'' contribution of
zero-skewness GPD. This obviously demonstrates that the
normalization condition for odd $j$ will {\em not} necessarily
imply the limit (\ref{Wro-Sta}) and so the conformal ratio
(\ref{r-ratio-con}) cannot be taken for granted.

In our modelling we also follow the suggestion of
\cite{Pol98,PolShu02} and expand conformal GPD moments in
SO(3) $t$-channel partial waves  \cite{KumMuePas07}
\begin{eqnarray}
\label{Exp-ConMom} F_{j}(\eta,t)=
\sum_{J=J_{\rm min} \atop {\rm even}}^{j+1}
F_{jJ}(t)\,  \eta^{j+1-J}   \hat{d}_{\cal F}^{J}(\eta)
\,,
\end{eqnarray}
labelled by the angular momentum $J$, where $\eta = -1/\cos\theta$ is
expressed in terms of the $t$-channel center-of-mass scattering angle%
\footnote{We are here in the first place interested in the
assignment of quantum numbers. Thus, we neglect some corrections
which die out either in the Bjorken limit or by setting the proton
velocity in the c.o.m.~frame to one. These corrections appear in
the partial waves and their amplitudes. However,  after completing
the Sommerfeld-Watson transform  they  will partly cancel each
other in the small-$x$ region. Hence, it is for our application
more convenient to neglect them from the beginning.}. The PW
amplitude $F_{jj+1}$ with angular momentum $J=j+1$ is nothing but
the Mellin moment of the zero-skewness GPD, while the PW
amplitudes with $J$ smaller than $j+1$ are suppressed by a factor
$\eta^\rho$ ($\rho = j+1-J$). The SO(3) partial waves
$\hat{d}_{\cal F}^{J}$, normalized to one at $\eta=0$,  are the
crossed version of Wigner's reduced rotation matrices
$d_{0,\lambda-\lambda^\prime}^{J}$ \cite{Die03a}. For the
$t$-channel helicity conserved `electric'
($\lambda=\lambda^\prime=1/2$) and helicity flip `magnetic'
($\lambda=-\lambda^\prime=1/2$)  GPD moments
\begin{eqnarray}
\label{CFF-ele+mag}
\mbox{\boldmath $H$}_j +  \frac{t}{4 M^2} \mbox{\boldmath $E$}_j\,,
\quad\mbox{and}\quad
\mbox{\boldmath $H$}_j +  \mbox{\boldmath $E$}_j\,,
\end{eqnarray}
respectively, they are expressed by  Legendre (Gegenbauer  with
index $\nu=1/2$) polynomials,
\begin{eqnarray}
\label{Def-d-1/2} \hat{d}_{\cal F}^J(\eta) &\!\!\! \propto \!\!\!&
\eta^J d_{0,0}^J(1/\eta)
\qquad\qquad\qquad\qquad\qquad\qquad\mbox{for}\quad {\cal F} = {\cal H} +
\frac{t}{4M^2} {\cal E} \;,
\\
&\!\!\! = \!\!\!&
\frac{\Gamma(1/2) \Gamma(J+1)}{2^{J}\Gamma(1/2+J)} \eta^{J}
C^{1/2}_{J}(1/\eta)\,,
\nonumber
\end{eqnarray}
with $J_{\rm min}=0$ and Gegenbauer  polynomials with index $\nu=3/2$,
\begin{eqnarray}
\label{Def-d-3/2}
\hat{d}_{\cal F}^J(\eta) &\!\!\! \propto \!\!\!& \eta^{J} \frac{z
\sqrt{J(J+1)}}{\sqrt{1-z^2}} d_{0,1}^J(z) \big|_{z=1/\eta}
\qquad\qquad\mbox{for}\quad {\cal F} = {\cal H} + {\cal E} \;,
\\
&\!\!\! = \!\!\!&  \frac{\Gamma(1/2) \Gamma (J)}{2^{J}
\Gamma(J+1/2)}  \eta^{J-1}  C_{J-1}^{3/2}(1/\eta) \;,
\nonumber
\end{eqnarray}
with $J_{\rm min}=2$. Although we here consider integral $J$, it
is already appropriate to think in terms of the Regge language,
providing us a guideline for the resummation of
(\ref{Exp-ConMom}). The quantum numbers associated with our
conformal GPD moments are associated with ``pomeron'' and parity- and
charge conjugation-even ``Reggeon exchanges''  \cite{JiLeb00,Die03a},
where the leading ones belong to the $f$--meson trajectory,
generically given as $\alpha(t) \sim 1/2 + t/{\rm GeV}^2$. Each
term in the sum (\ref{Exp-ConMom}) should therefore have a leading
pole at $J =\alpha(t)$ rather than at $j=\alpha(t)-1$. Note that
the former scenario, well established from the physical point of
view, yields a non-trivial skewness ratio, while the latter, more
formal and vague scenario, yields the conformal ratio
(\ref{Wro-Sta}). Polynomiality is completed, i.e., the
``magnetic'' and ``electric'' GPD moments are polynomials of order
$j+1$ and $j-1$, respectively, where $j$ is odd. Terms
proportional to $\eta^{j+1}$ contribute to the so-called $D$-term,
which finally builds up the subtraction constant in the ``dispersion
relation''. To this constant not only the $J=0$ $\sigma$-meson
contributes, but rather all exchanges with $ J \le j+1$. Still, in
our small-$\Bx$ DVCS application the ``pomeron trajectory''
dominates these ``Reggeon exchanges'', including the subtraction
constant, and so we shall hereafter ignore them in our small-$x$
GPD modelling.

As one realizes from Eq.~(\ref{Exp-ConMom}), for $t=0$ the proper
partial waves for $\mbox{\boldmath $H$}_j$ are Legendre
polynomials (\ref{Def-d-1/2}), while $\mbox{\boldmath $E$}_j$ can
be represented as a sum of $-\mbox{\boldmath $H$}_j$ and an
addendum, expanded with respect to Gegenbauer polynomials
(\ref{Def-d-3/2}). For the sake of simplicity we take these
assignments also for $t\neq 0$. Moreover, in the small-$\eta$
kinematics the different SO(3) partial waves have (for complex
valued $j$) the same asymptotic behavior. Hence, in the following
we will relax the requirement of the complete polynomiality.

The SO(3) PW expansion (\ref{Exp-ConMom})  can be written in the
form of the transformation (\ref{Ske-in-conSpa}). One immediately
reads off that the SO(3) PW expansion of the skewness  matrix is
given by
\begin{eqnarray}
r_{j}(\eta|F) =
\sum_{J=J_{\rm min} \atop {\rm even}}^{j+1}
\frac{F_{jJ}}{F_{jj+1}} \,  \eta^{j+1-J}   \hat{d}_{\cal F}^{J}(\eta)\,.
\end{eqnarray}
We now construct three different  models for the skewness dependence,
which will be employed for fitting of small-$\Bx$ DVCS data.
Showing also the corresponding ``dual'' model we have:
\begin{itemize}
\item[\em  i.\phantom{ii}]
a leading SO(3) partial wave (l-PW), i.e., minimalist ``dual''
($\rho=0$) model,
\item[\em  ii.\phantom{i}]
a leading and next-leading SO(3) partial wave (nl-PW), i.e.,
the minimal ``dual'' ($\rho = 0, 2$) model, and
\item[\em  iii.]
a model-dependent resummation of SO(3) partial waves ($\Sigma$-PW).
\end{itemize}

{\ }

\noindent {\em i.~l-PW model:} Taking in the  expansion
(\ref{Exp-ConMom}) only the leading SO(3) PW  $J=j+1$ into account
and representing the Legendre polynomials $C_{j+1}^{1/2}(z)$ by
hypergeometric functions, we  write this model in the form:
\begin{eqnarray}
\label{Def-Mod-1-exa}
\mbox{\boldmath $H$}^{({\rm l})}_j(\eta,t) =
\mbox{\boldmath $r$}^{({\rm l})}_j(\eta) \mbox{\boldmath $q$}_j(t)\,,
\quad
\mbox{\boldmath $r$}^{({\rm l})}_j(\eta) =
\frac{\Gamma(1/2) \Gamma(j+2)}{2^{j+1} \Gamma(3/2+j)} \eta^{j+1}
{_2F_1} \left(\!{-j-1,j+2 \atop 1}\bigg|\frac{\eta-1}{2\eta}\!\right)
\left(\!
\begin{array}{cc}
1 & 0 \\
0 & 1
\end{array}
\!
\right),
\nonumber\\
\end{eqnarray}
where the forward moments $\mbox{\boldmath $q$}_j$ are specified
in Eqs.~(\ref{Ans-MomSigm}) and (\ref{Ans-MomG}). This is the most
restrictive model, and its skewness ratio is fixed by the
conformal value (\ref{r-ratio-con}), as claimed in
Refs.~\cite{ShuBieMarRys99, MarNocRysShuTeu08}.

{\ }

\noindent {\em ii.~nl-PW model:}
To build a flexible model we
include also the next-leading partial wave amplitude with $J=j-1$.
For simplicity, we express this PW in terms of the leading one,
where, however, the conformal spin is shifted by two units:
\begin{eqnarray}
\label{Def-Mod-2-exa}
\mbox{\boldmath $H$}^{({\rm nl})}_j(\eta,t)  &\!\!\!=\!\!\!&
\mbox{\boldmath $H$}^{({\rm l})}_j(\eta,t)
+ \eta^2\,  \theta(j \ge 3)\, \mbox{\boldmath $S$}
\mbox{\boldmath $H$}^{({\rm l})}_{j-2}(\eta,t)\,,
\qquad
\mbox{\boldmath $S$} = \left(
\begin{array}{cc}
{s}^{\rm sea}  & 0 \\
0 & {s}^{\rm G}
\end{array}
\right)
\,.
\end{eqnarray}
The $\eta^2$-proportional term, having a  pole at
$J=j-1=\alpha(t)$, is crucial for adjustment of the normalization
of the DVCS amplitude, controlled by the entries $s^{\rm sea}$ and
$s^{\rm G}$ in the matrix $\mbox{\boldmath $S$}$. For complex
conformal spin the pole is at $j=2+\alpha-1$ within the constraint
$\Re{\rm e} j \ge 3 $ and might thus be denoted as spurious.

{\ }

\noindent {\em iii.~$\Sigma$-PW model:}
It turns out that the sum (\ref{Exp-ConMom}) of
SO(3) partial waves with
physically-motivated effective ``Regge poles'',
$$F_{jJ}\propto \frac{1}{J-\alpha},$$ yields
after analytic continuation in $j$ the same small-$\eta$ asymptotic
as implemented in our toy skewness moments (\ref{Mod-GPD-Reg-Mom-1}),
see below Eq.~(\ref{Mod-GPD-Reg-Mom-2}). To have a
flexible skewness ratio, we take the l-PW
model  and add the difference of the toy
model (\ref{Mod-GPD-Reg-Mom-1})  and the
l-PW one%
\footnote{Note that this second term will die out in the forward
limit, and so it is invisible in the zero-skewness GPD. Of course,
such a trick can be used in any representation.
}:
\begin{eqnarray}
\label{Def-Mod-3-exa}
\mbox{\boldmath $H$}^{(\Sigma)}_j(\eta,t)  =
\mbox{\boldmath $H$}^{({\rm l})}_j(\eta,t)
+ {\mbox{\boldmath $S$}}_j(\eta,t|\alpha)\, \mbox{\boldmath $q$}_j(t)
 \,,
\end{eqnarray}
where the entries of the skewness matrix
\begin{eqnarray}
\label{Def-Mod-3-exa-Sma}
\mbox{\boldmath $S$}_j(\eta,t) =
\left(\!
\begin{array}{cc}
{s}^{\rm sea}
S_j(\eta|\alpha_{\rm sea}(t))
& 0 \\
0 & {s}^{\rm G}  S_j(\eta|\alpha_{\rm G}(t))
\end{array}
\!
\right)
\end{eqnarray}
are defined in terms of the skewness moments
(\ref{Mod-GPD-Reg-Mom-1}) and (\ref{Def-Mod-1-exa}):
\begin{eqnarray}
\label{Def-Mod-3-exa-S}
S_j(\eta|\alpha(t))=
\frac{\Gamma(7+\alpha)}{\Gamma(7+\alpha(t))} \left(
r_j^{\rm toy}(\eta|\alpha(t)) -r_j^{({\rm l})}(\eta)\right)
\,.
\end{eqnarray}
For convenience, we included an additional factor
$\Gamma(7+\alpha)/\Gamma(7+\alpha(t))$ that makes the
$t$-dependence of skewness effect more flat.

{\ }

A few comments are in order.
Skewness parameters are denoted by the same symbols, $s^{\rm sea}$
and $s^{\rm G}$, in both nl-PW and $\Sigma$-PW models; however,
their normalizations are not related in an obvious way.
Still, for both models positive (negative) values of these
parameters imply increase (decrease) of the normalization of the
DVCS amplitude at $t=0$, compared to the conformal ratio situation
(\ref{r-ratio-con}) of the l-PW model.
The existence of both these flexible models explicitly shows that
the claim of Ref.~\cite{ShuBieMarRys99} about conformal ratio
cannot be derived from a conformal PW expansion (or equivalently
from the integral transformation \cite{Shu99}). The assumption of
Ref.~\cite{MarNocRysShuTeu08}, namely, that singularities in the
complex conformal $j$ plane with $\Re{\rm e} j > \alpha-1$ (where
$\alpha \sim 1$) are absent would exclude the chosen {\em
representation}, including the constraint $\Re{\rm e} j > 3$, for the nl-PW model.
To weaken the constraint, we will change the Mellin-Barnes integral
representation for  CFFs (or GPDs) by a variable shift $j\to j^\prime-2$,
as it is done below in Eq.~(\ref{Res-ImReCFF-nl}).
Then the spurious pole will be moved to the ``right'' position
$J=j^\prime+1 = \alpha(t)$. This is in accordance with our assumption that
the leading poles are associated with the angular momentum $J$ rather than with the conformal spin
$j$. Moreover, to have a mathematically {\em consistent}
 representation of the $\Sigma$-PW model we will perform below the
analytic continuation by means of Carlson's theorem and then the spurious
right half-plane poles in the complex $j$ plane are absent anyway in the conformal moments
(\ref{Def-Con-Pro}). Utilizing the small-$\eta$ expansion, we will also illuminate that
spurious poles should be considered as an artifact of this expansion.

Finally, we specify the models for the GPD $E$. Since in the
small-$\eta$ region the SO(3) partial waves (\ref{Def-d-1/2}) and
(\ref{Def-d-3/2}) behave rather smoothly and they can be safely
approximated by one, we do not need to be careful about specific
choices of expansion polynomials. Hence, for simplicity, we can
assume that the conformal moments of $\mbox{\boldmath $E$}_j$ are
proportional to those of $\mbox{\boldmath $H$}_j$:
\begin{eqnarray}
\label{Ans-E}
\mbox{\boldmath $E$}_j = \left(
\begin{array}{c}
\left({\cal B}^{\rm sea}/N^{\rm sea}\right) H^{\rm sea}_j + \cdots\\
\left({\cal B}^{\rm G}/N^{\rm G}\right) H^{\rm G}_j
\end{array}
\right),
\end{eqnarray}
where the ellipsis stands for valence contributions. The
normalization of the GPD moment $\mbox{\boldmath $E$}_j$ is  for
$j=1$ given by the anomalous gravitomagnetic moments, see, e.g.,
Refs.~\cite{KobOku62,Ter99,Ter06}, of flavor singlet quarks
and gluons
$$
{\cal B}^{\Sigma}= {\cal B}^{\rm sea} + {\cal B}^{\rm val}\quad\mbox{and}\quad {\cal B}^{\rm G}\,.
$$
Ji's decomposition%
\footnote{A overview on the common spin decomposition schemes can
be found in Sect.~5 of Ref.~\cite{BurMilNow08}.}
of the proton spin \cite{Ji96} reads then
\begin{eqnarray}
\label{AngMomSumRul}
{\cal J}^{\rm \Sigma} + {\cal J}^{\rm G} =
\frac{1}{2}\,,
\quad
{\cal J}^{\rm \Sigma} =
\frac{1}{2}\left(N^{\rm \Sigma}+ {\cal B}^{\rm \Sigma}\right)\,,
\quad {\cal J}^{\rm G} = \frac{1}{2}\left(N^{\rm G}+
{\cal B}^{\rm G}\right).
\end{eqnarray}
As a consequence of both momentum (\ref{MomSumRul}) and angular
momentum conservation (\ref{AngMomSumRul}), the sum rule
\begin{eqnarray}
\label{SumRul-graMom}
{\cal B}^{\rm \Sigma}  + {\cal B}^{\rm G}  =0 \;,
\end{eqnarray}
states that the anomalous gravitomagnetic moment of the nucleon
vanishes. Note, however, that the residue of $\mbox{\boldmath $E$}_j$ at
the ``pomeron pole'' $j=\alpha-1$ depends besides the normalization ${\cal B}$
also on the functional form with respect to $j$.

The flavor decomposition of the anomalous gravitomagnetic moment
is an important phenomenological goal, which we will address below
in Sect.~(\ref{SubSecSec-gra}) within the ansatz (\ref{Ans-E}) in
an ad hoc model-dependent manner.   Let us set up a few scenarios.
Generic thoughts within simplified models, utilizing large $x$
counting rules, isospin symmetry, and the values of the nucleon
magnetic moment, suggest that the sum of valence $u$ and $d$
contributions provide an almost vanishing valence-like anomalous
gravitomagnetic moment \cite{KumMuePas08a}. With this taken literally, we
would have the scenario:
\begin{eqnarray}
\label{Sze-A}
{\cal B}^{\rm G}= -{\cal B}^{\rm sea}\,, \quad
{\cal B}^{\rm val} \equiv {\cal B}^{u_{\rm val}}+{\cal B}^{d_{\rm val}} = 0\,.
\end{eqnarray}
On the other hand, lattice simulations \cite{Broetal07,Hagetal07}
suggest the scenario
\begin{eqnarray}
\label{Sze-B}
{\cal B}^{\rm \Sigma} =  0 \quad \Rightarrow
\quad {\cal B}^{\rm val}  = -{\cal B}^{\rm sea}\,,
\quad{\cal B}^{\rm G}  = 0\,,
\end{eqnarray}
which would suggest that in our model the chromomagnetic
``pomeron'' should be absent. However, disconnected contributions,
which are related to a gluonic $t$-channel exchange,  are not taken into
account in present lattice measurements
\cite{Broetal07,Hagetal07}, and one might be tempted to
reinterpret the lattice results as valence dominated, i.e., ${\cal
B}^{\rm val}=0$, supporting our generic thoughts (\ref{Sze-A}).

\subsection{Numerical evaluation of CFFs}
\label{SecSub-NumEva}

As motivated above, we approximate the CFFs ${\cal F} =\{ {\cal
H},{\cal E} \} $ in the small-$\Bx$ region by their flavor singlet
part
\begin{eqnarray}
{\cal F}(\xi,t,{\cal Q}^2) \approx e^2_{\rm S} \;
{^{\rm S}\! {\cal F}}(\xi,t,{\cal Q}^2)
\,, \qquad e^2_{\rm S} = \frac{1}{N_f} \sum_{q=u,d,s\cdots} e_q^2\,,
\end{eqnarray}
where the (averaged) fractional squared charge $e^2_{\rm S}$ is
for $N_f$ active quarks. In the factorization formula for the CFFs
we will set both renormalization and factorization scales equal to
the photon virtuality ${\cal Q}$. Instead of the momentum fraction
representation, we utilize the conformal partial wave expansion
for complex-valued conformal spin. The flavor singlet part of our
signature even CFFs is then evaluated via a Mellin-Barnes integral
\begin{eqnarray}
\label{Res-ImReCFF}
\left\{\!
{{^{\rm S}\! {\cal H}} \atop {^{\rm S}\!{\cal E}}}
\!\right\}\!(\xi,t,{\cal Q}^2)\!
= \frac{1}{2i}\int_{c-i \infty}^{c+ i \infty}\!
dj\,\xi^{-j-1} \left[i +\tan\left(\frac{\pi j}{2}\right)\! \right]
\! \big[
{\mbox{\boldmath ${\mathbb C}$}}\otimes{\mbox{\boldmath ${\mathbb E}$}}
\big]_{j}(\alpha_s({\cal Q}),\alpha_s({\cal Q}_0))
\left\{\!
{\mbox{\boldmath $H$}}_{j} \atop  {\mbox{\boldmath $E$}}_{j}
\!\right\}
(\xi,t,{\cal Q}_0^2)\,.
\nonumber\\
\end{eqnarray}
The coefficient functions $\mbox{${\mathbb C}$}$ and the evolution
operator $\mbox{${\mathbb E}$}$ are evaluated in perturbation
theory. For instance, in LO (hand-bag) approximation the hard
scattering amplitude  reads
\begin{eqnarray}
\label{CPW-LO}
{\mbox{\boldmath ${\mathbb C}$}}_j
\stackrel{\rm LO}{=}
\frac{2^{j+1}\Gamma(j+5/2)}{\Gamma(3/2)\Gamma(j+3)}\; \left(1,0 \right) \,,
\end{eqnarray}
while the evolution operator is the same as in unpolarized DIS.
The radiative corrections to the coefficient functions
${
\mbox{\boldmath ${\mathbb C}$}}$
and the evolution
operator ${ \mbox{\boldmath${\mathbb E}$}}$
are presented%
\footnote{For clarity, the perturbative quantities which were
denoted in \cite{KumMuePas07} as ${\mbox{\boldmath ${\cal
C}$}}$ and ${\mbox{\boldmath ${\cal E}$}}$, are renamed in the
present paper into ${\mbox{\boldmath ${\mathbb C}$}}$ and
${\mbox{\boldmath ${\mathbb E}$}}$.} in Ref.~\cite{KumMuePas07}
for the standard minimal subtraction (\ms) scheme in NLO and
for a special conformal scheme (\cs) up to NNLO.  We add that
in the momentum fraction representation the corresponding
expressions are known for the \ms\ scheme in NLO approximation%
\footnote{Since even at NLO a diagrammatic calculation of
evolution kernels has not been performed yet, except for the two
simplest ones in the quark sector
\cite{Sar82,DitRad81,MikRad85,MikVla08}, we do not expect that the
NNLO results in the \ms\ scheme will become available in near
future. }.

A few comments are in order. In the \cs~scheme conformal symmetry
is manifest, up to the breaking by the trace anomaly, also beyond
LO. Thus, the NNLO corrections can be obtained by conformal
mapping from the corresponding forward quantities, elaborated in
DIS; for details we refer to Ref.~\cite{KumMuePas07} and
references therein (for integral conformal spin see also
Refs.~\cite{MelMuePas02,BraKorMue03}). In the \cs\ scheme the
convolution $\left[{\mbox{\boldmath ${\mathbb
C}$}}\otimes{\mbox{\boldmath ${\mathbb E}$}}\right]_{j}$ reduces
to a simple multiplication, while in the \ms\ scheme an additional
summation over the conformal spin must be included.

In the evolution operator we resum only the leading logs and
perturbatively expand the non-leading ones. In the small-$\xi$
region the perturbative expansion of the universal evolution
operator is getting unstable, so that the difference to an
evolution operator in which also non-leading logs are resummed is
of some numerical importance. This difference is implicitly
absorbed by a reparameterization of sea quark and gluon PDFs or
GPDs at the given input scale.

The  continuation of coefficient functions and anomalous
dimensions from integral to complex-valued $j$ is, as in DIS,
straightforward. This is also the case for zero-skewness GPDs,
already given in Eqs.~(\ref{Ans-MomSigm}, \ref{Ans-MomG}) as
ratios of $\Gamma$ functions. The continuation of the skewness
moments, as polynomials of $\eta$ is more intricate. It is
required that these moments obey certain bounds for $j\to \infty$
with $|{\rm arg}(j)| \le \pi/2$, see Ref.~\cite{MueSch05}. Then
Carlson's theorem tells us that their analytic continuation is
unique. The conformal moments (\ref{Def-Mod-1-exa}) of our l-PW
model, expressed by hypergeometric functions $_2F_1$, and their
descendants, i.e., the hypergeometric functions $_3F_2$  appearing
in the skewness moments (\ref{Mod-GPD-Reg-Mom-1}) with argument
$(\eta-1)/2\eta <0$, satisfy the required condition. However,
their $\eta$-symmetrized counterparts, with argument
$(1+\eta)/2\eta >0$, are problematic. Such functions are now
evaluated on the branch cut in the complex $\eta$ plane.  The
correct treatment is to take the discontinuity over the cut, and
to add a term that restores polynomiality for odd $j$:
\begin{eqnarray}
\label{Def-Con-Pro}
&&\!\!\!\!\!\!\!
{_3F_2}\!\left(\!{-j, 3+j, 2-\alpha \atop 2, 3 - \alpha}\bigg|
\frac{1+\eta}{2 \eta }\!\right) \Rightarrow \hfill
\\
&&\!\!\!\!\!\!\!
\frac{
{_3F_2}\left(\cdots \bigg|\frac{1+\eta}{2 \eta } + i \epsilon\right)
-{_3F_2}\left(\cdots \bigg|\frac{1+\eta}{2 \eta }-i \epsilon\right)
}{2 i \sin(\pi j)}
+\left(\!\frac{2 \eta }{1+\eta}\!\right)^{2-\alpha}
\frac{\Gamma(1+j)\Gamma(3-\alpha)\Gamma(1+\alpha +j)}
     {\Gamma(3+j)\Gamma(\alpha) \Gamma (3-\alpha+j)} \;.
\nonumber
\end{eqnarray}
In the continuation procedure for odd $j$ moments the factor
$(-\eta)^j$ is replaced by $-\eta^j$. The Mellin moments, obtained
by this continuation procedure, do {\em not} possess singularities
for $\Re{\rm e}{j} > \alpha-1$ with $\alpha \sim 1$ and are
unproblematic for $\alpha < 2$. It has been already numerically
checked for $ \alpha < 1 $  in Ref.~\cite{MueSch05} that our toy
model in the conformal Mellin space is equivalent to its momentum
fraction space representation. We convinced ourselves that for
$\eta \neq \xi $ this holds true also for $ \alpha \sim 1.2 $. As
already mentioned, the singular behavior of the toy GPD at the
cross-over line is absent in our $\Sigma$-PW model.

For the numerical evaluation of the Mellin-Barnes integral
(\ref{Res-ImReCFF}) we choose as integration path a straight line
segment, parallel to the imaginary axis. Intercept $c$ of this
line and the real axis is taken to be $c\approx 0.35$, so that the
contour lies to the right of the ``pomeron pole'' and to the left of
the poles of tangent function in (\ref{Res-ImReCFF}). For the
nl-PW model the polynomiality condition of the
$\eta^2$-proportional term in Eq.~(\ref{Def-Mod-2-exa}) requires
us to take a separate Mellin-Barnes integral where the integration
path is shifted by two units to the right, i.e., the intercept of
the integration path is now $c+2$. Shifting now the integration
variable, $j\to j-2$, we combine both pieces into a single
Mellin-Barnes integral, schematically written as
\begin{eqnarray}
\label{Res-ImReCFF-nl}
\left\{\! {{^{\rm S}\! {\cal H}} \atop {^{\rm S}\!{\cal E}}} \!\right\}
= \frac{1}{2i}\int_{c-i \infty}^{c+ i \infty}\!
dj\,\xi^{-j-1} \left[i +\tan\left(\frac{\pi j}{2}\right) \right]
\left[
\left[
{\mbox{\boldmath ${\mathbb C}$}}\otimes{\mbox{\boldmath ${\mathbb E}$}}
\right]_{j}
+
\left[
{\mbox{\boldmath ${\mathbb C}$}}\otimes{\mbox{\boldmath ${\mathbb E}$}}
\right]_{j+2}
{\mbox{\boldmath ${S}$}}
\right]
\left\{\! {\mbox{\boldmath $H$}}_{j}^{(l)} \atop
{\mbox{\boldmath $E$}}_{j}^{(l)} \!\right\}\,,
\end{eqnarray}
for our nl-PW model. Obviously, the $\eta^2$-suppressed term in
integral conformal moments contributes to the leading Regge
behavior and influences the value of the residue function. We emphasize again
that the spurious pole in the nl-PW model is now moved  to the
``right'' position $J = j+1=\alpha$ and that the new
integration variable can now be viewed as the (shifted) $t$-channel angular momentum, i.e.,
$j=J-1$ .

For the small-$\xi$ kinematics, we are interested in, we can speed
up numerics by expanding  the conformal GPD moments in the
vicinity of $\eta=0$. For l-PW model (\ref{Def-Mod-1-exa}), using
well-known expansion of Legendre functions \cite{AbrSte} one gets
\begin{eqnarray}
\label{Def-Mod-1-app}
\mbox{\boldmath $r$}^{({\rm l})}_j(\eta) =
\left(\!
\begin{array}{cc}
1 & 0 \\
0 & 1
\end{array}
\!
\right) \left[1 + O(\eta^2)\right]  + O(\eta^{2j+3})
\quad
\Rightarrow
\quad
\mbox{\boldmath $H$}^{({\rm l})}_j(\eta,t) =
\mbox{\boldmath $q$}_j(t)\,.
\end{eqnarray}
With  our choice for the integration path, i.e., $\Re{\rm e} j =c
> 0$, the $O(\eta^{2j+3})$ term is negligibly small and the
conformal GPD moments reduce to the Mellin moments of
zero-skewness GPDs. The same approximation we can use in
Mellin-Barnes integral (\ref{Res-ImReCFF-nl}) for the nl-PW model.
The small-$\eta$ expansion of the skewness moments
(\ref{Mod-GPD-Reg-Mom-1}) for the $\Sigma$-PW model, for complex
valued $j$, obtained within the prescription (\ref{Def-Con-Pro}),
was worked out in Ref.~\cite{MueSch05}:
\begin{eqnarray}
\label{Mod-GPD-Reg-Mom-2}
r^{(\rm toy)}_j \! =1 -
\left(\frac{\eta}{2}\right)^{j+1-\alpha} \frac{\Gamma(1/2)
\Gamma(1+j) }{\Gamma(3/2+j) }\,
\frac{ \Gamma(1-\alpha ) \Gamma (1+\alpha+j)}{4^{\alpha} \Gamma (\alpha )
\Gamma (1-\alpha +j) }\, \frac{2 \left[ 1+  O(\eta^2) \right]
}{1+\tan \left(\frac{j \pi }{2}\right) \tan
\left(\frac{\pi  \alpha }{2}\right)} +O(\eta^{2j+3})\,.\!
\nonumber\\
\end{eqnarray}
This expression has a series of ``conformal sibling poles'' in the
left half-plane at $j= -\alpha-1 - k$ with $k=0,1,2,\cdots$. The
series of spurious poles in the right half-plane at $j= \alpha+1 + 2 k$
with $k=0,1,2,\cdots$ are absent in the exact result, and they
appear here as an artifact of the small-$\eta$ expansion. The
function $r_j(\eta)$ has also a zero at $j=\alpha-1$ that cancels
the ``pomeron pole'' in the Mellin moment; however, the
corresponding CFF still grows as $\xi^{-\alpha}$ at small $\xi$.
We have numerically checked that the exact and approximate results
agree well even for larger values of $\eta \sim 0.2$; an example
for $\alpha=1/2$ was presented in Ref.~\cite{MueSch05}. In contrast
to ``Regge poles'', the ``conformal siblings'' will move to the right
with increasing $-t$. Fortunately, as long as the condition
\begin{eqnarray}
\label{Mod-Con}
-t < \frac{2 + \alpha}{\alpha^\prime}
\end{eqnarray}
is satisfied, it is
guaranteed that they all lie to the left of the lowest integral
value $j=1$. All of the `conformal' trajectories contribute to the
residuum of the leading ``Regge pole'' at $J=\alpha(t)$ and this
provides the possibility to adjust the normalization of CFFs.

Plugging the approximations (\ref{Def-Mod-1-app}) and
(\ref{Mod-GPD-Reg-Mom-2}) into the $S$-function
(\ref{Def-Mod-3-exa-S}) of the $\Sigma$-PW model, provides us with
the expression convenient for numerical treatment
\begin{eqnarray}
\label{Def-SkeFunMel} S_j(\eta|\alpha(t)) =
-
\left(\frac{\eta}{2}\right)^{j+1-\alpha(t)}
\frac{
\sqrt{\pi}\Gamma(1+j) }{\Gamma(3/2+j)
}\,
\frac{
\Gamma(1-\alpha(t) )\Gamma (1+\alpha(t)+j)
}{
4^{\alpha(t)}\Gamma (\alpha(t) )\Gamma (1-\alpha(t) +j) }\,
\frac{2\Gamma(7+\alpha)/\Gamma(7+\alpha(t))}{1+\tan
\left(
\frac{j \pi }{2}\right) \tan \left(\frac{\pi  \alpha(t) }{2}
\right)}.
\end{eqnarray}
Note that the factor $\xi^{j+1-\alpha}$ for $\eta=\xi$  is
partially cancelled in the Mellin-Barnes integral (\ref{Res-ImReCFF-nl}) by
$\xi^{-j-1}$, which leads to the $\xi^{-\alpha}$  behavior  of
CFFs. Hence, the Regge behavior in conformal space is not
entirely related to a $j=\alpha-1$ pole.
This is natural from our point of view that
the ``Regge'' behavior is associated with a pole in the complex
$J$ plane, and, consequently, the residue is then expanded in terms of conformal
PWs.

Finally, we have numerically checked that the small-$\eta$
approximation of the skewness moments works rather well, i.e., for
$\xi < 0.01$ ($\xi \sim 0.1$) we find a per mil (below 1\%) effect
for the $\Sigma$-PW model.   The resulting CFFs satisfy also the
``dispersion relation'' (\ref{Def-DisRel}) for $ 0 < \xi < 0.01$ ($\xi
\sim 0.1$) on the per mil (4\%) level.   As expected, the medium-
and large-$\xi$ regions are unimportant in the dispersion integral
for the real part at small $\xi$. The small-$\eta$ approximation
of the SO(3) PWs has been often considered, e.g., in
Ref.~\cite{KumMuePas07}, and is even more unproblematic
than the approximation of the $\Sigma$-PW model.

\section{GPD interpretation of DVCS data in collider kinematics}
\label{Sec-Fits}

We present now a detailed model study of the small-$\Bx$ DVCS
measurements of the H1 and ZEUS Collaborations
\cite{Aktas:2005ty,Aaretal07,Chekanov:2003ya,Sch07} by means of
least square fits at LO, NLO in the $\overline{\rm MS}$ and
$\overline{\rm CS}$, and at NNLO in the $\overline{\rm CS}$
scheme. The fitting procedure is described in
Sect.~\ref{SubSec-FitStrPar}. Thereby, we aim for a partonic or
GPD interpretation; however, the nature of the DVCS process
precludes a coverage of the whole space of possible GPD models
and, therefore, we are unable to determine the subspace of models
that correctly describe the DVCS data. Instead, we use three
representative models defined in Sect.~\ref{subsec-Fits} in an
attempt to illuminate all facets of the problem. In
Sect.~\ref{SubSecSec-LO} we give specific emphasis to LO fits,
since the outcome might be conveniently employed in studies of
fixed target experiments. GPD reparameterization effects due to
radiative corrections are considered in Sect.~\ref{SubSecSec-NLO}.
In Sect.~\ref{SecSubSub-TraDis} we extract the transverse
distribution of partons from our fits.  Finally, we address in
Sect.~\ref{SubSecSec-gra} the question of whether measurements of
beam charge asymmetry allow one to access the GPD $E$ and,
consequently, to get a handle on the chromomagnetic ``pomeron''
and the anomalous gravitomagnetic moment. The lessons from the
fits are listed in Sect.~\ref{SubSec-Les}.

\subsection{Fitting strategies and parameters}
\label{SubSec-FitStrPar}

Our numerical studies are based on the twist-two approximation of
the DVCS cross section (\ref{Def-CroSec}) and BCA (\ref{Def-Asy})
without gluon transversity, the models from
Sect.~\ref{subsec-Fits}, the Mellin-Barnes integral
(\ref{Res-ImReCFF}), and on the least square fitting routine
MINUIT \cite{JamRos75}. We fit to the
DVCS cross section and to the DIS $F_{2}$ structure function data, where
we add statistical and
systematical errors in quadrature and look for the values of GPD
model parameters that minimize the $\chi^2$ function. For
instance, for the DVCS cross section $\sigma$ we have
\begin{eqnarray}
\chi^2 =\sum_{d}
\frac{(\sigma^{\rm exp}_d-\sigma_d^{\rm the})^2
}{
(\delta \sigma_d)^2}\,,
\quad
\delta \sigma_d =\sqrt{
(\delta_{\rm sys} \sigma_d)^2 + (\delta_{\rm stat} \sigma_d)^2
}\,,
\end{eqnarray}
where $d$ enumerates the experimental data points. The quality of
fits is assessed by the $\chi^2$ value divided by the number of degrees of
freedom (d.o.f.) and by probability of such and larger $\chi^2$
values. As bad fits we consider those where $\chi^2/{\rm d.o.f}$
is considerably larger than one, or, more precisely, where its
probability is smaller than 0.2. We will also separately present the
$\chi^2$ values contributed by $t$-, $W$- and ${\cal Q}^2$-dependence. Since
there is no one-to-one correspondence between individual fit
parameters and the particular kinematic parameters of the CFF, we
cannot determine the exact number of the d.o.f. Therefore, we
simply give then $\chi^{2}$ per number of data points (n.o.p.).
The difference is small for large number of points; so this should
provide good idea about the quality of the description of
dependence on the individual variables. An error analysis of the
resulting parameters is an intricate task, which we postpone.

We set the input scale, at which the GPD models and parameters are
specified, to
\begin{eqnarray}
\label{Set-InpSca}
{\cal Q}_0^2 = 4\,\GeV^2\,.
\end{eqnarray}
We consider then the charm quark as massless and set $N_f=4$ and
$e^2_{\rm S} = 5/18$.  Numerical effects of $b$ quarks, known to
be small in DIS, will be neglected here.
Alternatively, the production of heavy quarks might be described
within perturbation theory \cite{Nor03}, worked out at NLO \cite{Nor00}.
The flavor scheme choices we consider to be foremostly related to different
partonic interpretations. The value of the QCD running coupling is
specified as
\begin{eqnarray}
\frac{\alpha_s({\cal Q}_0=\sqrt{2.5}\,\GeV)}{2\pi} =
\left\{
\begin{array}{c}
  0.0606 \\
  0.0518 \\
  0.0488
\end{array}
\right\}
\quad
\mbox{for}
\quad
\left\{
\begin{array}{c}
  {\rm LO} \\
   {\rm NLO} \\
   {\rm NNLO}
\end{array}
\right\}
\end{eqnarray}
and corresponds to its phenomenological value $\alpha_s(M_{Z^0}=
91.18\, \GeV)=0.114$ at the reference scale $M_{Z^0}$, using the
standard evolution prescription \cite{CheKuhSte00}. However, we
perform forward evolution over the $\mathcal{Q}^2$ intervals
of interest keeping a fixed number of four active quarks.

\begin{figure}[t]
\centerline{\includegraphics[scale=0.6]{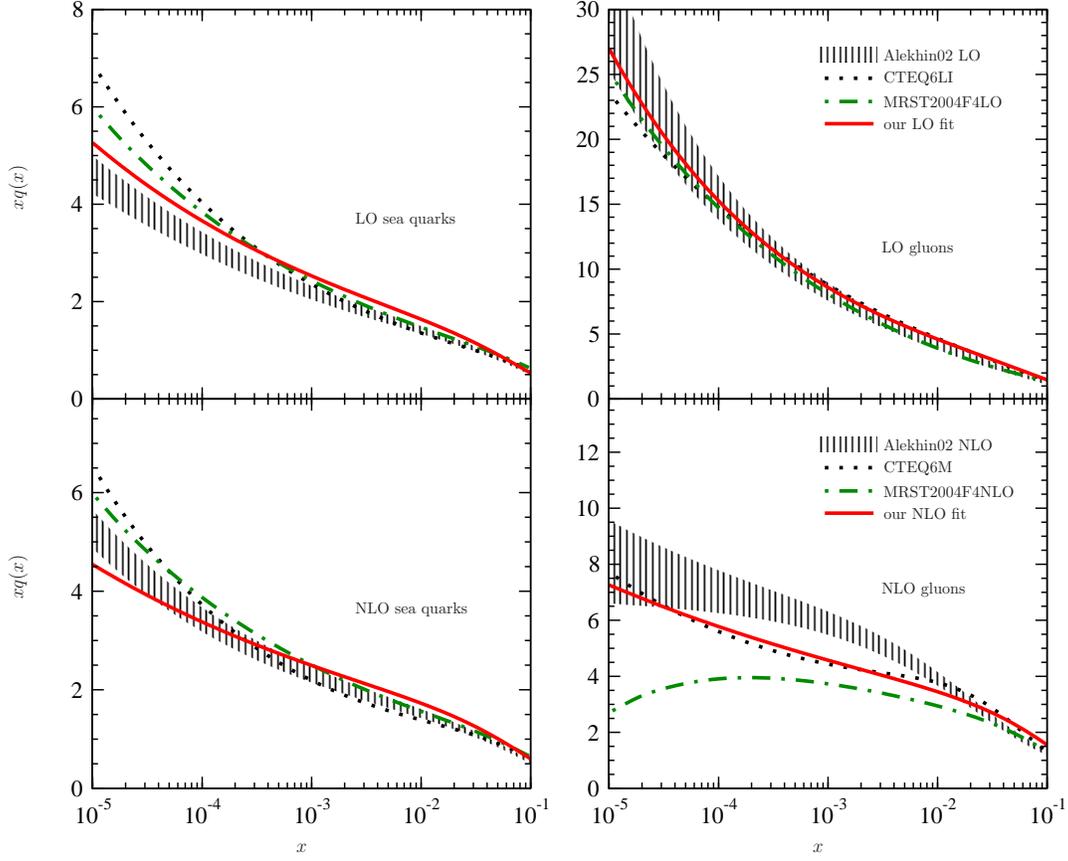}}
\caption{\label{FigFit-DIS} Parton densities at LO (up) and NLO (down)
are shown for quarks (left) and gluons (right) at the input scale ${\cal Q}_0^2 =
4\,\GeV^2$ (solid). For comparison we also show Alekhin's
para\-meterization \cite{Ale02} with errors  (vertically dashed band) and
some other standard PDFs \cite{MarStiTho06,Pumetal02} (dotted, dash-dotted). }
\end{figure}

Let us recall that our GPD $H$ models are parameterized by the
averaged  momentum fraction $N$, ``pomeron trajectory''
$\alpha(t)$, large-$j$ parameter $\beta$, skewness parameter $s$,
and one parameter that controls the residue function
$\mbox{\boldmath $\beta$}(t)$, e.g., the $\Sigma$-PW model reads
\begin{eqnarray}
H^{\rm I}_j(\eta)=  N^{\rm I} \left[1 + s^{\rm I}\, S_j(\eta)\right]
\frac{B(1-\alpha^{\rm I}+j,\beta^{\rm I} +1)}{B(2-\alpha^{\rm I},\beta^{\rm I}+1)}\,
\frac{1-\alpha^{\rm I}+j}{1-\alpha^{\rm I}(t)+j}\, \mbox{\boldmath $\beta$}^{\rm I}(t)
\end{eqnarray}
for ${\rm I}=\{{\rm sea}, {\rm G}\}$.
In our previous studies \cite{KumMuePas07} we relied on the ad hoc
l-PW model ($s^{\rm I}=0$) and performed a simultaneous fit to DIS and DVCS
data, where the momentum sum rule was not imposed. We utilize now
this constraint to fix the gluon normalization,
\begin{eqnarray}
\label{DIS-Par0}
N^{\rm G} =1-N^{\rm sea}-N^{\rm val}\,, \quad N^{\rm val} = 0.4\,,
\end{eqnarray}
where the momentum fraction for valence quarks $N^{\rm val}$
is compatible with values
obtained from standard PDF fits. Our flexible GPD models
could be fitted just to DVCS data, where PDF normalization and
``Regge intercepts''
\begin{eqnarray}
\label{DIS-Par1}
N^{\rm sea}\,,\quad \alpha^{\rm sea}\,,\quad  \alpha^{\rm G}\,
\end{eqnarray}
at the input scale would be taken from some of the standard parameterizations%
\footnote{Due to the small-$\Bx$ collider kinematics, we
neglect the ``Reggeon'' contributions to the flavor singlet part.}.
However, instead of this, we mostly perform \emph{two-step} fits.
First we fit to the DIS data to extract the PDF parameters
(\ref{DIS-Par1}). There the $\beta$ parameters for sea quark and
gluon PDFs are taken as
\begin{eqnarray}
\label{DIS-Par2}
\beta^{\rm sea} = 8,\,\quad  \beta^{\rm G} =6\,,
\end{eqnarray}
which is slightly larger than their canonical values from counting
rules, accounting thus for the increase with resolution scale.
Inclusion of DIS fit in our procedure is convenient because it
ensures identical scheme conventions and approximations for the
evolution operator in DIS and DVCS. The DIS data used come from H1
data set \cite{Aidetal96} for the DIS structure function $F_2$ in
the region $$2.5\,\GeV^2 \le Q^2 \le 90\,\GeV^2\;.$$ The choice of
these 85 data points is sufficient for our studies and at the input
scale (\ref{Set-InpSca}) our resulting PDFs are consistent with
standard parameterizations, as demonstrated at LO and NLO accuracy
in Fig.~\ref{FigFit-DIS}.
Note that the PDFs in the regions $10^{-5}\le x
\lesssim 10^{-4}$ and $10^{-2}\lesssim  x$ (mainly) arise from
extrapolation within our functional pomeron-like ansatz and that
we use our conventions, e.g.,  perturbative NLO corrections
are consequently expanded in powers of $\alpha_s$ \cite{KumMuePas07}.

The remaining free model parameters are the skewness parameters,
the cut-off masses and power behavior in the residue function
(\ref{Ans-bet-dip}), [or exponential slope parameters in
(\ref{Ans-bet-exp})], and the ``Regge slopes''
\begin{equation}
\label{Fit-ParSet}
s^{\rm sea}\,,\quad s^{\rm G}\,;
\qquad\qquad
M^{\rm sea}\,,\;\; p^{\rm sea}\;\; [B^{\rm sea}] \,,
\quad
M^{\rm G}\,,\;\;p^{\rm G}\;\; [B^{\rm G}]\,;
\qquad\qquad
\alpha^{\prime\, \rm sea} \,,\quad \alpha^{\prime\, \rm G}\,,
\end{equation}
which control the normalization, $t$-dependence, and the shrinkage
of the diffractive forward peak in the DVCS amplitude,
respectively. These parameters are to be determined by fits to
DVCS data in the second step of the procedure. However, it turns
out that from our fits we cannot simultaneously pin down the
cut-off masses and powers $p$ in the ansaetze (\ref{Ans-t-dip}),
so we fix the latter.
Large $-t$ counting rules would suggest to take for sea quarks
$p^{\rm sea}=4$ and for gluons $p^{\rm G}=3$. To have the possibility
of direct comparison of the
fitting results with the characteristic size of the
proton, given by the cut-off mass in the dipole parameterization
of the Sachs form factors, we choose a dipole ansatz:
\begin{equation}
\label{Fit-ParSet-1}
p^{\rm sea}=p^{\rm G} =2\,.
\end{equation}

Moreover, since gluons are not directly accessible and a full
separation of quarks and gluons through the evolution effects is
not yet possible, we must further reduce the parameter space. In
our previous ad hoc model studies we learned that fitting routine
tends to use the gluonic $t$-slope to adjust the normalization of
the total DVCS amplitude. Here we get rid of this unwanted
complication by fixing the gluonic cut-off mass (or slope) and
taking its value from the analysis of elastic $J/\psi$ production.
According to Ref.~\cite{FraKoeStr95}, this process is dominated by
the two-gluon $t$-channel exchange and the charm quark mass
already provides an internal hard scale that translates to an
effective $J/\psi$ photoproduction scale of ${\cal Q}_{\rm
eff}^2\approx 3\, \GeV^2$. Therefore, one may even view the
$J/\psi$ photoproduction as a hard process with access to the
gluon GPD \cite{FraStrWei05}. From the differential cross section
measurements \cite{Adletal00,Cheetal02,Cheetal04,Aktetal05} the
``pomeron trajectory'' (here defined with respect to $W$
rather than $\xi$) and the residual $t$-slope were extracted by a
fit with an exponential ansatz \cite{Aktetal05}:
\begin{eqnarray}
\label{Par-fix-Gslo}
\alpha(t) &\!\!\!=\!\!\!&
1.224\pm 0.010\pm 0.012 + (0.164\pm 0.028\pm 0.030)\,t/{\rm GeV}^2\,,
\nonumber
\\
b &\!\!\!=\!\!\!& 2 B =
\left(4.630 \pm 0.060^{+0.043}_{-0.163}\right)/{\rm GeV}^2\,.
\end{eqnarray}
Also, the measurements strongly favor an exponential
$t$-dependence of the differential cross section, where the
$t$-slope $b$ is quoted in Eq.~(\ref{Par-fix-Gslo}) for $W=90\,\GeV$.
Still, at the GPD level%
\footnote{Whether the measurement of the longitudinal part of the
$J/\psi$ electroproduction  cross section, known to much smaller
accuracy than the photoproduction one \cite{Aktetal05}, is then
consistent with the collinear factorization approach, elaborated
in Ref.~\cite{IvaSchSzyKra04} to NLO,  is to our best knowledge
not yet investigated.}, we consider it more natural to take a
power-like form for the residual $t$-dependence, as argued in
Sect.~\ref{SurSec-GPD-mod}. Therefore, the dipole ansatz
will be used for the majority of our fits, with some fits
using exponential ansatz performed for comparison. The value of
the ``Regge slope'' parameter $\alpha^\prime$  is not fully pinned
down. The measurements of the ZEUS Collaboration \cite{Cheetal02}
give a slightly lower mean value as quoted in
Eq.~(\ref{Par-fix-Gslo}) and the electroproduction measurements
\cite{Cheetal04,Aktetal05}  indicate that $\alpha^\prime$ might be
compatible with zero at a larger resolution scale. The
photoproduction fit (\ref{Par-fix-Gslo}) serves us to fix the
slope of the gluonic residue function, while for the ``Regge
slope'' at the input scale (\ref{Set-InpSca}) we will consider two
values. To maximize model differences, we choose the combinations,
\begin{eqnarray}
\label{Ext-Glu-Par}
\{ M^{\rm G}= \sqrt{0.7}\, {\rm GeV}, \;
\alpha^\prime_{\rm G}=0.15/{\GeV}^2 \}
\qquad
\mbox{and}
\qquad
\{B^{\rm G} = 2.32/{\GeV}^2,\;  \alpha^\prime_{\rm G}=0\}\,,
\end{eqnarray}
for dipole and exponential ansatz, respectively. Here, the values of
$M^{\rm G}$ and $B^{\rm G}$ correspond to each other  for
$\alpha^\prime_{\rm G}=0$.

Finally,  we equate the quark and gluon ``Regge slope''
parameters, $\alpha^\prime_{\rm sea} = \alpha^\prime_{\rm G}$.
Then our parameter sets for the DVCS fits read
\begin{align}
  \{s^{\rm sea},\, s^{\rm G},\, M^{\rm sea}\}\,,
  &\quad \text{with fixed} \quad M^{\rm G} = \sqrt{0.7}\,\GeV,\;
  \alpha^\prime_{\rm sea} = \alpha^\prime_{\rm G} =0.15/\GeV^2\,,
\label{Ans-t-dip}
\\
\intertext{for the dipole, and}
  \{s^{\rm sea},\, s^{\rm G},\, B^{\rm sea}\}\,,
 &\quad \text{with fixed} \quad  B^{\rm G} = 2.32/{\rm GeV}^2,\;
 \alpha^\prime_{\rm sea} = \alpha^\prime_{\rm G} =0\,,
\label{Ans-t-exp}
\end{align}
for exponential ansatz.

{\ }

The small-$\Bx$ DVCS cross section measurements of the H1 and ZEUS
Collaborations are published in
Refs.~\cite{Aktas:2005ty,Aaretal07} and \cite{Chekanov:2003ya},
respectively. The kinematics covers the intervals
$$
3\, \GeV^2  \lesssim  {\cal Q}^2 \lesssim 80\, \GeV^2\,,
\quad
45\, \GeV \lesssim  W  \lesssim 145\, \GeV\,,
\quad
0.1\, \GeV^2 \lesssim -t \lesssim 0.8\, \GeV^2\,.
$$
We do not include in our fits one ZEUS data set%
\footnote{The excluded data is for the cross section, integrated
over $t$, versus $W$ for fixed ${\cal Q}^2=9.6\, \GeV^2$.
Including this set and the analogous one from H1 would unavoidably
increase the $\chi^2$ value of our fit. These two data sets are
mutually compatible; however, we give here preference to H1 data
set, since it possesses a smoother $W$-dependence.}, H1 data
versus $\Bx$, and model-dependent extractions of skewness ratio
and $t$-slope given by those Collaborations. Note that the more
recent H1 measurement \cite{Aaretal07}, providing a larger data
set on the $t$-dependence, was not used in our previous GPD study
\cite{KumMuePas07}. Altogether, the H1 and ZEUS measurements
provide us with 101 DVCS data points.

\begin{figure}[t]
\begin{center}
\includegraphics[scale=1.0]{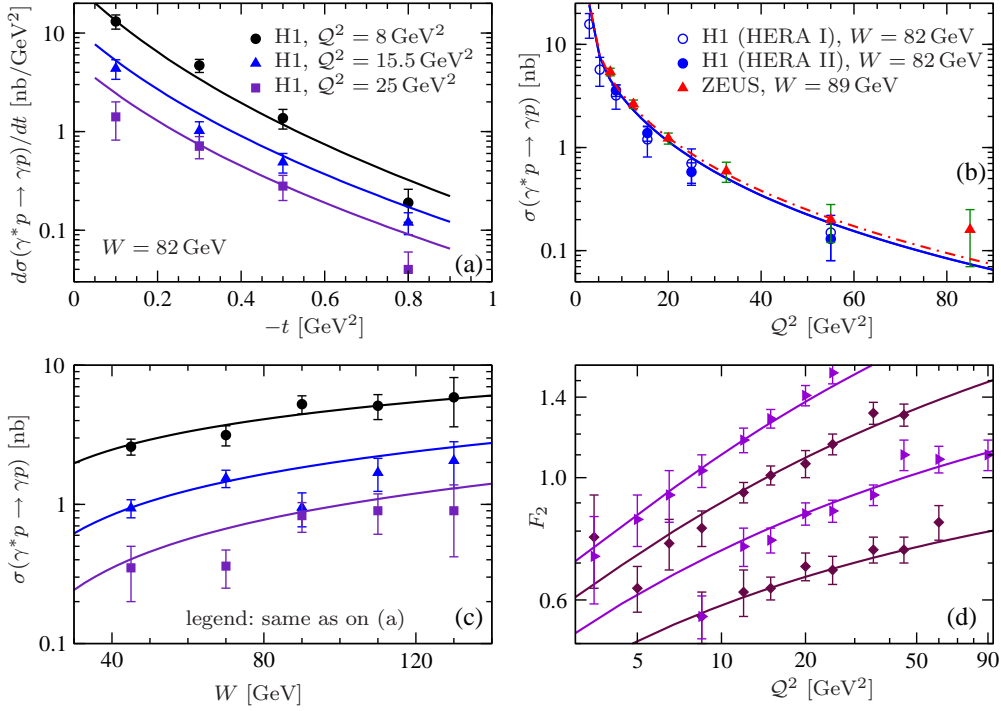}
\end{center}
\caption{\small Two-step fit at LO to DVCS
\cite{Aktas:2005ty,Aaretal07,Chekanov:2003ya} and DIS
\cite{Aidetal96} data with the $\Sigma$-PW model and the
dipole ansatz (\ref{Ans-t-dip}):
(a) differential DVCS cross section versus $-t$ for three values
of ${\cal Q}^2$  at $W=82\, \GeV$ \cite{Aaretal07},
(b) total DVCS cross section versus ${\cal Q}^2$ at
$W=82\, \GeV$ (circles, solid) \cite{Aktas:2005ty,Aaretal07}
and at $W=89\, \GeV$ (triangles, dash-dotted) \cite{Chekanov:2003ya},
(c) total DVCS cross section versus $W$ at the same
${\cal Q}^2$ values as on (a) \cite{Aaretal07}, and
(d) DIS structure function $F_2(\Bx,{\cal Q}^2)$
versus ${\cal Q}^2$ for
$\Bx=\{8\cdot 10^{-3}, 3.2\cdot 10^{-3}, 1.3\cdot
10^{-3}, 5\cdot 10^{-4}\}$ \cite{Aidetal96}.
} \label{fig:LOfitres}
\end{figure}
We give a LO fitting example for the $\Sigma$-PW model in
Fig.~\ref{fig:LOfitres}. Here the first panel shows the
$t$-dependence, the second and third the $W$- and ${\cal
Q}^2$-dependence, respectively, and the fourth one our fit to the
DIS structure function $F_2$.

\subsection{Leading order fits to DVCS cross section measurements}
\label{SubSecSec-LO}

To simplify our analysis of DVCS cross section measurements we
will neglect the helicity non-conserved CFF ${\cal E}$.  According
to the formula (\ref{Def-CroSec}) for the DVCS cross section, this
is justified if the condition
$$
\frac{-t}{4 M^2} |{\cal E}|^2 \ll |{\cal H}|^2
$$
holds true. Since the average value of $-t$ is $\sim 0.2\,
\GeV^2$, there will be at the most a few percent contamination by
${\cal E}$ as long as the modulus of this CFF is comparable to
$|{\cal H}|$. In the case that $|{\cal E}|$ turns out to be
considerably larger than $|{\cal H}|$, our findings, or better to
say our interpretation, will become questionable.

In Sect.~\ref{SurSec-GPD-mod} we already mentioned that a
reasonable description of DVCS data at LO accuracy in the
small-$\Bx$ region could not be achieved in previous ad hoc GPD
model studies \cite{FreMcD01a,KumMuePas07,GuzTec08}.  On the
other hand, one expects that this should be possible using more
flexible GPD models.
\begin{table}[th]
\centering
\begin{tabular}{|cl|ccccc|}
\hline
model & $\alpha_{s}$ & $\chi^2/{\rm d.o.f.}$ DIS & $\chi^2/{\rm d.o.f.}$ DVCS  &
$\chi^2_t/{\rm n.o.p.}$ & $\chi^2_W/{\rm n.o.p.}$ & $\chi^2_{{\cal Q}^2}/{\rm n.o.p.} $ \\
\hline\hline
\phantom{n}l, dipole\phantom{.} &   LO      &  49.7/82 & {\bf 280./100} & {\bf 181./56} & {\bf 63.6/29} & {\bf 36.2/16}  \\
\phantom{n}l, exp.\phantom{ole} &   LO        &  49.7/82 & {\bf 316./100} & {\bf 192./56} & {\bf 79./29} & {\bf 44.9/16}  \\
 nl, dipole\phantom{.} &   LO      &  49.7/82 & 95.9/98 & 53.2/56 & 27./29 & 15.8/16  \\
 nl, exp.\phantom{ole}    &  LO      &  49.7/82 & 97.9/98 & 49.1/56 & 31.2/29 & 17.7/16  \\
$\Sigma$, dipole\phantom{.} &  LO  &  49.7/82 & 101./98 & 57.7/56 & 27.4/29 & 16./16  \\
$\Sigma$, exp.\phantom{ole} &  LO    &  49.7/82 & 102./98 &  51./56 & 32.3/29  & 18.6/16 \\
\hline
l, dipole\phantom{.} &   LO      & \multicolumn{2}{c}{\bf 321./182} & {\bf 189./56} & {\bf 51.1/29} & {\bf 27.9/16} \\
\hline
\end{tabular}
\caption{
\small
$\chi^2$ values and their individual contributions,
coming from experimental data on $t$-, $W$-, and ${\cal Q}^{2}$-dependence,
for various models with a residual dipole (\ref{Ans-t-dip}) and
exponential (\ref{Ans-t-exp}) $t$-dependence for two-step fits
(first DIS, then DVCS; first six rows), and for simultaneous fit
(DIS+DVCS; last row). Boldface numbers indicate bad fits,
as defined in Sect.~\ref{SubSec-FitStrPar}.
}
\label{tab:conschiLO}
\end{table}
These statements are quantified in Table~\ref{tab:conschiLO}, where
for various models we list the total DIS and DVCS $\chi^{2}$ over the
number of d.o.f., as well as the partial
contributions to $\chi^{2}$ from various subsets of data, over
the corresponding  n.o.p.
Our ad hoc model (l-PW) is with
$\chi^2/{\rm d.o.f.}\approx 3$ highly disfavored at LO accuracy in
both fitting strategies (two-step and simultaneous), while flexible
(nl- and $\Sigma$-PW)
GPD models correctly describe both DIS and DVCS data (for them we display only
results of two-step fits). For both flexible models, and with either
dipole (\ref{Ans-t-dip})  or exponential
(\ref{Ans-t-exp}) residual $t$-dependence, we have
$\chi^2/{\rm d.o.f.} \approx 1$, and partial $\chi^2$
values indicate a good description for the $t$-, $W$-,  and ${\cal
Q}^2$-dependence.
The values of our fitting parameters are listed in Table \ref{tab:consparsLO}.
\begin{table}[th]
\begin{center}
\begin{tabular}{|cl|cccrcrcrr|} \hline
model & $\alpha_{s}$ & $N^{\rm sea}$ & $\alpha^{\rm sea}(0)$ & $(M^{\rm sea})^2$ &
 $s^{\rm sea}$ & $\alpha^{\rm G}(0)$ & $s^{\rm G}$ & $B^{\rm sea}$  &
$b^{\rm eff}$ & BCA \\
 &  &  &  & \footnotesize [GeV$^2$] & & &
 & \footnotesize [GeV$^{-2}$]& \footnotesize [GeV$^{-2}$] & \\
\hline \hline
\phantom{n}l, dipole\phantom{.} & LO &
{\bf 0.152} & {\bf 1.158} & {\bf 0.062} &  & {\bf 1.247}  &  &
                {\bf 33.} & {\bf 5.7} & {\bf 0.19} \\
\phantom{n}l, exp.\phantom{ole} & LO &
{\bf 0.152} & {\bf 1.158}  &   &  &  {\bf 1.247} &    &
               {\bf 29.}    & {\bf 5.1} & {\bf 0.23} \\
nl, dipole\phantom{.}  & LO &
0.152  & 1.158  & 0.48 & -0.15  & 1.247  & -0.81   & 4.8 & 5.5  & 0.13 \\
nl, exp.\phantom{ole} & LO &
0.152  & 1.158  &   & -0.18  & 1.247   & -0.86 & 3.1  & 5.8 & 0.14 \\
$\Sigma$, dipole\phantom{.}  & LO &
0.152  & 1.158   & 0.42 & -11.  & 1.247   & -32.   & 5.4 & 5.5 & 0.14 \\
$\Sigma$, exp.\phantom{ole} & LO &
0.152  & 1.158   &   & -13.  & 1.247   &  -34.  & 3.1 & 5.8 & 0.15
\\
\hline
\end{tabular}
\caption{ \small Model parameters, as obtained by two-step fits
from Table~\ref{tab:conschiLO}, together with quark GPD $H$
$t$-slope $B^{\rm sea}$ at $x=10^{-3}$ and  ${\cal Q}^2 =4\, \GeV^2$, CFF
$\mathcal{H}$ $t$-slope (\ref{Def-beff}) $b^{\rm eff}$ at $W=82\,
\GeV$ and ${\cal Q}^2 =10\, \GeV^2$, both in GeV$^{-2}$, and
resulting BCA (\ref{Def-Asy}). For the dipole and
exponential (exp.) $t$-dependence the fixed variables are given
in Eq.~(\ref{Ans-t-dip}) and (\ref{Ans-t-exp}), respectively.
Again, boldface numbers arise from bad fits.
}
\label{tab:consparsLO}
\end{center}
\end{table}

In Section~\ref{SecSubSub-fai-sma-x} we reveal the reason for
the failure of ad hoc models.  Then, in Section~\ref{SecSubSub-Ske-t-cro},
we provide some insight into skewness- and $t$-dependence and
their cross-talk for our flexible models.

\subsubsection{The failure of the small-$\boldsymbol x$ conformal
skewness ratio at LO}
\label{SecSubSub-fai-sma-x}

A whole class of GPD models has the same small-$x$ behavior as the
l-PW model, characterized by their skewness ratio being equal to
its conformal value (\ref{r-ratio-con}). As explained in
Sect.~\ref{SurSec-GPD-mod}, for the ``pomeron'' case, of interest
here, we can also include the  ``dual'' model of
Ref.~\cite{GuzTec08} and the original RDDA ($b=1\approx \alpha$)
in this class. We explained in Sect.~\ref{subsec-Fits} that,
contrarily to the claim of
Refs.~\cite{ShuBieMarRys99,MarNocRysShuTeu08}, the conformal ratio
(\ref{r-ratio-con}) cannot be a general GPD property. It is
important to clarify the phenomenological status of this small-$x$
claim. We shall now have a closer look at its failure by utilizing
the l-PW model.

The l-PW model implies a normalization of the CFF that generally
overshoots the experimental data, which is also manifested in the
large skewness effect (\ref{r-ratio-con}), i.e., $r\sim 1.6$. To
compensate for the too large normalization in the fitting process
the $t$-slope gets increased via a very low cut-off mass parameter
$(M^{\rm sea})^2 \approx 0.05\, \GeV^2$ or a very large slope
parameter $B^{\rm sea}\approx 30/\GeV^2$, see Table
\ref{tab:consparsLO}. As a consequence, the $t$-dependence of the
cross section, with $\chi^2_t/{\rm n.o.p.}\approx 3$, is
particularly badly described. A simultaneous DIS/DVCS fit also
does not help, see Ref.~\cite{KumMuePas07} or last row of Table
\ref{tab:conschiLO}. We add that in such a fitting strategy the
structure function $F_2$ is well described, where the gluon PDF
comes out slightly softer than in a two-step fit.

We exemplify now that the failure of the l-PW model cannot be
cured by modifying the $t$-dependence. From the second line of
Table~\ref{tab:conschiLO} we see that a purely factorized
$t$-dependent ansatz, namely, the exponential one
(\ref{Ans-t-exp}), is with $\chi^2/{\rm d.o.f.} \approx 3.2$  even
more disfavored than the dipole ansatz (\ref{Ans-t-dip}). Such an
exponential ansatz was also employed in the ``dual'' model of
Ref.~\cite{GuzTec06}. Our poor description of the DVCS data
contradicts the statement of Ref.~\cite{GuzTec06} that such a
model is compatible with the DVCS data of the HERA I run
\cite{Aktas:2005ty,Chekanov:2003ya}. We repeated the analysis of
Ref.~\cite{GuzTec06} taking a pure Regge ansatz with no residual
$t$-dependence, but with large ``Regge slope'' instead:
\begin{eqnarray}
\label{GT-alpha-prime}
\alpha^\prime_{\rm sea} = 0.9/\GeV^2
\quad\mbox{and}\quad
\alpha^\prime_{\rm G} = 0.5/\GeV^2\,.
\end{eqnarray}
One might expect that such a choice
lowers the normalization of the CFF,
$$
\frac{{\cal H}^{\rm Regge}(\xi,t)}{ {\cal H}^{\rm exp.}(\xi,t)}
\approx
\frac{
\Gamma(3/2+\alpha(t))\Gamma(2+\alpha)
}{
\Gamma(3/2+\alpha)\Gamma(2+\alpha(t))
}
\exp\left\{-|t| (\alpha^\prime \ln(2/\xi) -B)\right\}
\lesssim 1 \quad \mbox{for small } \xi,
$$
in such a way that DVCS data are better described. In contrast to
this expectation, we have found total disagreement
with the data, e.g.,
$
\chi^2/{\rm d.o.f.} = {\bf 2100/101}
\,.
$
Fortunately, the inconsistency of our findings
with those of Ref.~\cite{GuzTec06}  has been resolved in
Ref.~\cite{GuzTec08} and our statement that l-PW models (or
minimalist ``dual'' model and the small-$x$ claim
\cite{ShuBieMarRys99,MarNocRysShuTeu08}) are disfavored at LO
\cite{KumMuePas07} holds  true for a pure Regge ansatz
(\ref{GT-alpha-prime}), too. Note that the value of
$\alpha^\prime_{\rm G}$ as large as in Eq.~(\ref{GT-alpha-prime}) is already
excluded by the H1 and ZEUS electroproduction measurements of
vector mesons, which is dominated by two-gluon exchanges.

We also recall that in a previous LO investigation \cite{FreMcDStr02},
performed in momentum fraction space, both the quark and
gluonic $r$-ratios were taken to be one at the input scale.
What happens then is that the large amount of gluons,
with their constant skewness function,
drives the quark GPD to rapidly approach its conformal skewness ratio
with increasing ${\cal Q}^2$, see Ref.~\cite{DieKug07a}.
The failure of this model shows that the skewness ratio for gluons is
also not equal to the conformal value $r^{\rm G}\approx 1$.
To investigate this some more, we used the l-PW
model for gluons, where the skewness ratio is close to one, see
Eq.~(\ref{r-ratio-con}), and allowed a flexible
parameterization of the sea quark GPD, e.g., the $\Sigma$-PW model
with dipole ansatz. We indeed obtained a bad fit
\begin{eqnarray}
\label{Res-fix-glu-LO}
\chi^2/{\rm d.o.f.} = {\bf 278/99}\,, \quad
\chi_t^2/{\rm n.o.p.} = {\bf 168/56}\,, \quad
\chi^2_W/{\rm n.o.p.} = {\bf 70/29}\,, \quad
\chi^2_{{\cal Q}^2}/{\rm n.o.p.} =  {\bf 40/16}\,,
\end{eqnarray}
verifying our expectations.

It is  popular in LO model descriptions of hard exclusive vector meson
production \cite{GolKro05,GolKro07,MarRysTeu99,MarNocRysTeu07} to
reduce the amount of gluons by taking the NLO PDF parametrization
rather the LO one, since NLO gluon PDFs are by a factor of two or
so smaller than LO ones, see error bands in
Fig.~\ref{FigFit-DIS}(b). This recipe has been also used in
Refs.~\cite{MarRysTeu99,MarNocRysTeu07} to conclude that the
exclusive $J/\Psi$ production is phenomenologically consistent
with the LO small-$x$ claim of Ref.~\cite{ShuBieMarRys99}:
\begin{eqnarray}
\label{Def-rG-Rat-0-Shu}
R^{\rm G} =
2^{\alpha_G-1} r^{\rm G}
> 1.
\end{eqnarray}
In our opinion this finding does not necessarily support the
small-$x$ claim, obtained from a LO analysis. The recipe is only
justified in the phenomenological context of modelling the  hard
exclusive production of vector mesons
\cite{GolKro05,GolKro07,MarRysTeu99,MarNocRysTeu07}; however, in a
simultaneous description of DIS and DVCS data this recipe is
obviously in conflict with the counting scheme of the collinear
factorization approach. Moreover, if we take at the input
scale NLO PDFs  instead of LO ones in the model of the previous paragraph,
we found that although
\begin{eqnarray}
\label{Res-fix-glu-NLO}
\chi^2/{\rm d.o.f.} =  {\bf  154/99} \,, \quad
\chi_t^2/{\rm n.o.p.} = {\bf 95/56}\,, \quad
\chi^2_W/{\rm n.o.p.} = {\bf 38/29}\,,   \quad
\chi^2_{{\cal Q}^2}/{\rm n.o.p.} =  {\bf  21/16}
\end{eqnarray}
improve, compared to Eq.~(\ref{Res-fix-glu-NLO}), the DVCS
fits are still unacceptable.

\subsubsection{Skewness ratio, $\boldsymbol t$-dependence, and their cross-talk}
\label{SecSubSub-Ske-t-cro}

Our flexible models resolve the normalization problem and fits are
unproblematic in both two-step and simultaneously fitting
strategies, see Table \ref{tab:conschiLO} and
Fig.~\ref{fig:LOfitres}. As expected, we find negative values of
the $s$-parameters, listed in Table~\ref{tab:consparsLO}. This can
be also viewed as the reduction of the large skewness effect of
the l-PW model. We recall that the definition of these
$s$-parameters does not {\em directly} allow to read off from them
the size of the skewness effect, or to compare their size in two
models.

We quantify the skewness effect in the momentum fraction space by
the $r$-ratio (\ref{Def-r-Rat-0}). An analogous quantity,
emphasizing the physical rather than partonic aspect, can be
defined as the ratio of the imaginary parts of amplitudes for DVCS
and forward Compton scattering, which can be expressed in terms of
the differential DVCS cross section at $t=0$ and the  DIS cross
section $\sigma_T(\gamma^\ast p \to \mathbf{X})$ for transversally
polarized photon exchange, respectively. Assuming an exponential
$t$-dependence, the skewness effect is revealed by utilizing the
total DVCS cross section \cite{Aaretal07}:
\begin{eqnarray}
\label{Def-R-Rat}
R(W,{\cal Q}^2)=
\frac{
\sqrt{ 16 \pi \sigma_{\rm DVCS}\, b({\cal Q}^2)
/(1+\rho^2)}}{
\sigma_T(\gamma^\ast p \to \mathbf{X})
}
\stackrel{\rm LO}{=}
\frac{
H^{\rm sea}(x,\eta=x,t=0,{\cal Q}^2)
}{
q^{\rm sea}(X,{\cal Q}^2
)}\Big|_{X=2x/(1+x)}\,.
\end{eqnarray}
Here the  $t$-slope $b({\cal Q}^2)$ is  extracted from a fit
with $\alpha^\prime=0$ \cite{Aaretal07},
\begin{equation}
b({\cal Q}^2) = A\left[1-B \log(Q^2/2\,\GeV^2)\right],
\quad A= 6.98 \pm 0.54\, \GeV^{-2}\,,\; B=0.12\pm 0.03 \;,
\label{eq:fitAB}
\end{equation}
and the ratio of real to imaginary part of the DVCS
amplitude might be set to $\rho\! =\! -\cot(\alpha({\cal Q}^2) \pi/2)$.
At LO this $R$-ratio can also be expressed in terms of sea quark
GPD, where the momentum fraction for the PDF (or zero-skewness GPD at $t=0$)
is  $X=2x/(1+x)$. 
The relation between skewness ratios (\ref{Def-r-Rat-0}) and
(\ref{Def-R-Rat}), considered now as a function of $x$ rather than $W$,
follows  from their definitions:
\begin{eqnarray}
\label{Rel-R2r}
R(x,{\cal Q}^2) &\!\!\!\stackrel{\rm LO}{=}\!\!\!&
2^{\alpha({\cal Q}^2)} r({\cal Q}^2)
\qquad\mbox{for small }  x
\,,
\end{eqnarray}
where we set $q^{\rm sea}(x,{\cal Q}^2)/q^{\rm sea}(X,{\cal Q}^2)=
2^{\alpha({\cal Q}^2)}$. In the kinematical region considered here
$\alpha\sim 1.2$ and so $R \sim 2 r$.

\begin{figure}[t]
\begin{center}
\includegraphics[scale=0.7]{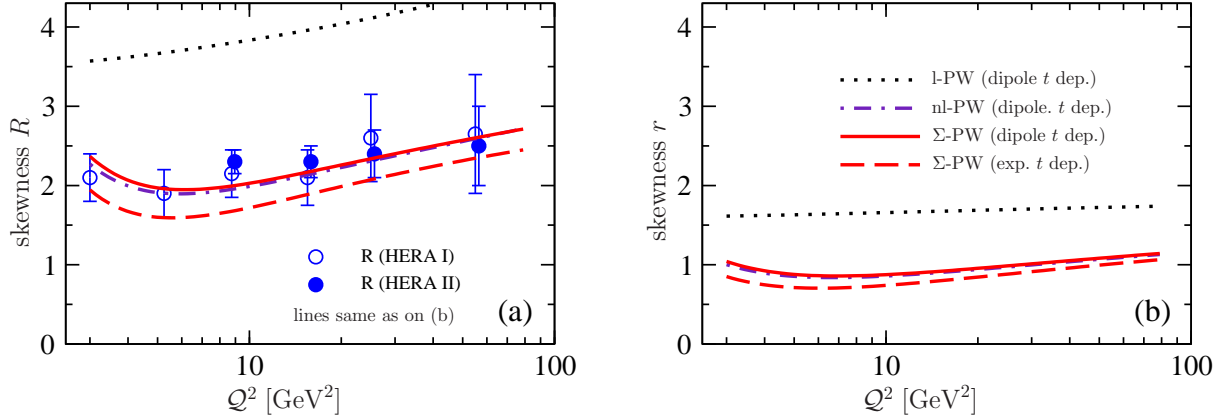}
\end{center}
\caption{\small (a)  skewness ratio R (\ref{Def-R-Rat}) for $W =
82$ GeV$^2$ compared to H1 data and (b) quark skewness ratio $r$
(\ref{Def-r-Rat-0}) for $x=10^{-3}$.  Model parameters, obtained
from fits as in Fig.~\ref{fig:LOfitres}, are specified in the main
text and in Table~\ref{tab:consparsLO} for l-PW (dotted), nl-PW
(dot-dashed), $\Sigma$-PW (solid) with dipole ansatz and
$\Sigma$-PW with exponential $t$-dependence (dashed).
}
\label{fig:r}
\end{figure}

In panels (a) and (b) of Fig.~\ref{fig:r} we display  $R$ and $r$
for fixed $W$ and for fixed $x$, respectively, as functions of
${\cal Q}^2$.   Needless to say, the l-PW model, obeying
the small-$x$ claim \cite{ShuBieMarRys99,MarNocRysShuTeu08}
of conformal ratio for quarks, is in conflict
with experimental measurements (dotted curves). The experimentally
measured $R$-ratio is well reproduced by both the nl-PW
(dot-dashed) and the $\Sigma$-PW (solid) models within a dipole
ansatz (\ref{Ans-t-dip}), which are barely distinguishable. One
sees that the $R$-ratio is slightly larger than 2, while the
$r$-ratio is generically $r\approx 1$, with their ratio as expected from
Eq.~(\ref{Rel-R2r}). This experimental value of $R\sim 2$ is
sometimes interpreted as a large skewness effect; however,
from the GPD perspective, i.e., $r\sim 1$, it is much more appropriate
to consider it as a manifestation of a small or zero skewness effect.

We also display the result for a $\Sigma$-PW model with
exponential (\ref{Ans-t-exp}) $t$-dependence (dashed). Compared to
the dipole ansatz (\ref{Ans-t-dip}), it provides only a slightly
smaller skewness ratio, i.e., the modulus of its negative
$s$-parameters is slightly larger. In other words, the
normalization of CFF $${\cal H}(\xi, t=0,{\cal Q}^2_0)$$ at $t=0$
is larger for CFF with a dipole $t$-dependence in order to
compensate for initial faster decrease with $t$. This can be also
observed by comparing the parameters $B^{\rm sea}$ and $s$ in the
third and fourth (or fifth and sixth) rows of
Table~\ref{tab:consparsLO}. Thus, the quite drastic correlation of
skewness effect and $t$-dependence for the disfavored l-PW model,
observed in the previous section, appears also for flexible GPD
models, but in a much milder form. Note that the  CFF
$${\cal H}(\xi, \langle t \rangle \approx 0.2\, \GeV^2,{\cal
Q}^2_0)$$ at the mean value $\langle t \rangle \approx 0.2\,
\GeV^2$, and the effective slope over the accessible $t$-interval,
see Eq.~(\ref{Def-beff}) below, are described equally good with
both ansaetze.

\begin{figure}[t]
\begin{center}
\includegraphics[scale=0.7]{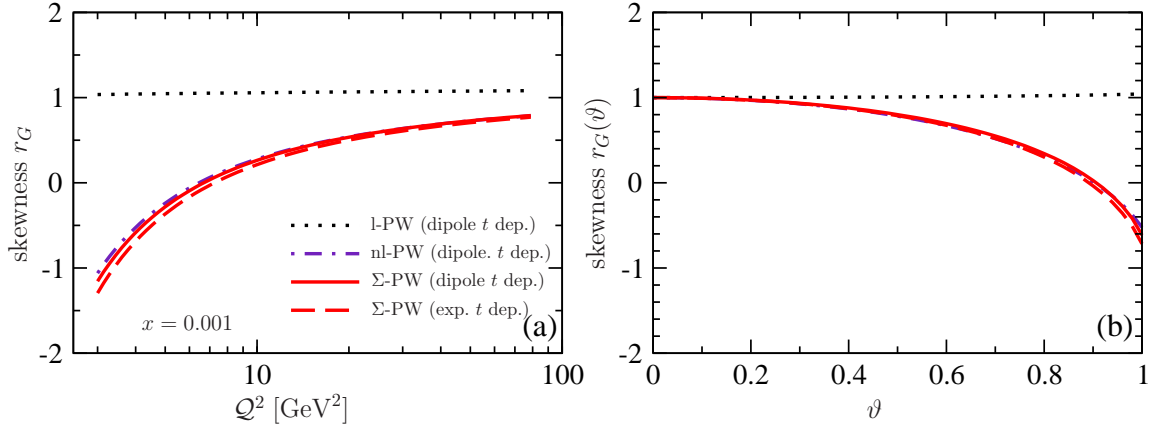}
\end{center}
\caption{ \small
Gluonic skewness ratio $r^{\rm G}$ (\ref{Def-rG-Rat-0})
versus ${\cal Q}^2$ for $x=10^{-3}$ (a) and skewness function
$r(\theta)$ (b). Models are the same as in  Fig.~\ref{fig:r}.
}
\label{fig:skewLOG}
\end{figure}
Let us also have a closer look at the gluonic skewness effect. The
large negative skewness parameters $s^{\rm G}$ in
Table~\ref{tab:consparsLO} indicate that the gluonic $r$-ratio is
much smaller than one. This is seen in Fig.~\ref{fig:skewLOG}(a)
for both flexible models, which are again barely distinguishable,
and whose gluon skewness is even negative at the input scale. If
the skewness ratio at the input scale is constrained to be
positive, we  could not obtain acceptable LO fits. Although the
GPD on the cross-over line $\eta=x$ has no probabilistic
interpretation, and negative values are thus not forbidden, a
gluonic GPD model with a zero is suspicious. We consider this zero
as an artifact related to an improper modelling of the skewness
function $r(\vartheta)$, defined in Eq.~(\ref{GPD-ans}), at the
initial scale. The fact that with increasing ${\cal Q}^2$ the
gluonic $r$-ratios of both models approach in the \emph{same} way
the conformal ratio $r_{\rm G}\approx 1$, reveals that their
skewness functions are almost the same, see
Fig.~\ref{fig:skewLOG}(b), although their conformal moments look
quite different. We add that the same turns out to be true for the
quark GPD models. Hence, finally one realizes that the $\Sigma$-PW
model is \emph{effectively} equivalent to the nl-PW one.

Within our models, the $t$-dependence is well described at LO
accuracy, too, and  we cannot discriminate between the two types
of functional dependence. As argued in Sect.~\ref{SurSec-GPD-mod},
we prefer the dipole ansatz (\ref{Ans-t-dip}) with a small value
of $\alpha_{\rm sea}^\prime = \alpha_{\rm G}^\prime=0.15$. In
panel (a) of Fig.~\ref{fig:LOfitres} a fitting example is shown
and the $\chi^2$ values, listed in Table~\ref{tab:conschiLO},
confirm a good description of the DVCS data set.

\begin{figure}[t]
\begin{center}
\includegraphics[scale=0.7]{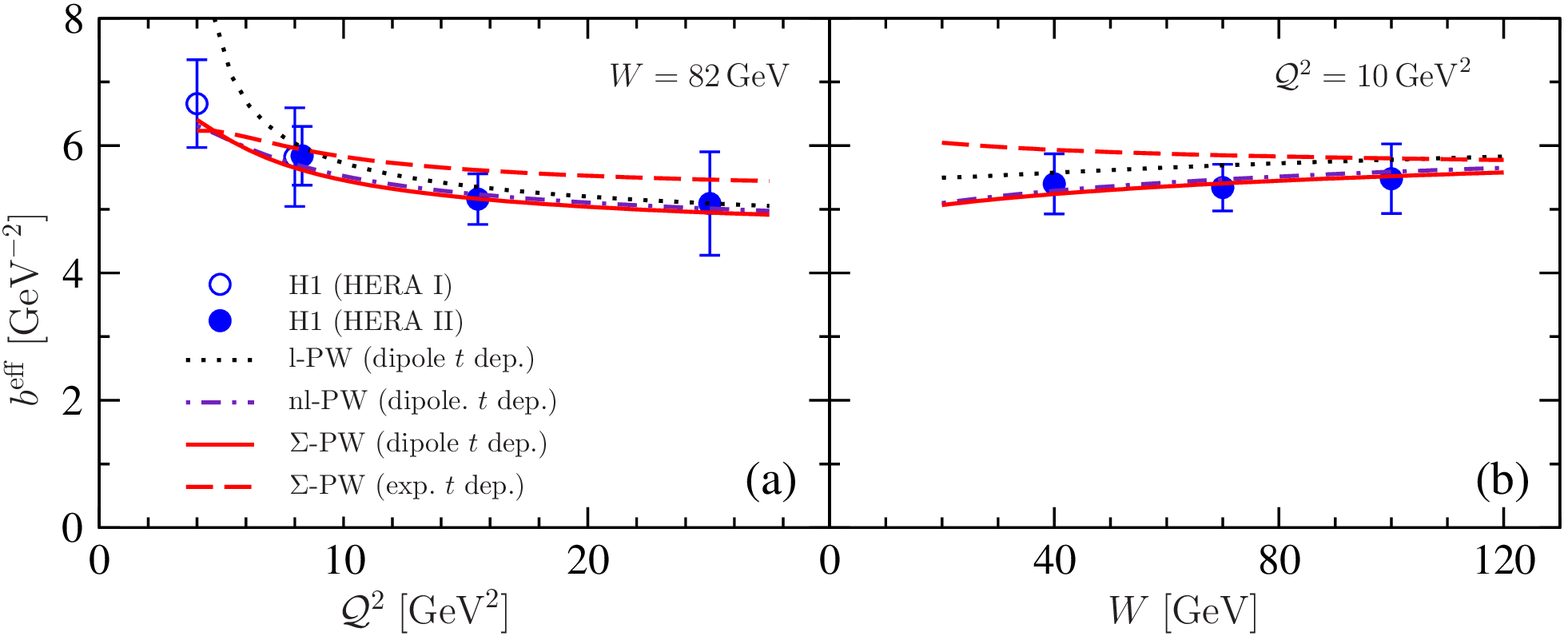}
\end{center}
\caption{ \small
Effective t-slope
$b^{\rm eff}$ (\ref{Def-beff}) versus ${\cal Q}^2$ for
$W = 82\, \GeV$ (a) and versus $W$ for $\mathcal{Q}^2=10\, \GeV^2$ (b),
compared to H1 data \cite{Aaretal07}.  Models are the same as in Fig.~\ref{fig:r}.
}
\label{fig:beff}
\end{figure}
To convince the reader in a more obvious way that our dipole GPD
parameterization (\ref{Ans-t-dip}) is compatible with the
exponential fit (\ref{eq:fitAB}) of the H1 Collaboration, we
evaluate the effective exponential slope
\begin{eqnarray}
\label{Def-beff}
b^{\rm eff} =
\frac{1}{t_1-t_2} \ln
\frac{\frac{d\sigma_{\rm DVCS}}{dt}(W,t_1,{\cal Q}^2)
}{
\frac{d\sigma_{\rm DVCS}}{dt}(W,t_2,{\cal Q}^2)}\,.
\end{eqnarray}
Thereby, we consider the experimentally accessible interval, i.e.,
we set $$t_1= - 0.1\, {\rm GeV}^2\quad\mbox{and}\quad t_2= - 0.8\,
{\rm GeV}^2$$ and evaluate $b^{\rm eff}$ from our GPD models that
are fitted {\em only} to the cross section measurements.

In Fig.~\ref{fig:beff} we show the effective slope $b^{\rm eff}$
versus ${\cal Q}^2$ for $W=82\,\GeV$ and versus $W$ for ${\cal
Q}^2=10\,\GeV^2$. One clearly realizes that the solid and
dash-dot-dotted curves, which result from the $\Sigma$- and nl-PW
model with a dipole ansatz (\ref{Ans-t-dip}), respectively,
describe the experimental H1 data very well. In the left panel
the ${\cal Q}^2$ evolution of  $b^{\rm eff}$ follows the
experimental data and so the perturbative evolution of these GPD
models is fully compatible with the measurements. Also the
flatness of the $W$-dependence of the data, indicating the absence of
a shrinkage of the diffractive forward peak, is well reproduced
with the input values $\alpha^\prime_{\rm sea}=\alpha^\prime_{\rm
G}=0.15\, \GeV^2$ , see Fig.~\ref{fig:beff}(b) Consequently, one
cannot conclude that the DVCS measurements indicate that
$\alpha^\prime$ is zero at a lower scale.

The quark dipole cut-off mass parameter $M^{\rm sea} \approx
0.67\,\GeV$ in both of our models is rather similar and somewhat
smaller than the fixed gluonic cut-off mass $M^{\rm G} \approx
0.85 \,\GeV$. They essentially coincide with the characteristic
scale of the nucleon, as it appears in the dipole parameterization
of its electromagnetic form factors.
Note that the  dipole masses and the resulting quark slope $B^{\rm sea}$
at the input scale obviously differ $10-15\%$ between the two flexible
GPD models, compare third and fifth rows in
Table~\ref{tab:consparsLO}. This essentially reflects the residual
$t$-dependence of the skewness effect at the input scale
and establishes a cross-talk between $t$-dependence and skewness.

Alternatively, we can also fit the experimental data set by an
exponential $t$-dependence (\ref{Ans-t-exp}) with $\alpha_{\rm
sea}^\prime = \alpha_{\rm G}^\prime=0$. As already mentioned, the
quality of fits are equally good, compare third (fifth) with forth
(sixth) line in Table \ref{tab:conschiLO}. Our exponential
$t$-parametrization  resembles the one used by the H1
Collaboration \cite{Aaretal07} in the fit (\ref{eq:fitAB}) of the
differential DVCS cross section measurements. Indeed, at the input
scale ${\cal Q} = 2\, \GeV$ our LO $\Sigma$-PW fit gives $b=
6.2/\GeV^2$  which is consistent with the H1 value
(\ref{eq:fitAB}),  $b= (6.4 \pm 0.5)/\GeV^2$. As in the H1 fit
(\ref{eq:fitAB}), the evolution to higher scales decreases the
slope parameter. However, in our GPD parameterization this
reduction is slightly weaker than in the H1 fit, displayed below
in Fig.~\ref{fig:dipexp}, together with the NLO prediction of
the $\Sigma$-PW.  Nevertheless, we conclude that the
decoration of the exponential $t$-slope parameter in GPDs by an
additional ${\cal Q}^2$ dependence as in Refs.
\cite{FreMcD01a,GuzTec06,GuzTec08} is a redundant complication,
which spoils the perturbative evolution.
The exponential ansatz
(\ref{Ans-bet-exp}) describes the measurements well, too, where
the small offset between GPD description and H1 measurements is a bit
larger than for the dipole ansatz, see also H1 fit
below in Fig.~\ref{fig:dipexp}.

Let us finally have a closer look at the ``Regge slope''
parameters. A  small value of $\alpha^\prime_{\rm G} <
0.25/\GeV^2$ can be deduced from $J/\Psi$ data. However, one might
wonder whether a large value of $\alpha^\prime_{\rm sea} \sim
1/\GeV^2$, as considered in Ref.~\cite{DieKug07a} and utilized in
Ref.~\cite{GuzTec06}, plays some role for H1/ZEUS kinematics.
{F}rom the Regge theory point of view, see discussion in
Sect.~\ref{subsec-Fits}, such a large slope parameter belongs to
``Reggeon exchanges'', that are suppressed relatively to the
``pomeron'' one by $\xi^{\lambda}$ with $\lambda \gtrsim 1/2$.
Therefore, we expect that our fits should disfavor a large
$\alpha^\prime$ value for the quarks.

\begin{figure}[t]
\begin{center}
\includegraphics[scale=0.7,clip]{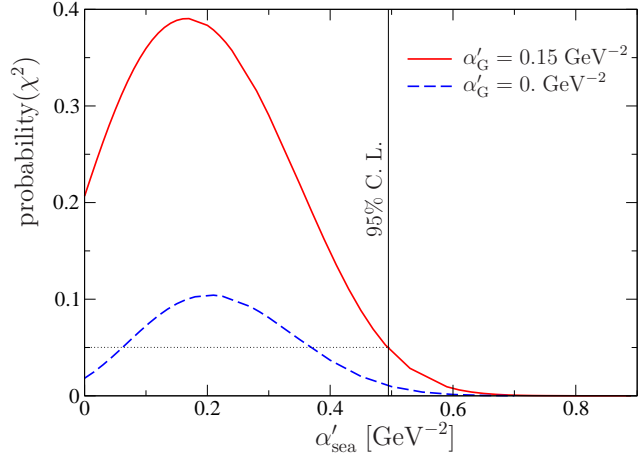}
\end{center}
\caption{\small Probability of (such and larger)
$\chi^2$ obtained by fitting the $\Sigma$-PW
model and the dipole  ansatz with $\alpha^\prime_{\rm G}=0.15/\GeV^2$  (solid)
and $\alpha_{\rm G}^\prime=0$ (dashed), with respect to the `
`Regge slope'' parameter $\alpha^\prime_{\rm sea}$, which is held
fixed during fitting procedure.
}
\label{fig:AlpS}
\end{figure}

To investigate this, we have performed a series of fits with the
$\Sigma$-PW model, accompanied by the dipole ansatz, with
different fixed value of $\alpha^\prime_{\rm sea}$ parameter.
Probability of the $\chi^2$ of these fits is plotted in
Fig~\ref{fig:AlpS}.
One notices that the preferred value is indeed around
$\alpha^{\prime}_{\rm sea} \approx 0.15\,\GeV^{-2}$, while values
beyond $0.5\,{\rm GeV}^{-2}$ are excluded at 95\% confidence
level.  This confirms our model assumption that ``Reggeon''
contributions are not dominant in the H1/ZEUS kinematics. Note
also that the choice $\alpha^\prime_{\rm G}=0$, shown as dashed
curve, is disfavored in the dipole ansatz.

\subsection{Beyond leading order fits}%
\label{SubSecSec-NLO}

\begin{table}[ht]
\centering
\begin{tabular}{|cl|ccccc|}
\hline model & $\alpha_{s}$ & $\chi^2/{\rm d.o.f}$ DIS &
$\chi^2/{\rm d.o.f}$ DVCS  &
$\chi^2_t/{\rm n.o.p}$ &$\chi^2_W/{\rm n.o.p}$ &$\chi^2_{{\cal Q}^2}/{\rm n.o.p}$ \\
\hline\hline \phantom{n} l &  NLO(\ms)
& 71.6/82 & {\bf 148./100} & {\bf 77.6/56} & {\bf 36.8/29} & {\bf 33.9/16}  \\
\phantom{n} l &  NLO(\cs)
& 71.6/82 & 105./100 & 62.9/56 & 25.1/29 & 17./16  \\
 nl           &  NLO(\ms)
 & 71.6/82 & 102./98 & 60.2/56 & 23.9/29 & 17.5/16  \\
 nl           &  NLO(\cs)
 & 71.6/82 & 104./98 & 61.4/56 & 24.9/29 & 18.1/16  \\
$\Sigma$      & NLO(\ms)
& 71.6/82 & 101./98 & 60./56 & 23.9/29 & 17.5/16  \\
$\Sigma$      & NLO(\cs)
& 71.6/82 & 104./98 & 61.5/56 & 24.9/29 & 18.1/16  \\
 \hline
\end{tabular}
\caption{\small $\chi^2$ values as in Table~\ref{tab:conschiLO} of
NLO DIS and DVCS fits for various models in the \ms{} and \cs{}
scheme. } \label{tab:conschiNLO}
\end{table}
We shall now pursue the model dependence and GPD reparameterization
in DVCS fits beyond LO. At NLO the gluons become a part of the
hard-scattering process, and they may induce large radiative
corrections. The inclusion of radiative corrections in a DVCS fit
is compensated by the reparameterization of the GPDs at the
initial scale. Within a flexible GPD parameterization we have good
fits beyond LO in both the \ms\ and \cs\ schemes. In
Table~\ref{tab:conschiNLO} we list the $\chi^2$ values for the
dipole ansatz.

It has been demonstrated in Ref.~\cite{KumMuePas07} that beyond LO
a good description of HERA I DVCS data
\cite{Aktas:2005ty,Chekanov:2003ya} was possible also within the l-PW
model, where, however, the dipole cut-off masses were used to
adjust the normalization of the DVCS cross section, see Sect.~7.2
and Fig.~16 of Ref.~\cite{KumMuePas07}. After including the HERA
II DVCS data \cite{Aaretal07} with many more data points for the
$t$-dependence, it is, however, not necessarily true that the
l-PW model still works. In particular for the \ms\ scheme $\chi^2$
is significantly large, see first line in
Table~\ref{tab:conschiNLO}, while in the \cs\ scheme such a model fit
is acceptable within the dipole ansatz. In contrast, within an
exponential ansatz the fits in the \ms\ scheme are acceptable,
however, in the \cs\ they are now disfavored:
$$
\chi^2/{\rm d.o.f.} = \left\{ {95/100 \atop {\bf 155/100}  }
\right\} \,, \quad \chi_t^2/{\rm n.o.p.} =  \left\{ {50/56 \atop
{\bf 108/56}  }  \right\}
\quad\mbox{for}\quad \left\{ { \overline{\rm MS}  \atop
\overline{\rm CS}  } \right\}\,,
$$
where in both cases $\chi^2_W/{\rm n.o.p.}$ and $\chi^2_{{\cal
Q}^2}/{\rm n.o.p.}$ values are fine.

These findings illustrate the intricate interplay of the
functional form of a given ad hoc ansatz and radiative
corrections. If one would have relied on the claim of
Ref.~\cite{ShuBieMarRys99} about the conformal skewness ratio, one
might have wondered whether the outcome of the NLO fitting favors
the l-PW model with the exponential ansatz in the \ms\ scheme  or
with the dipole ansatz in the \cs\ scheme, see, e.g., the
discussion in Ref.~\cite{GuzTec06}. We should recall here that the
difference between the \ms\ and \cs\ schemes is entirely related
to `non-diagonal' rotation effects%
\footnote{Obviously, this rotation is numerically significant in
the DVCS kinematics, i.e., $\eta=\xi$.  This shows that the
statement of Ref.~\cite{MarNocRysShuTeu08}, namely, that such
$\eta$-proportional effects are negligibly small and so the  LO
claim \cite{ShuBieMarRys99} about the conformal skewness ratio remains
valid at NLO, is amiss, too. The differences between both
schemes have also been studied in Ref.~\cite{KumMuePas07}.} that
vanish in the  $\eta = 0$ case. In other words, this scheme
rotation is nothing but a skewness effect and we observed in fact
an interplay of skewness and $t$-dependence, discussed in the
previous section in context of flexible models. So we emphasize
again that an `overinterpretation' of fit results within an ad hoc
GPD model is inappropriate.

\begin{figure}[ht]
\begin{center}
\includegraphics[scale=0.70]{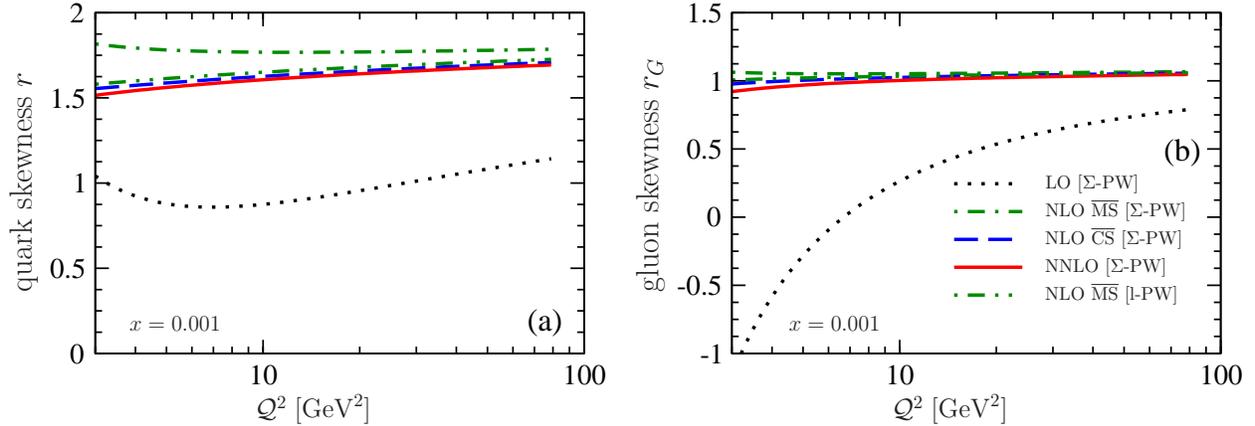}
\end{center}
\caption{\small Quark skewness $r$ (\ref{Def-r-Rat-0}) (a) and
gluon skewness $r_G$ (\ref{Def-rG-Rat-0}) (b) versus ${\cal Q}^2$
for $x=10^{-3}$, extracted from a two-step fit. Dash-dot-dotted
(green) line is l-PW model at NLO(\ms), while other lines are for
$\Sigma$-PW model at LO (dotted, black), NLO(\ms) (dash-dotted,
green), NLO(\cs) (dashed, blue) and NNLO (solid, red). In all
models we employ the dipole ansatz (\ref{Ans-bet-dip}). }
\label{fig:skew}
\end{figure}
The discussion of the previous paragraph becomes superfluous, if
one utilizes flexible GPD models. Then we can provide also in the
\ms\ scheme good two-step fits and we can reveal the skewness
effect. The extracted NLO fitting parameters for the \ms\ and
\cs\ schemes are listed in Table~\ref{tab:consparsNLO} for the
dipole ansatz.
\begin{table}[ht]
\centering
\begin{tabular}{|cl|cccrccccc|} \hline
model & $\alpha_{s}$ & $N^{\rm sea}$ & $\alpha^{\rm sea}(0)$ &
$(M^{\rm sea})^2$ & $s^{\rm sea}$ & $\alpha^{\rm G}(0)$  & $s^{\rm
G}$  & $B^{\rm sea}$ & $b^{\rm eff}$  & BCA
\\ \hline\hline
\phantom{n}l & NLO(\ms) & {\bf 0.168} & {\bf 1.128} & {\bf 0.71} &
& {\bf 1.099} & & {\bf 3.5} & {\bf 5.0} & {\bf 0.10}
\\
\phantom{n}l & NLO(\cs) & 0.168 & 1.128 & 0.57 &      & 1.099 &
& 4.2 & 5.7 & 0.09
\\
nl & NLO(\ms) & 0.168 & 1.128 & 0.59 & 0.04 & 1.099 & 0.02 & 4.0 &
5.6 & 0.09
\\
nl & NLO(\cs) & 0.168 & 1.128 & 0.58 &-0.01 & 1.099 &-0.01 & 4.1 &
5.6 & 0.09
\\
$\Sigma$ & NLO(\ms) & 0.168 & 1.128 & 0.60 & 3.10 & 1.099 & 1.10 &
4.0 & 5.7 & 0.09
\\
$\Sigma$ & NLO(\cs) & 0.168 & 1.128 & 0.58 & -0.42& 1.099 &-0.58 &
4.1 & 5.6 & 0.09
\\ \hline
\end{tabular}
\caption{\small Model parameters as specified in
Table~\ref{tab:consparsLO} obtained by fits from
Table~\ref{tab:conschiNLO}. } \label{tab:consparsNLO}
\end{table}
If one goes from LO to NLO, the most drastic changes appear in the
skewness parameters. They mutate from large negative values to
moderate positive ones in the \ms\ scheme. This qualitative jump
is also illustrated in Fig.~\ref{fig:skew}, compare LO fit (dotted
curve) with the others, where we show the skewness ratios for
$x=10^{-3}$ versus ${\cal Q}^2$. (This ratio for a fixed ${\cal
Q}^2$ is rather flat over the experimentally accessible interval
$x\in\{10^{-4},10^{-2}\}$.) One realizes in the left panel that
the $\Sigma$-PW  quark model at NLO \ms\ (dot-dashed) overshoots
now the conformal ratio (\ref{r-ratio-con}), i.e., $r^{\rm sea} \approx
1.6$. In the right panel it is illustrated that the gluon ratio at
NLO matches the conformal ratio (\ref{r-ratio-con}), i.e., $r^{\rm
G}\approx 1$. We recall that the gluon PDF, extracted from DIS, is in
LO approximation twice as large as in NLO, see
Fig.~\ref{FigFit-DIS}(b), and  so one might consider $r^{\rm
G}\sim 1/2$ as a realistic LO skewness ratio.

In the \cs\ scheme the skewness parameters $s^{\rm sea,G}$ are
small and negative for the dipole ansatz (\ref{Ans-t-dip}) and so
all models are compatible with the conformal  ratio
(\ref{r-ratio-con}), obtained at LO. This is displayed for the
$\Sigma$-PW model by the dashed curves in Fig.~\ref{fig:skew}.
The same holds true if we include in this scheme NNLO corrections,
shown as solid curve.
\begin{table}
\centering
\begin{tabular}{|cl|ccccc|} \hline
model & $\alpha_{s}$ &  $\chi^2/{\rm d.o.f}$ DIS & $\chi^2/{\rm
d.o.f}$ DVCS  & $\chi^2_t/{\rm n.o.p}$ &$\chi^2_W/{\rm n.o.p}$
&$\chi^2_{{\cal Q}^2}/{\rm n.o.p}$
\\ \hline\hline
\phantom{n}l &  NNLO(\cs)     & 81./82  & 80.1/100 & 49.2/56  & 19.1/29  & 11.9/16 \\
 nl &  NNLO(\cs)    & 81./82  & 78.8/98  & 46.7/56  & 18.9/29  & 13.2/16 \\
$\Sigma$ & NNLO(\cs) & 81./82  & 78.8/98  & 46.7/56  & 18.9/29  & 13.2/16 \\
 \hline
\end{tabular}
\caption{\small $\chi^2$ values as in Table~\ref{tab:conschiLO}
for NNLO DIS and DVCS fits within  various models in the \cs\
scheme.} \label{tab:conschiNNLO}
\end{table}
Table~\ref{tab:conschiNNLO} states that all models provide similar
and good  $\chi^2$ values.  The values of the $s$-parameters,
listed in Table~\ref{tab:consparsNNLO}, are small and negative and
with a slightly larger modulus than at NLO for the \cs\ scheme,
cf. Table~\ref{tab:consparsNLO}.
\begin{table}[h]
\centering
\begin{tabular}{|cl|cccrccccc|} \hline
model & $\alpha_{s}$ & $N^{\rm sea}$ & $\alpha^{\rm sea}(0)$ &
$(M^{\rm sea})^2$ & $s^{\rm sea}$ & $\alpha^{\rm G}(0)$  & $s^{\rm
G}$ & $B^{\rm sea}$ & $b^{\rm eff}$ & BCA
\\ \hline \hline
l-PW & NNLO(\cs)& 0.172 & 1.125 & 0.56 &       &  1.104 &       &
4.2 & 5.6 & 0.12
\\
nl-PW & NNLO(\cs) & 0.172 & 1.125 & 0.57 & -0.01 &  1.104 & -0.04
& 4.2 & 5.6 & 0.12
\\
$\Sigma$-PW & NNLO(\cs) & 0.172 & 1.125 & 0.57 &  -0.89 & 1.104 &
-1.80 & 4.2 & 5.6 & 0.12
\\
\hline
\end{tabular}
\caption{\small Model parameters as specified in
Table~\ref{tab:consparsLO} obtained by fits from
Table~\ref{tab:conschiNNLO}. } \label{tab:consparsNNLO}
\end{table}
Hence at NNLO in the \cs~scheme, same as at NLO, non-leading SO(3)
partial waves do not play essential role. The differences between
the three models are so small that we obtain the same values for
the effective and partonic $t$-slope parameters.

\begin{figure}[t]
\centerline{\includegraphics[scale=0.75]{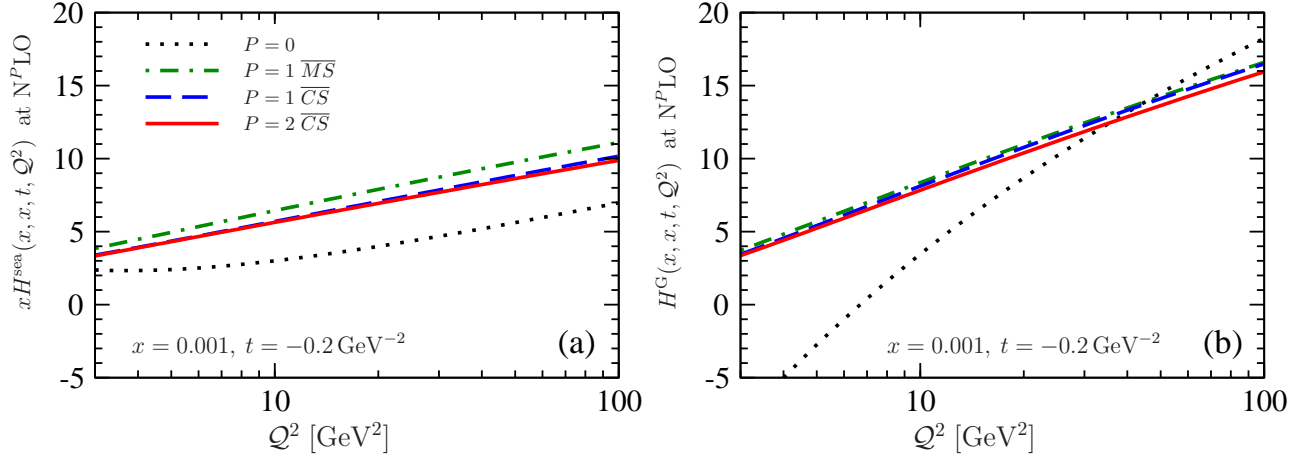}}
\caption{ \label{Reparam} Sea quark (a) and  gluon (b) GPDs within
dipole ansatz (\ref{Ans-bet-dip}) as they result from a LO
(dotted), NLO  \ms~scheme (dash-dotted) and \cs~scheme (dashed) as
well as  NNLO  \cs~scheme (solid) fit. }
\end{figure}
The reparameterization of sea quark and gluon GPDs can be directly
read off from Fig.~\ref{Reparam}. Here the GPDs on the cross-over
line are shown for the $\Sigma$-PW model with dipole ansatz as they
arise from our fits at various orders and schemes. It is again
obvious that the most drastic effect appears if we go from the LO
(dotted) to NLO description. Quark GPDs at NLO, compared to the LO
ones, are enhanced by a factor of two or so while the LO gluons
suffers from our model artifact. The sizable reparameterization
effects for quarks of about 100\% or even more are in view of the
quoted corrections in Ref.~\cite{KumMuePas07}%
\footnote{There a given l-PW model has been employed to estimate
the radiative corrections for the CFF $\cal H$.} surprising.
Naively, we would have expected as in DIS moderate correction on
the level of 20\% or so. The very large NLO corrections in the
quark sector might be connected to the artifacts of our gluon
model, discussed above, and we would here not exclude the
possibility that a fully flexible GPD model possess in the quark
sector only moderate NLO corrections. Consequently, for such a
model the quark skewness ratio must then be $\sim 1$ also beyond
LO, i.e., much smaller then the conformal ratio $r^{\rm sea}
\approx 1.6$. The difference at NLO between the \ms~(dash-dotted)
and \cs~(dashed) scheme are clearly visible for the sea quark GPD, while
this skewness-induced effect is tiny for gluons. This simply
reflects the properties of the off-diagonal scheme transformation,
which is set by conventions. As already observed in
Ref.~\cite{KumMuePas07} within the l-PW model and a simultaneous
fit, the NNLO corrections lead only to a slight change of the
parameters, obtained at NLO in the \cs\ scheme, compare the dashed
and solid curves in Figs.~\ref{Reparam} and also \ref{fig:skew} or
the corresponding entries in Table~\ref{tab:consparsNLO} with
Table~\ref{tab:consparsNNLO}.

One might wonder whether the observation that the \cs\ (N)NLO
skewness ratio approaches the LO conformal value
(\ref{r-ratio-con}) is a definite model independent feature or
simply an accident. Naively, one might expect that a reduction of
gluonic skewness ratio will also decrease the quark one, since
they contribute with a different sign to the CFF. In this way one
might realize a quark skewness ratio which would be closer to the
LO findings.  However, taking a more negative value of $s^{\rm
G}$, like that occurring in LO fits, provides bad NLO fits. As in
our unsuccessful attempt in the previous section (to have a
positive gluon GPD at the cross-over line at LO), we consider also
this finding as a model artifact. Namely, our models do not allow
adjustment of the scale dependence of the skewness ratio in a
flexible way.

\subsection{Transverse distribution of partons}
\label{SecSubSub-TraDis}

We would now like to deliver a partonic interpretation of our
analysis of the $t$-dependence of the DVCS cross section. We stick
to the probabilistic interpretation of zero-skewness GPDs in the
infinite momentum frame \cite{Bur02}. In this frame the proton
might be viewed as a disc with a radius of
\begin{eqnarray}
\label{Rad-ProDis}
\sqrt{4 \frac{d}{d t} \ln
F_1(t)\Big|_{t=0}} \approx 0.6\,\mbox{fm}\,,
\end{eqnarray}
arising from the  parameterization of the Dirac form factor in
terms of the common dipole expressions for the Sachs form factors
within a cut-off mass of $0.71\, \GeV^2$. We add that the electric
charge radius $\sqrt{\langle r^2 \rangle} = 0.875$  fm of the
proton, quoted in the Review of Particle Physics \cite{RPP08},  is about
ten percent larger than that from the dipole parameterization.
Hence, also our proton disc radius (\ref{Rad-ProDis}) would then increase to
about $0.66$ fm.

The transverse width of parton distribution, i.e., the average
distance $\sqrt{\langle \vec{b}^2 \rangle}$ of the struck parton
from the proton center, is directly given by the $t$-slope of the
zero-skewness GPD
\begin{eqnarray}
\label{Def-tra-wid}
\langle \vec{b}^2 \rangle(x,{\cal Q}^2) =
4\frac{d}{d t} \ln H(x,\eta=0,t,{\cal Q}^2)\Big|_{t=0}\,.
\end{eqnarray}

\begin{figure}
\begin{center}
\includegraphics[scale=0.7]{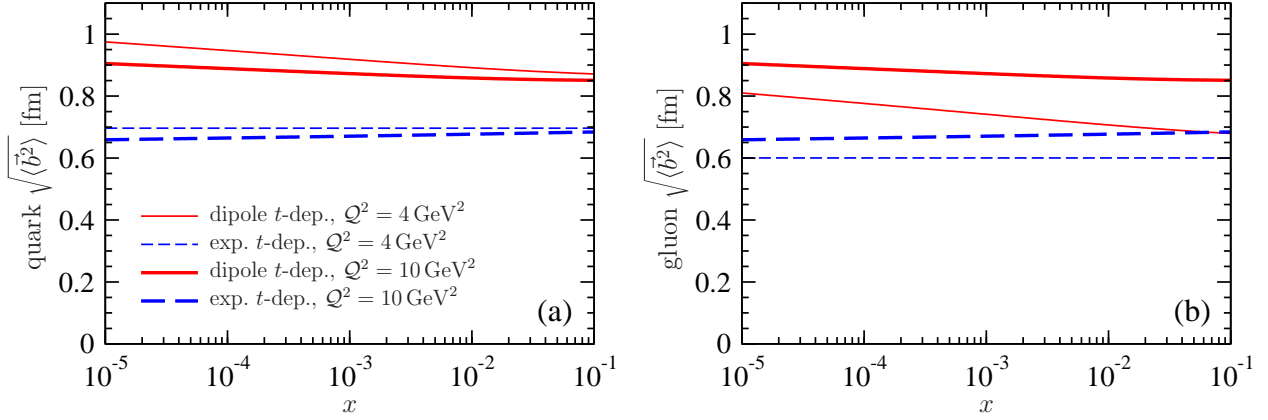}
\end{center}
\caption{ \small
Transverse width $\sqrt{\vec{b}^2}$ of sea quark (a) and gluon (b)
distributions  versus $x$  for ${\cal Q}^2=4\, \GeV^2$ (thin) and
${\cal Q}^2=10\, \GeV^2$ (thick), resulting from the LO fit of
$\Sigma$-PW model with dipole (solid) and exponential  (dashed)
$t$-dependence ansatz, specified in Eqs.~(\ref{Ans-t-dip}) and
(\ref{Ans-t-exp}), respectively.
}
\label{fig:b}
\end{figure}
Let us first consider the LO interpretation, where we take the
$\Sigma$-PW model as it is specified in
Table~\ref{tab:consparsLO}. As discussed in
Sect.~\ref{SecSubSub-Ske-t-cro}, the skewness effect will talk
back to the $t$-slope parameters on the $10\%-15\%$ level, which
implies a $\lesssim 7\% $ model uncertainty for the transverse
width. The rigidity, still present in our flexible models, does
not allow us to reveal from LO fits to the DVCS data the
transverse distribution of gluons. Therefore, we take the
parameters as before for the dipole (\ref{Ans-t-dip}) and
exponential (\ref{Ans-t-exp}) $t$-dependence, where the gluon
$t$-slope parameters are fixed from the $J/\psi$ photoproduction.

The value of the transverse width, obtained from fitted GPDs, also
depends on the functional form of the residual $t$-dependence,
which defines the extrapolation from the accessible $t$ interval
to $t=0$. For the exponential ansatz our value for gluons is
quoted in Ref.~\cite{StrWei03} and coincide with the proton disc
radius of $0.6$ fm. For $\alpha^\prime=0$, the dipole ansatz would
provide a $10\%$ larger value than the exponential one.  Also a
non-vanishing value of $\alpha^\prime$ leads to an increase of the
transverse width.  Taking $\alpha^\prime=0.15$ in exponential
ansatz, the gluonic transverse width increases, e.g.,  to $\approx
0.72$ fm for $x=10^{-3}$, coming rather close to the one in our
dipole ansatz (\ref{Ans-bet-dip}) with $\approx 0.77$ fm. We
might conclude that even for an exponential $t$-dependence, the
experimental uncertainties of the $\alpha^\prime$ measurement in
$J/\psi$ production do not exclude a gluon transverse width at small
$x$ that is larger than the proton disc radius (\ref{Rad-ProDis}).

The gluon transverse width, obtained at the input scale, is
depicted in the right panel of Fig.~\ref{fig:b} as thin solid
(dipole) and thin dashed  (exponential) curves. Our LO fit states
then that the quark transverse width at the input scale  is
slightly larger than for gluons, where the characteristic 
$x$-dependence, related to the partonic shrinkage effect, appears. We
infer that our LO quark interpretation strongly reflects our
assumptions and that we cannot discriminate, e.g., for $x\approx
10^{-3}$ and ${\cal Q}^2=4\,\GeV^2$, between
\begin{eqnarray}
\label{TraWid-LO-Inp}
\sqrt{\langle \vec{b}^2\rangle}_{\rm sea}
\approx  \left\{ {0.9 \atop   0.7}\right\} \mbox{fm}
\,,\quad
\sqrt{\langle \vec{b}^2\rangle}_{\rm G}
\approx  \left\{ {0.8 \atop   0.6}\right\} \mbox{fm}
\qquad\mbox{for}\;
\left\{
{ \text{dipole}\: (\ref{Ans-t-dip})
\atop
\text{exponential}\: (\ref{Ans-t-exp}) }\right\} \;
\text{ansatz}\,.
\end{eqnarray}
That at the input scale (thin curves) gluons are more centralized
than sea quarks has been also found in Ref.~\cite{Mue06} by a
fine-tuning procedure. Thereby, the latter were dynamically
generated within the double log approximation at LO from gluons
with a (soft) pomeron trajectory (\ref{Def-PomTra}). We emphasize
that there the differences between a tripole and an exponential
ansatz were hardly visible in impact space for distances smaller
than $1.5$ fm and that the larger value of the transverse
width for a tripole ansatz arose due to its long tail in impact
space, see illustrative examples in Ref.~\cite{Mue06}.

At larger resolution scale, e.g.,  ${\cal Q}^2 =10\,\GeV^2$, the
partonic shrinkage effect is practically washed out, see the
flatness of the thick solid curves in Fig.~\ref{fig:b}. This
indicates that $\alpha^\prime$ rapidly approaches zero with
growing ${\cal Q}^2$.  Also the initial differences between the
transverse widths of quarks and gluons diminish by a slight
decrease of the former and increase of the latter. However, the
`typical' value associated with the particular $t$-dependence
ansatz is robust. Finally, our model findings  for ${\cal Q}^2
=10\,\GeV^2$,
\begin{eqnarray}
\label{TraWid-LO-10}
\sqrt{\langle \vec{b}^2\rangle}_{\rm sea}  \approx
\sqrt{\langle \vec{b}^2\rangle}_{\rm G} \approx
\left\{ {0.9 \atop   0.65}\right\} \mbox{fm}
\qquad\mbox{for}\;
\left\{{ \text{dipole}\: (\ref{Ans-t-dip})
   \atop \text{exponential}\: (\ref{Ans-t-exp}) }\right\} \; \text{ansatz}\,.
\end{eqnarray}
are compatible with those of Ref.~\cite{Mue06}, where, e.g., a value of
$\approx 0.85$ fm was quoted  for a tripole ansatz.

We would also like to emphasize that for the exponential ansatz
the gluon transverse width is at the input scale the same as
quoted in Ref.~\cite{FraStrWei05}.  It perturbatively evolves to a
slightly larger value with increasing scale. In Fig.~\ref{fig:b}
we also extrapolate the transverse widths obtained from fitted
GPDs to larger $x \in [10^{-2},10^{-1}]$. This extrapolation is
not supported by the interpretation of $J/\psi$ data from fixed
target experiments in Ref.~\cite{FraStrWei05}, where within a
dipole ansatz a value of $\approx 0.53$ fm for $x\sim 0.1$  was
quoted. Interestingly, the increase of this value by about $20\% -
30\%$ to $\sim 0.6$ fm has been explained with chiral dynamics
\cite{StrWei03}. We add that in realistic GPD models the
$t$-dependence dies out at $x \to 1$, e.g., seen in lattice
simulations of Ref.~\cite{Hagetal07}, and so the partons are
entirely concentrated in the center of the proton. This feature
can be simply implemented in our models by decorating the cut-off
mass (or the exponential slope parameter) with $j$-dependence.

Next we shall show that our LO findings with fixed gluon slope
parameters are only moderately influenced by perturbative
corrections and scheme conventions, or even by the release of the
gluon slope parameters at NLO (or NNLO). Note that this was not
the case in our previous investigations, where we employed the
l-PW model \cite{KumMuePas07}. We mainly work in the \ms\ scheme
to NLO accuracy, which is appropriate for a possible global GPD
analysis of deeply exclusive electroproduction processes in
momentum fraction space, and we use the $\Sigma$-PW model. We
might have employed the nl-PW model as well, however, beyond LO
there is only very little model uncertainty induced by the
skewness dependence, see $B^{\rm sea}$ values in Tables
\ref{tab:conschiNLO} and \ref{tab:conschiNNLO}.

\begin{figure}[t]
\begin{center}
\includegraphics[scale=0.82]{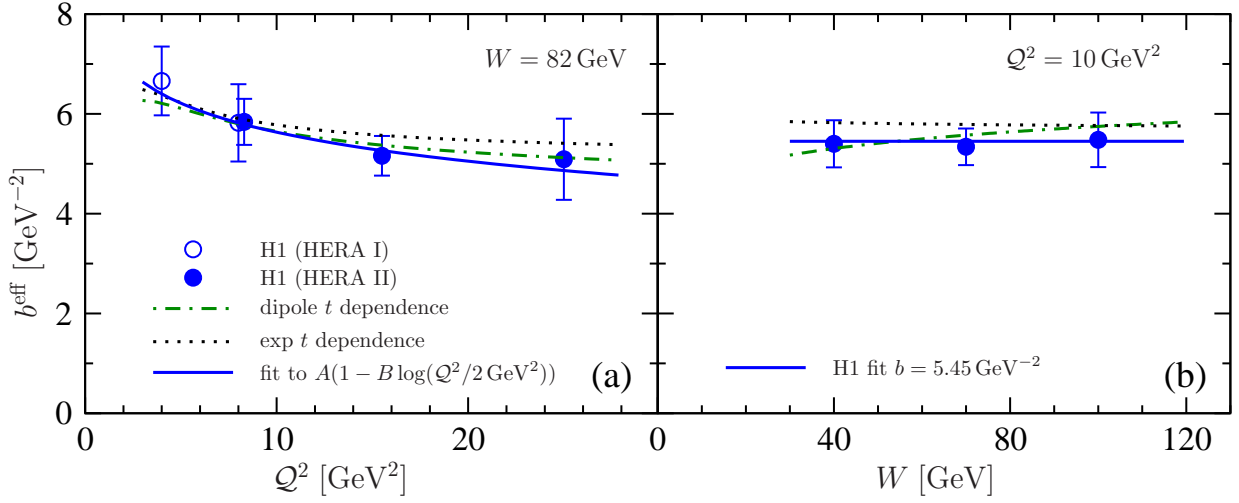}
\end{center}
\caption{\small Effective slope parameter   $b^{\rm eff}$, defined in Eq.~(\ref{Def-beff}),
obtained from  the $\Sigma$-PW model
with dipole $(1 - t/M^2)^{-2}$ ($\alpha^\prime =0.15/\GeV^2$) (dot-dashed) and
exponential $\exp(t/2b)$ ($\alpha^\prime =0$) (dotted)  $t$-dependence.
Parameters where
obtained by a NLO (\ms)  fit and  are specified
in Table \ref{tab:consparsNLO}.  Solid line is the fit (\ref{eq:fitAB})
of the H1 Collaboration.
}
\label{fig:dipexp}
\end{figure}
We first demonstrate in Fig.~\ref{fig:dipexp} that at NLO the
description of the measured $t$-slope is for fixed gluon slope
parameters fully analogous to our LO findings in
Fig.~\ref{fig:beff}. Again, the effective slope parameter
(\ref{Def-beff}) is evaluated from the outcome of our fits to the
DVCS cross section with the dipole (dot-dashed) and exponential
(dotted) ansatz, where the NLO (\ms) parameters are given in Table
\ref{tab:consparsNLO}. Comparing the corresponding curves in both
figures one can barely see a difference. We also display the H1
fit  (\ref{eq:fitAB}) as solid curve. Obviously, only if
experimental errors could be very drastically reduced, one might
be able to discriminate between the three curves --- for a more detailed
discussion see Sect.~\ref{SecSubSub-Ske-t-cro}.

\begin{figure}[ht]
\begin{center}
\includegraphics[scale=0.70]{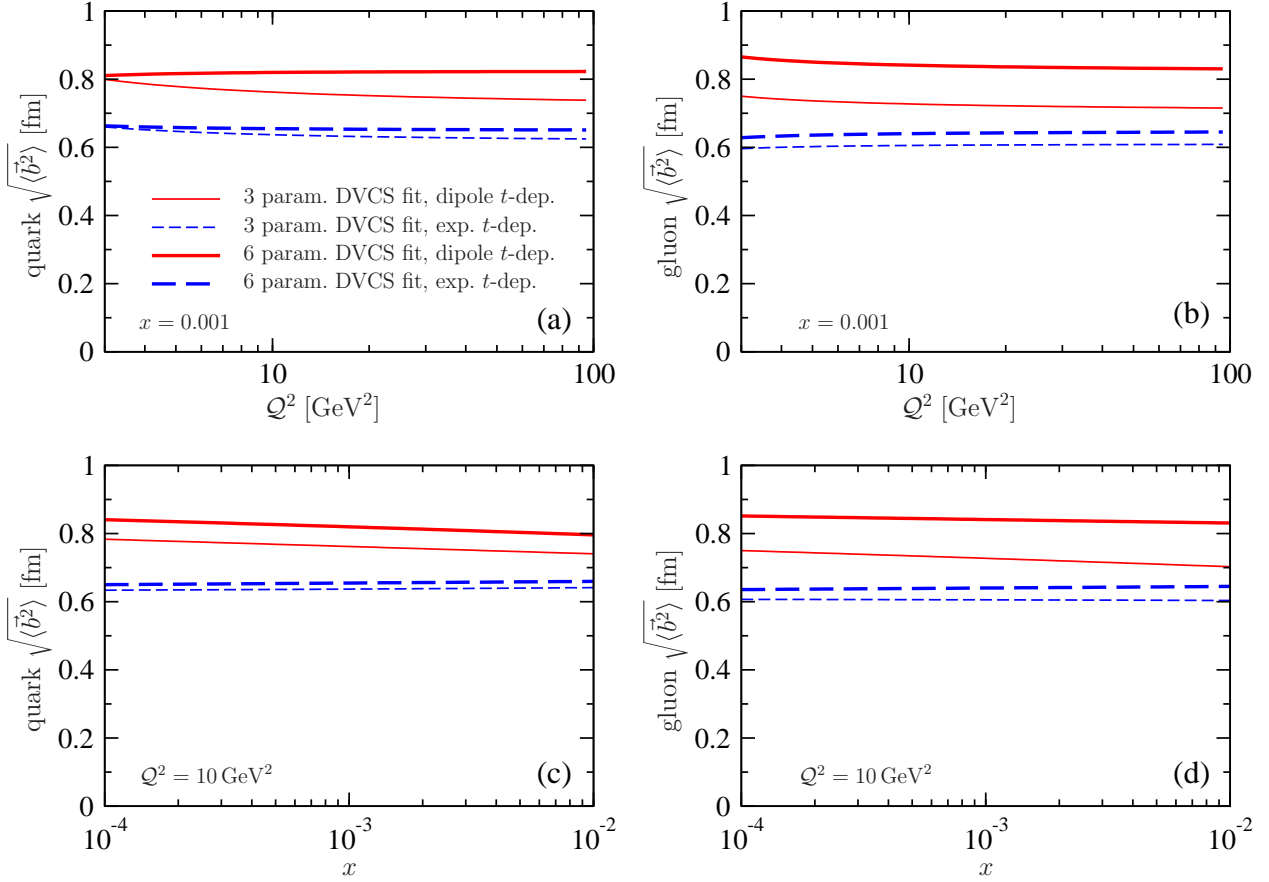}
\end{center}
\caption{\small
Quark and gluon transverse distribution
widths $\sqrt{\langle\vec{b}^2\rangle}$ with
respect to ${\cal Q}^2$ for $x=10^{-3}$ (a, b)
and with respect to $x$ for ${\cal Q}^2 = 10\, \GeV^2$ (c, d),
obtained from NLO \ms\ DVCS fits of
$\Sigma$-PW model with dipole  (solid) and
exponential (dashed) $t$-dependence.
Thin lines correspond to parameter choice from (\ref{Ans-t-dip})
or (\ref{Ans-t-exp}), while for thick lines first
the gluon residual $t$-slope parameter $M^G$ or $B^G$ and then both
$\alpha^{\prime}_{\rm sea}$ and $\alpha^{\prime}_{\rm G}$
parameters were released.
}
\label{fig:TrWidthNLO}
\end{figure}

A simultaneous release of all four $t$-related parameters ($B^{\rm
sea,G}, \alpha^{\prime}_{\rm sea,G}$) leads to a non-convergent
search for the minimal $\chi^2$. Therefore, to pin down also the
``Regge slope'' parameters, we use a three-step fitting procedure
(fitting to DIS data being the first step). We first release the
gluonic residue $t$-slope parameter, which yields for the
exponential $t$-dependence
\begin{eqnarray}
B^{\rm sea} =  2.85\, \GeV^{-2}\,,\;\;
s^{\rm sea} = 0.71\,,
\qquad
B^{\rm G} = 2.71\, \GeV^{-2}\,,\;\; s^{\rm G} = 0.36\,.
\end{eqnarray}
Now releasing the $\alpha_{\rm sea}^\prime=\alpha_{\rm
G}^\prime=0$ parameters (thick dashed line in
Fig.~\ref{fig:TrWidthNLO}), we find that they essentially stay at
zero. Compared with our three-parameter DVCS fit ansatz
(\ref{Ans-t-exp}) (thin dashed), where fixed parameters were taken
from $J/\Psi$ photoproduction, the gluonic $t$-slope only slightly
increases ($17\%$). Hence, we consider the two-gluon GPD models,
obtained from DVCS and $J/\Psi$ photoproduction, as compatible.
Moreover, the small values of the skewness parameters indicate
that the exponential ansatz in the \ms~scheme is mostly compatible
with the l-PW model, as pointed out in Sect.~\ref{SubSecSec-NLO}.

For the dipole ansatz the parameter set after the second fitting
step reads
\begin{eqnarray}
(M^{\rm sea})^2 =  0.54\,\GeV^2\,,\;\;
s^{\rm sea} = 5.24\,,
\qquad
(M^{\rm G})^2 = 0.47\,\GeV^2\,,\;\;
s^{\rm G} = 8.71
\end{eqnarray}
with $\alpha^\prime_{\rm sea}=\alpha^\prime_{\rm
G}=0.15/\GeV^{2}$. We realize that the quark skewness and cut-off
mass parameters only slightly change, cf.~Table
\ref{tab:consparsNLO}. The gluon cut-off mass moderately decreases
and the skewness parameter increases, compensating each other in
the CFF and so the cross section is well described. Now
releasing the $\alpha^\prime$ parameters we essentially find that
for the dipole ansatz (thick solid) they slightly reduce for
quarks and more significantly for gluons:
\begin{eqnarray}
\alpha^\prime_{\rm sea}= 0.15  \, \to  \,  0.12\;, \qquad
\alpha^\prime_{\rm G}= 0.15 \,   \to \,   0.08 \,.
\end{eqnarray}
The quality of fits, compared to those shown in Table
\ref{tab:conschiNLO} for fixed gluon slope parameters, does not
change.

Fig.~\ref{fig:TrWidthNLO} summarizes our NLO (\ms) findings for
the quark and gluon transverse width
$\sqrt{\langle\vec{b}^2\rangle}$, defined in
Eq.~(\ref{Def-tra-wid}). We note that for fixed gluon slope parameters
(thin lines) the gluonic transverse width, shown  in the right panel,
would fully coincide with our LO curves. The only moderate
difference appears in the dipole ansatz (thin solid), where the quark
transverse width decreases from  $\approx 0.9$ fm,
quoted in Eqs.~(\ref{TraWid-LO-Inp},\ref{TraWid-LO-10}), to
$\approx 0.8$ fm.  If we take the gluonic parameters from
the three-step fit, we see that the gluonic
transverse width for both the  dipole (solid) and
exponential (dashed) ansatz increases from
$\approx 0.7$ fm and  $\approx 0.6$ fm  to $\approx 0.85$ fm
and  $\approx 0.65$ fm, respectively.
However, this moderate difference for the gluons affects only slightly
the quark transverse width, see the left panel.
At NLO we observe already at the input scale that quark and gluon
transverse width are mostly the same. The values we can
quote
\begin{eqnarray}
\label{TraWid-NLO}
\sqrt{\langle \vec{b}^2\rangle}_{\rm sea}  \approx
\sqrt{\langle \vec{b}^2\rangle}_{\rm G} \approx
\left\{ {0.75-0.80 \atop   0.60-0.65}\right\} \mbox{fm}
\qquad\mbox{for}\;
\left\{{ \text{dipole}\: (\ref{Ans-t-dip})
   \atop \text{exponential}\: (\ref{Ans-t-exp}) }\right\} \; \text{ansatz}\,.
\end{eqnarray}
are stable under evolution.
Our results are fully robust with respect to scheme conventions or
radiative NNLO corrections.

\begin{figure}[ht]
\begin{center}
\includegraphics[scale=0.8]{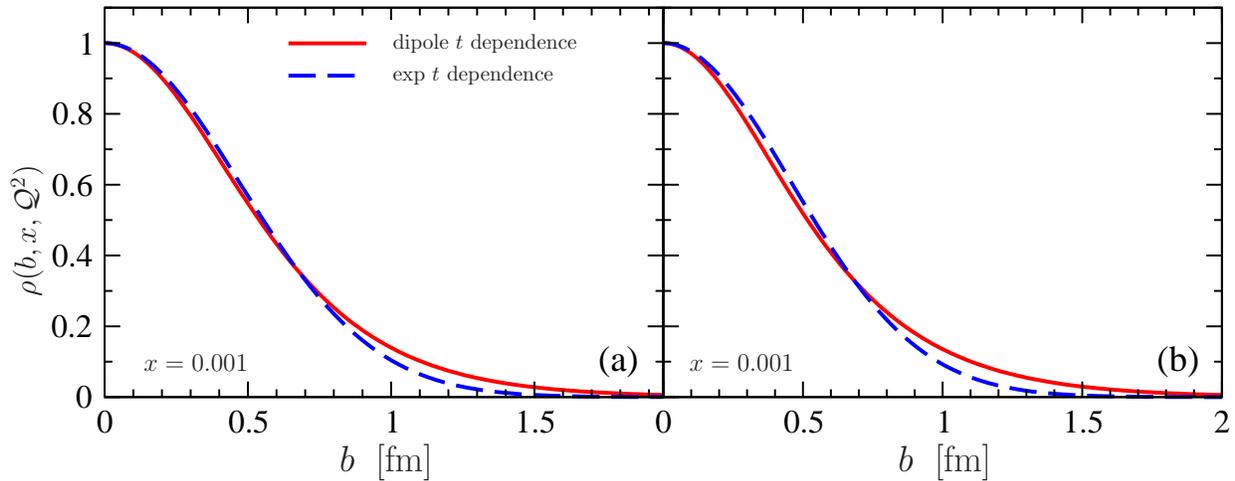}
\end{center}
\caption{\small
Quark (a) and gluon (b) transverse  profile function
(\ref{eq:TraProFun}) for ${\cal Q}^2=4\,\GeV^2$ and $x=10^{-3}$
is obtained from a six parameter DVCS fit, as in
Fig.~\ref{fig:TrWidthNLO}.
}
\label{fig:TraDis}
\end{figure}
We conclude that within our GPD models, having a flexible skewness
ratio and the same functional form of $t$-dependence for sea
quarks and gluons, the resolution of the transverse distribution
of partons is robust and that at small-$x$ (sea) quarks and gluons
have the same transverse width. For the exponential ansatz the
DVCS result $\approx 0.65$ fm is compatible with the one quoted in
Ref.~\cite{FraStrWei05}, namely $0.65$ fm. For the dipole ansatz,
we find a larger value $\approx 0.8$ fm, which simply reflects the
functional form we have employed. It is illustrated in
Fig.~\ref{fig:TraDis} that in impact space, the (normalized)
transverse profile function
\begin{eqnarray}
\label{eq:TraProFun}
\rho(b,x,{\cal Q}^2) =
\frac{
\int_{-\infty}^{\infty}\! d^2\vec{\Delta}\;
e^{i \vec{\Delta} \vec{b}} H(x,\eta=0,t=-\vec{\Delta}^2,{\cal Q}^2)
}{
\int_{-\infty}^{\infty}\! d^2\vec{\Delta}\;
H(x,\eta=0,t=-\vec{\Delta}^2,{\cal Q}^2)
}
\end{eqnarray}
for dipole and exponential $t$-dependence mainly differ for
distances larger than the disc radius of the proton, i.e., for $b>
0.6$ fm. Hence, the larger value of the transverse width for the
dipole ansatz arises from the long range tail of the profile
function, see the solid curve. Note that the model uncertainty in the
extrapolation of the GPD to $t=0$ corresponds to the uncertainty
in the long range tail.

\subsection{Is the anomalous gravitomagnetic moment accessible?}
\label{SubSecSec-gra}

Let us first adopt the ``classical'' Regge point of view in which
the chromomagnetic ``pomeron'' is absent, i.e.,  we can neglect as
above the CFF ${\cal E}$. Hence, the BCA  (\ref{Def-Asy}) is
already predicted by the outcome of our cross section fits
from the previous sections. A preliminary BCA measurement of the
H1 Collaboration with uncorrected acceptance effects has been
reported in Ref.~\cite{Sch07} for the kinematics
\begin{eqnarray}
\label{Kin-BCA}
0.05\,\GeV^2 \le |t| \lesssim 1\,\GeV^2   \,,\quad \langle
W\rangle =82\,\GeV\,, \quad \mbox{and} \quad \langle {\cal
Q}^2\rangle =8\,\GeV^2\,.
\end{eqnarray}
To evaluate the BCA (\ref{Def-Asy}), we integrate over the
$t$-interval and take the given mean values.
\begin{figure}[t]
\begin{center}
\includegraphics[scale=0.65]{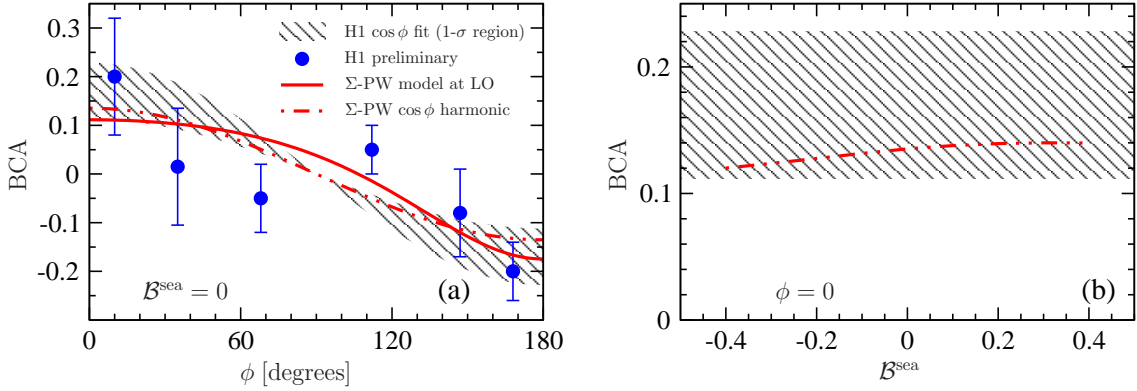}
\end{center}
\caption{$\cos\phi$ harmonic (dash-dot-dotted) of beam charge
asymmetry (BCA) for $\Sigma$-PW model fit at LO as a function of
azimuthal angle $\phi$ (a) or of anomalous gravitomagnetic moment
of sea quarks ${\cal B}^{\rm sea}=-{\cal B}^{\rm G}$ (b), compared
to preliminary H1 data \cite{Sch07}. Gray band is 1-$\sigma$
region of  H1 fit to $p_1 \cos\phi$, $p_1=0.17 \pm 0.3 \pm 0.5$,
with errors added in quadrature } \label{fig:BCA}
\end{figure}
In Fig.~\ref{fig:BCA} (a) we confront our LO prediction (solid
curve) for the $\Sigma$-PW model (\ref{Ans-t-dip}) with the
preliminary BCA measurement. Having in mind that we neglected the
$\cos(3 \phi)$ harmonics, induced by gluon transversity, and the
presumable small twist-three   $\cos(2\phi)$ harmonics, we can
safely state that our prediction is fully compatible with the
preliminary H1 measurement.

As explained in Sect.~\ref{Sec-DVCS}, accessing the twist-two CFFs
through the $\cos(\phi)$ harmonic in the Fourier decomposition
(\ref{BCA-FC}) is in the considered kinematics relatively clean.
We recall that this method can only diminish the contamination
from other CFFs.  The preliminary H1 fit,
\begin{eqnarray}
\label{BCA-H1-fit}
 p_1=0.17 \pm 0.3 \pm 0.5\,,
\end{eqnarray}
is shown in Fig.~\ref{fig:BCA}(a) as error band and agrees
well with our prediction (dashed curve), evaluated from the
expression (\ref{Def-Asy}). The deviation of the dashed curve from the
solid one is of kinematical origin and arises from the BH terms.
In the absence of large genuine twist-three contributions and gluon
transversity or its twist-four contamination, all other harmonics in
the Fourier decomposition (\ref{BCA-FC}) of the BCA  are predicted
to be small \cite{BelMueKir01}:
\begin{eqnarray}
0 < p_0 \lesssim  10^{-2} \,,
\quad
p_2 \sim -5 \cdot 10^{-2}\,,
\quad
0 < p_3 \lesssim  10^{-2} \,,
\end{eqnarray}
where higher harmonics for $i \ge 4$ can be considered as
negligible. We emphasize that a significant deviation from these
numbers would indicate that genuine twist-three and gluon
transversity (or twist-four corrections) are addressable in the
small-$\Bx$ kinematics, where it is assumed that electromagnetic
radiative corrections will not alter the angular dependence of the
BCA (\ref{Def-ChaAsy}).

To have a closer look to the parameter dependence of our
prediction, we write the approximation (\ref{Def-Asy-p1}) for
$t$-integrated data (valid for $|t_1| \gg -t_{\rm min} \approx
\Bx^2 M_p^2$) as
$$
p_1 \sim n
\frac{
\xi \int_{t_1}^{t_2}\! \frac{dt}{\sqrt{-t}} F_1  \Re{\rm e}{\cal H}
}{
h + \xi^2 \int_{t_1}^{t_2}\! dt  |{\cal H}|^2
}\Bigg|_{W= 82\,\GeV\atop {\cal Q}^2 = 8\, \GeV^2}\,,
\quad\mbox{with }\quad
t_1 = - 0.05\,\GeV^2\,,\;\;\; t_2 = -1\,\GeV^2\,,
$$
where $n$ and $h$ are two kinematical factors. Apart from the
skewness effect,  the size of $p_1$ is essentially governed by the
``pomeron trajectory''
$$
p_1 \propto \cot(\pi \alpha(\langle t\rangle,\langle {\cal Q}^2\rangle)/2)
\approx -\frac{\pi}{2}
\left(\alpha(\langle {\cal Q}^2\rangle)-1\right)\,,
$$
which for the mean values $\langle -t \rangle \approx 0.2\,\GeV^2$
and $\langle {\cal Q}^2\rangle=8\,\GeV^2$ can be replaced by the
intercept $\alpha(\langle {\cal Q}^2\rangle)$. The justification
for neglecting the $\alpha^\prime$ parameter can also be read off
from the last column in Table~\ref{tab:consparsLO}, which shows
that both ansaetze, with $\alpha^\prime = 0.15/\GeV^2$ and
$\alpha^\prime =0$, yield almost the same value for the BCA.
Hence,  in our Regge-inspired GPD framework the BCA prediction
arises to a great extent from the LO DIS fit. This is also
accompanied by a large uncertainty, namely, a small absolute error
of the DIS intercept $\alpha-1$ will induce a large relative one
for the BCA. Certainly, it would not be appropriate to quote here
the width of $\chi^2$ minimization curve as an error estimate,
since our model simplifications, e.g., neglecting ``Reggeon''
contributions, might influence the fit result for $\alpha-1$, too.
At NLO (NNLO) we observe that the BCA decreases to $p_1=0.9
(0.12)$, see last column in Table \ref{tab:consparsNLO}
(\ref{tab:consparsNNLO}), which corresponds to a 1-$\sigma$
deviation from the preliminary H1 mean value (\ref{BCA-H1-fit}).
The relative drop of the BCA size at NLO, compared to LO, seems
indeed to be correlated with that of the intercept,
cf.~corresponding columns in Tables \ref{tab:consparsLO} and
\ref{tab:consparsNLO}.

We mentioned in Sect.~\ref{Sec-DVCS} that a measurement of
$$
p_1 \sim n
\frac{
\xi \int_{t_1}^{t_2}\! \frac{dt}{\sqrt{-t}}
\left(
F_1  \Re{\rm e}{\cal H} - \frac{t}{4 M_p^2} F_2 \Re{\rm e}{\cal E}
\right)
}{
h + \xi^2 \int_{t_1}^{t_2}\! dt
\left(|{\cal H}|^2 - \frac{t}{4 M_p^2} |{\cal E}|^2\right)
}\Bigg|_{W= 82\,\GeV\atop {\cal Q}^2 = 8\, \GeV^2}\,,
$$
supplemented by the DVCS cross section measurement, should in
principle  allow to separate the CFFs ${\cal H}$ and ${\cal E}$.
It is obvious from the formula that a negative real part in $\cal
E $ will more strongly influence our BCA ``prediction'' than a
positive one.

To quantify the sensitivity of the BCA  in dependence  on ${\cal
B}^{\rm sea}$, we adopt the GPD $E$ model (\ref{Ans-E}) and rely
on the scenario ${\cal B}^{\rm G}=-{\cal B}^{\rm sea}$. We vary
${\cal B}^{\rm sea}$ at the input scale from  $-0.4\cdots 0.4$, which
corresponds to a variation of
$$
  -0.13 \le {\cal J}^{\rm sea}  \le 0.28
  \quad (0.43 \ge {\cal J}^{\rm G}\ge 0.02)
  \qquad \mbox{with} \quad {\cal J}^{\rm val} =0.2.
$$
In the right panel of Fig.~\ref{fig:BCA} we show the dependence of
the BCA (at $\phi=0$) on the ${\cal B}^{\rm sea}$ parameter for
the $\Sigma$-PW model at LO accuracy. We refitted for each given
value of ${\cal B}_{\rm sea}$ the model parameters of the GPD $H$
so that the DVCS cross section is described. As one realizes,
$p_1$ depends only slightly on the parameter ${\cal B}^{\rm sea}$;
its relative change is
$$
\delta p_1/p_1 \sim -15\% \cdots +5\% \quad\mbox{for}
\quad
{\cal B}^{\rm sea} = -0.4 \cdots 0.4.$$
We obtain rather similar findings for a ${\cal B}^{\rm G}=0$
scenario and within the nl-PW model. Contrarily to our
hopes, we conclude that both the model uncertainties of
$\cal H$ and the experimental errors are too large to obtain a
bound for ${\cal B}^{\rm sea}$.

\subsection{Lessons from fits}
\label{SubSec-Les}

We would like now to summarize the lessons from our fits,
presented in the previous sections. We recall that our DVCS
description relied on Regge-inspired GPD models, set up at a (low)
input scale,  and the collinear factorization framework  at
leading twist-two in LO and beyond. Both the factorization and
renormalization scale were set equal to the virtuality of the
incoming photon. Concerning strategies for fitting to the unpolarized
DIS structure function and DVCS cross section measurements we have
seen that
\begin{itemize}
\item GPD models with a flexible skewness ratio describe well the
 small-$\Bx$ DVCS cross section measurements at LO and beyond,
\item fitting parameters obtained from simultaneous (DVCS+DIS)
 and two-step (first DIS, than DVCS) fits are rather similar,
\item and that the partonic shrinkage effect can be addressed
by a three stage fitting strategy.
\end{itemize}
Our Regge-inspired modelling within the collinear factorization
framework, including the corresponding approximations, is
consistent with experimental measurements:
\begin{itemize}
\item small ``Regge slope'' parameters
$\alpha^\prime < 0.25/\GeV^2$ are favored and large ones for
``pomeron'' related quarks,
e.g., $\alpha^\prime \sim 1/\GeV^2$, are excluded,
\item the scale dependence of both the residual
$t$-dependence and ``Regge slope''  are compatible with pQCD,
\item and the real part of the amplitude, following from
DIS and DVCS cross section fits, predicts
the preliminary BCA measurement.
\end{itemize}

Relying on the assumption that the CFF ${\cal H}$ is dominant and
that higher twist contributions are unimportant, we conclude from
our fits in Sect.~\ref{SubSecSec-LO} that realistic GPD
models \emph{at LO} have the following properties:
\begin{itemize}
\item quark GPD models possess a skewness ratio $r^{\rm sea}\approx 1$
\item and to ensure this for a large ${\cal Q}^2$ lever arm,
 gluon GPDs have a skewness ratio $r^{\rm G} < 1$.
\end{itemize}
We emphasize that our finding $r^{\rm sea}\approx 1$ was predicted by
the aligned-jet model and it is realized in an RDDA for $b^{\rm
sea} \gg 1$. However, all popular gluonic GPD  models, possessing
$r^{\rm G}\approx 1$, are incompatible with the small-$\Bx$ DVCS data
(and presumably also in a pure LO description of data on hard
vector meson electroproduction, see, e.g., the dashed curves in
Fig.~5 of Ref.~\cite{GolKro07} that refer to the collinear
factorization prediction and are included there
for illustration).
The negative skewness ratio of our gluonic GPD models at the input
scale and the feature that it approaches the conformal ratio for
growing ${\cal Q}^2$ indicate that the parameterization of the
skewness function is rigid. We can safely state that
\begin{itemize}
\item the small-$x$ claim of Refs.~\cite{ShuBieMarRys99,MarNocRysShuTeu08}
 that the skewness ratio is equal to its conformal value is ruled out at LO
\item and a flexible parameterization of evolution has not been achieved so far.
\end{itemize}
The reader might be surprised by this second conclusion, which
contradicts opinions and model-dependent findings in the
literature. Nevertheless, we are not aware of a distinct investigation that
addresses the evolution of the residue function in dependence of
the initial condition, i.e., the skewness function in momentum fraction space,
the  {\em series} of conformal PWs in $J$ space,
the {\em  series} of forward-like functions $Q_\rho(z)$  in the ``dual model'',
or the {\em  series} of ``conformal sibling poles'' in Mellin space.

Including radiative corrections, we observed in Sect.~\ref{SubSecSec-NLO} that
NLO GPDs qualitatively differ from the LO ones:
\begin{itemize}
\item the quark skewness ratio in the \ms\ scheme  can be larger
  than the conformal one,
\item the quark skewness ratio in the \cs\ scheme matches the conformal one,
\item the scheme dependence at NLO is absorbed by a sizeable GPD reparameterization,
\item and the gluonic skewness ratio approaches in both schemes the conformal one.
\end{itemize}
The large reparameterization effects, in particular for quarks,
might be a consequence of the remaining rigidity of the skewness function in
our flexible models. We expect that this can be overcome by taking
into account two \emph{effective} non-leading SO(3)-partial waves
rather than one.

We would also like to emphasize that the scheme dependence at NLO, we
pursued in our studies, is entirely related to a `non-diagonal'
rotation, which shows that
\begin{itemize}
\item
the small-$x$ claim \cite{ShuBieMarRys99,MarNocRysShuTeu08} of
conformal skewness ratio can be sustained beyond LO only in
special schemes, however, presumably not in the \ms~one.
\end{itemize}
We add that our \ms~fit at NLO for GPD models with an exponential
$t$-dependence is consistent with the LO conformal skewness ratio.
In our opinion this fact does not necessarily support the
``logic'' of the small-$x$ claim, since it
relies on tree-level conformal symmetry
\cite{ShuBieMarRys99,MarNocRysShuTeu08} that is beyond LO not
explicitly implemented in the \ms~scheme. Hence, the ``predicted''
skewness ratio in the \ms~scheme at NLO differs from the LO one.

The stability of the perturbative approach can be presently
studied only in the \cs~scheme, and there
\begin{itemize}
\item the inclusion of NNLO corrections yields only a small change
of fitting parameters.
\end{itemize}
Thereby, the skewness ratio in the \cs~scheme is essentially given
by the conformal ratio (\ref{r-ratio-con}). Thus, in this scheme
beyond LO one might be tempted to conjecture that a GPD is indeed tied to
the corresponding PDF by conformal mapping. One might wonder
whether it is accidental that this hypothetical `holographical
principle' is realized within the group SO(2,1) \cite{KumMuePas08}.

The functional form of $t$-dependence cannot be pinned down from
present DVCS data. However, we can definitely state that:
\begin{itemize}
\item  ``Regge slope'' parameters are small at the input scale,
however, do not necessarily vanish,
\item non-vanishing ``Regge slope'' parameters rapidly
decrease with growing scale,
\item both residual dipole and exponential $t$-dependence
are compatible with present DVCS data,
\item within our models, there is a cross-talk between skewness and
  $t$-dependence at LO,
\item and beyond LO the skewness and $t$-dependence start to decouple
in our flexible models.
\end{itemize}
In a three-step fitting strategy we also observed  a correlation
between the functional form of the residue function and the
partonic shrinkage effect. Namely, an exponential $t$-dependence of
the former is quite consistent with a zero shrinkage effect, i.e.,
$\alpha^\prime_{\rm sea}=\alpha^\prime_{\rm G}=0$.

Our partonic interpretation of the transverse degrees of freedom,
given in Sect.~\ref{SecSubSub-TraDis}, is rather robust. Relying
on the decoupling of $t$- and skewness-dependence, we have found
for dipole (exponential) ansatz that
\begin{itemize}
\item sea quarks and gluons have roughly the same transverse
width $\sim 0.75 (0.65)$   fm,
\item the transverse width is rather stable under evolution,
\item and a possible partonic shrinkage effect at a
lower scale is rapidly washed out with growing  scale.
\end{itemize}

Unfortunately, we saw in Sect.~\ref{SubSecSec-gra} that both
theoretical uncertainties and experimental errors do {\em not allow}
to address
\begin{itemize}
\item the chromomagnetic ``pomeron'' or angular momentum of sea quarks,
\item and gluon transversity.
\end{itemize}

\section{Small-$\boldsymbol \Bx$ fit results as input for ``dispersion relation'' fits}
\label{Sec-DisRelFit}

The reader might wonder why have we repeatedly stressed that the
claim \cite{ShuBieMarRys99,MarNocRysShuTeu08} that at small-$x$ skewness ratio
has a conformal value (\ref{r-ratio-con}) is excluded at LO,
when this value seems to be realized in the \cs~scheme beyond LO.
However, this is entirely related to the ambiguities present in our
factorization conventions and partonic interpretations.
For instance, in a DVCS scheme, where the gluons are not
resolved in the hard-scattering subprocess, above LO findings at
the input scale would not be essentially altered by radiative corrections,
which only modify the  pQCD evolution predictions.

Such an interpretation is favored among model builders and also in
the GPD phenomenology of fixed target kinematics. We now utilize
our LO analysis of the small-$\Bx$ data for a ``dispersion relation''
fit in fixed target kinematics, where real part of CFF is taken to
be determined by imaginary part and subtraction constant, as
described in Sect.~\ref{Sec-DVCS} (see also
Ref.~\cite{KumMuePas08} for more sophisticated strategies and
Ref.~\cite{Gui08a} for an alternative approach). This substantially
reduces the model uncertainties in common GPD model description.
For fixed target kinematics, where the ${\cal Q}^2$ lever arm  is
rather limited, one may additionally rely on the so-called scaling
hypothesis, i.e., on the assumption that the GPD does not evolve
under the change of the photon virtuality. The primary goal of
such fits is to reveal the shape of the dominant GPD $H$ on its
cross-over line ($\eta=x$) from DVCS measurements on unpolarized
proton target.

The framework, as described in Sect.~\ref{Sec-DVCS} for
small $\Bx$ can be easily adapted. It is beyond the scope of this
paper to present a detailed description of our fits; however, we
would like to demonstrate that GPD phenomenology for DVCS can be
straightforwardly set up. We consider now three active quarks and
write the partonic decomposition of $\Im{\rm m}\mathcal{H}$ as
\begin{eqnarray}
\label{SpeFunH-ParDec}
\Im{\rm m}{\cal H}(\xi,t) =
\pi \left[
\frac{4}{9} H^{u_{\rm val}}(\xi,\xi,t)
+
\frac{1}{9} H^{d_{\rm val}} (\xi,\xi,t)
+
\frac{2}{9} H^{\rm sea}(\xi,\xi,t)
\right]\,,
\end{eqnarray}
taken at an ``input scale'' of ${\cal Q}^2 = 2\,\GeV^2$.
We model the GPD on the cross-over line using the DD
representation (\ref{Def-DD}),
\begin{eqnarray}
\label{DD-Rep2GPD} F(x, x,t) =  \frac{2}{1+x} \int_{0}^1\!
du\, f\!\left(\!\frac{u x}{1+x},\frac{1-2 u +x}{1+ x},t\!\right)\,,
\end{eqnarray}
and take the $t$-dependence from the quark spectator model \cite{HwaMue07}.
This  suggests the following functional form:
\begin{eqnarray}
\label{GPD-Ans}
H(x,x,t)  =
\frac{n\, r}{1+x}
\left(\frac{2 x}{1+x}\right)^{-\alpha(t)}
\left(\frac{1-x}{1+x}\right)^{b}
\frac{1}{\left(1-  \frac{1-x}{1+x} \frac{t}{M^2}\right)^{p}}\,.
\end{eqnarray}
Here $n$ is the residue normalization of the PDF, which can be
taken from PDF fits, $r$ is the skewness ratio at small $x$,
$\alpha(t)$ is the ``Regge trajectory'', which can be borrowed
from Regge phenomenology, $b$ controls the  large $x$ behavior,
which according to counting rules \cite{Yua03} should be different
from that of PDFs, and both $p$ and $M$ control the
$t$-dependence. The functional dependence on $t$ and $x$ in the
$t$-dependent part of the ansatz is specifically motivated by the
spectator model \cite{HwaMue07}. The small-$x$ behavior of the sea
quarks is taken from our small-$\Bx$ DVCS fit at LO, evolved
backwards. The corresponding GPD is described by the model
(\ref{GPD-Ans}) within the parameters
\begin{eqnarray}
\label{eq:GloFit-fix-sea}
\alpha^{\rm sea}(t) = 1.13 + 0.15\, t/\GeV^2\,,
\quad
n^{\rm sea} =1.5\,, \quad r^{\rm sea} =1\,,
\quad
(M^{\rm sea})^2=0.5\,\GeV^2\,,
\quad   p^{\rm sea}= 2\,.
\end{eqnarray}

For valence quarks, in accordance with our Regge-inspired
modelling, which implies that we take the $\rho$ and $\omega$
trajectories, and standard PDF parameterizations, e.g.,
Ref.~\cite{Ale02}, we fix quark parameters to be:
\begin{eqnarray}
\label{eq:GloFit-fix-val}
\alpha^{\rm val}(t) =  0.43 + 0.85\, t/\GeV^2\,,
\quad
n^{\rm val} =1.35\,,
\quad
(M^{\rm val})^2=0.64\,\GeV^2\,,
\quad
 p^{\rm val}=1\,,
\end{eqnarray}
i.e., we take a monopole ansatz with fixed cut-off mass for the
residual $t$-dependence. Let us argue that our  model
for the GPD on the cross-over line is generically consistent with some full
valence-quark GPD model that satisfies the form factor sum rule%
\footnote{Note that in popular Regge-inspired GPD models,
used in DVCS phenomenology, the form factor constraint, which is already
lost on generic level, is not implemented. Unfortunately, such models are
sometimes employed to ``constrain'' quark angular momentum.}.
Our choice $p^{\rm val}=1$ is motivated by the dimensional counting rule for the Dirac form
factor, which states that this form factor behaves as $(-t)^{-2}$
for $-t \to \infty$. Loosely spoken, the missing power here would
arise in our GPD model from the $x$-integration for fixed $\eta$,
e.g., $\eta=0$, providing us another monopole with cut-off mass
$$
\sqrt{\frac{1-\alpha^{\rm val}}{\alpha^{\prime{\rm val}}}}\approx
M^{\rm val}= 0.8\, \GeV .
$$
This choice is consistent with the
characteristic nucleon scale and our ansatz is roughly associated
with an {\em effective} dipole parameterization for the Dirac form
factor with a cut-off mass of $M^{\rm val} \approx 0.8\,\GeV.$
Such an effective parameterization of the Dirac form factor is for
small $-t \lesssim 0.5\,\GeV^2$ region, on which we are
interested, at the 20\% level consistent with the standard dipole
parameterization, used above in Eq.~(\ref{Rad-ProDis}).

Furthermore, we take into account the pion-pole contribution \cite{PenPolGoe99},
displayed in Eq.~(153) of Ref.~\cite{BelMueKir01}, which, however, is a
non-dominant contribution in DVCS. More important is the subtraction constant
(\ref{Def-DisRel})
\begin{eqnarray}
{\cal C}(t) =  \frac{C}{\left(1- \frac{t}{(M^{\rm sub})^2}\right)^2},
\end{eqnarray}
where the normalization $C$ and dipole cut-off mass are fitting
parameters. According to the ``dispersion relation'' (\ref{Def-DisRel}),
we also include this constant in the non-dominant CFF ${\cal E} = {\cal C}(t)$,
however, set its imaginary part as well as the CFF $\widetilde {\cal H}$ to zero.

The five fitting parameters,
$$
b^{\rm sea}\,,\qquad
r^{\rm val}\,,\;\; b^{\rm val}\,, \qquad C\,,\;\; M^{\rm sub},
$$
control the large- and small-$x$ behavior of the imaginary and the
real part of ${\cal H}$. Note that by a ``dispersion relation''
fitting procedure one accesses even the large, experimentally
inaccessible $x$ region. We also recall that the subtraction
constant (\ref{Def-DisRel}) can be evaluated from the so-called
$D$-term and it should not be confused with a so-called $J=0$
(fixed) pole \cite{BroCloGun71,BroCloGuo72} in the context of
merging parton model and Regge phenomenology, see footnote
\ref{foonot-J=0}, the discussion below Eq.~(\ref{Def-d-3/2}) in
Sect.~\ref{subsec-Fits}, and for more details
Refs.~\cite{KumMuePas07,KumMuePas08}.

In the fitting procedure we utilize the formula set of
Ref.~\cite{BelMueKir01}, which are based on $1/{\cal Q}^2$
approximation for the squared amplitude. Unfortunately, this
approximation is inappropriate for JLAB kinematics; still, it can
be improved in the unpolarized target case by a `hot fix', taken
from a refined evaluation for a scalar target \cite{BelMue08}. The
available fixed target data for unpolarized proton are the BCA
from HERMES \cite{Airetal06,Airetal08}, the beam spin asymmetry (BSA) from
CLAS \cite{Giretal07}, and the  electroproduction cross section measurements from
HALL A \cite{Cametal06}. The covered kinematics in $\xi(=x)$ is
depicted below in panel (a) of Fig.~\ref{fig:Glo}.

\begin{figure}[t]
\includegraphics[clip,scale=0.85]{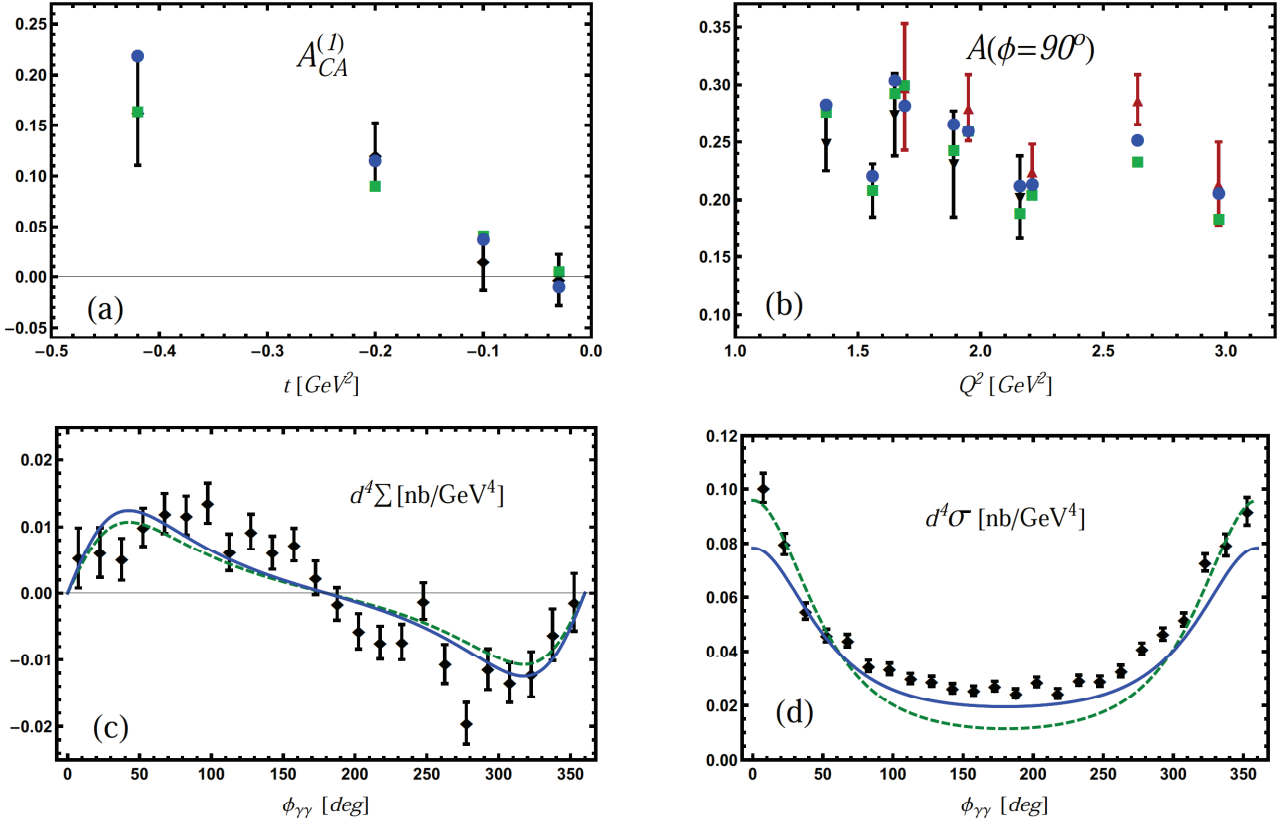}
\caption{\small Description of data by global model fits to HERMES \cite{Airetal08}
and CLAS \cite{Giretal07} data (squares, dashed line) and additionally including
the HALL A measurements (circles, solid line): (a) BCA
$A^{(1)}_{\rm CA}$ from HERMES versus $t$ (diamonds), (b) BSA
$A(\phi=\pi/2)$ from CLAS versus ${\cal Q}^2$ for $t \sim
-0.15\,\GeV^{2}$ (upside triangles) and $t \sim -0.3\, \GeV^{2}$
(triangles), (c) helicity-dependent and (d) unpolarized differential cross
section from HALL A for $\Bx=0.36$, $-t=0.33\, \GeV^2$, and ${\cal Q}^2=2.3\,\GeV^2$
(diamonds). Note that in (d) the $\cos(2\phi)$ harmonic of the
interference term is not included.}
\label{fig:Glo0}
\end{figure}
Within the hypothesis that $\cal H$ is the dominant CFF we can
describe HERMES and CLAS data, where in the later case only small
$-t  \ll {\cal Q}^2$ data were taken into account. Altogether we
included 36 data points and found with $$\chi^2/{\rm d.o.f.}
\approx 32/31$$ an acceptable fit, see squares in panels (a) and
(b) of Fig.~\ref{fig:Glo0}. The fit also predicts a
preliminary BSA measurement from the HERMES
Collaboration \cite{Ell07} (not shown) as well as, to some extent, the
helicity-dependent cross
section%
\footnote{We adopt here the terminology of HALL A
\cite{Cametal06}, where this observable is defined as half of the
cross section difference for positive and negative electron beam
helicity.} measurement from the Hall A Collaboration, however, not
the unpolarized (helicity-independent) one at $-t=0.33\, \GeV^2$,
displayed by dashed curves in panels (c) and (d) of
Fig.~\ref{fig:Glo0}, respectively.  It provides the following
parameter set for our GPD model:
\begin{eqnarray}
\label{FitA}
b^{\rm sea} =3.09\,,
\qquad
r^{\rm val} =0.95\,,\;\; b^{\rm val}=0.45\,,
\qquad
 C=0.24\,,\;\; M^{\rm sub}=0.5\,\GeV\,.
\end{eqnarray}
The $b$ values are smaller than the corresponding $\beta$ ones
for PDFs, which is in accordance with Ref.~\cite{Yua03}. The
skewness ratio is smaller than the conformal one, which is in
this case $\approx 1.2$. However, one should
bear in mind that we did not include ``Reggeon exchanges'' in the
flavor singlet sector. The parameter of $C$ is here smaller than
suggested by chiral quark soliton model
\cite{GoePolVan01,SchBofRad02,Wak07,Goeetal07} or lattice
\cite{Hagetal07} calculations and the $t$-dependence of the
subtraction constant is rather steep.

The hypothesis of the $H$-dominance is within our model
(\ref{GPD-Ans}) no longer valid for the cross section measurements
of HALL A \cite{Cametal06}, performed at large $\Bx=0.36$, ${\cal Q}^2=2.3\,\GeV^2$, and for
four $t$ values, $$-t=\{0.17\,\GeV^2,0.23\,\GeV^2, 0.28\,\GeV^2,
0.33\,\GeV^2 \}.$$ To describe the {\em full} set of these HALL A
data, it is required that the real part of the DVCS amplitude
steeply varies in this $t$ interval and becomes large at
$-t=0.33\, \GeV^2$. Both of these features are not implemented in
common GPD models~\cite{PolVan08,Gui08}. We will now  explore whether these features
of the real part arise from the inclusion of the GPD $\widetilde
H$. To be specific, we take a model for the CFF
$$
\Im{\rm m}\widetilde {\cal H}(\xi,t) = \pi \left(2 \frac{4}{9}  + \frac{1}{9}\right) \frac{ \widetilde{n}}{1+\xi}
\left(\frac{2 \xi}{1+\xi}\right)^{-\alpha^{\rm val}(t)}
\frac{1}{1-  \frac{1-\xi}{1+\xi} \frac{t}{(M^{\rm val})^2}} \left(\frac{1-\xi}{1+\xi}\right)^{3/2}\,,
$$
where for simplicity we employ the parameter set
(\ref{eq:GloFit-fix-val}) and fixed $b=3/2$. Hence, only the
normalization  $\widetilde{n}$ appears as a new fitting parameter.
If we take this parameter of order one, our ansatz is to some
extent consistent with standard GPD models.

In our second global fit  we take the DVCS data set from above and include four values of
the azimuthal angle asymmetry
$$
\int_{0}^{2\pi}\!{dw}\, \cos(\phi) \frac{d^4\sigma(\phi,\cdots)}{d\phi d\Bx dt d{\cal Q}^2}
\Big/
\int_{0}^{2\pi}\!{dw}\,  \frac{d^4\sigma(\phi,\cdots)}{d\phi d\Bx dt d{\cal Q}^2}\,,
$$
which are obtained from the four unpolarized cross section
measurements of the  HALL A Collaboration. Here, the integral
measure $dw$, defined in Eq.~(103) of Ref.~\cite{BelMueKir01},
compensates the $\phi$ dependence of the Bethe-Heitler propagator
product. In other words, in our fit only the relative changes of
the {\em unpolarized} cross section with respect to a $\cos\phi$
modulation, dominated on amplitude level by twist-two CFFs, is
essential. Within such a strategy the absolute normalization of
the cross section is irrelevant.

Compared to the previous global DVCS fit, this provides us four
data points and one degree of freedom (normalization of $\widetilde{\cal H}$) more. The quality of fit
$$\chi^2/{\rm d.o.f.} \approx 33/34$$ is good and the parameters
for the GPD $H$ are
\begin{eqnarray}
\label{FitB}
b^{\rm sea} =4.6\,,
\qquad
r^{\rm val} =1.11\,,\;\; b^{\rm val}=2.4\,,
\qquad C=6.0\,,\;\; M^{\rm sub}=1.5 \,\GeV\,.
\end{eqnarray}
The value of the subtraction constant agrees now qualitatively
with the original chiral quark soliton model estimate
\cite{GoePolVan01}, quoted as $C\approx 5$. For the normalization
of $\widetilde {\cal H}$  we find the value $\widetilde n \approx
3$. This corresponds to  a huge skewness effect $r_{\widetilde
{\cal H}} \sim 5$. We consider this finding as an {\em effective}
parameterization of some GPD contribution that reveals our lack of
partonic understanding of the unpolarized cross section
measurement of the HALL A Collaboration. We also like to stress
that the real part of the DVCS amplitude at larger values of $\Bx$
is an intricate quantity that cannot be simply related to a
specific GPD property. Note also that a closer look reveals that
essentially only the HALL A data point at $-t= 0.33\,\GeV^2$
forces us to explore the scenario of non-standard GPD modelling.

In this way, our model (\ref{FitB}) describes present DVCS data, including the
helicity-dependent cross section \cite{Cametal06} and preliminary
BSA \cite{Ell07} data which have not been employed in the fits.
This is exemplified by the circles and the solid curves in
Fig.~\ref{fig:Glo0}. Note that in panel (d) the deviation of our
fit result from the data points  might be attributed to a
$\cos(2\phi)$ harmonic in the interference term. However, the
extraction of GPDs from a DVCS measurement is highly nontrivial
and we relied here on model assumptions and employed a simple
least square fit. Thus, we like to spell out a clear warning,
namely, the result we presented does not exclude a successful
description of the data within a rather different set of
parameters:
\begin{itemize}
\item
a literal interpretation
of these (first) model-dependent fit results is not appropriate.
\end{itemize}
Nevertheless, the parameters (\ref{FitB}) look rather reasonable
from the generic point of view, as stated above, and they might be
used to set up an $H$ GPD model in any favored representation.

\begin{figure}[t]
\centerline{\includegraphics[clip,scale=0.45]{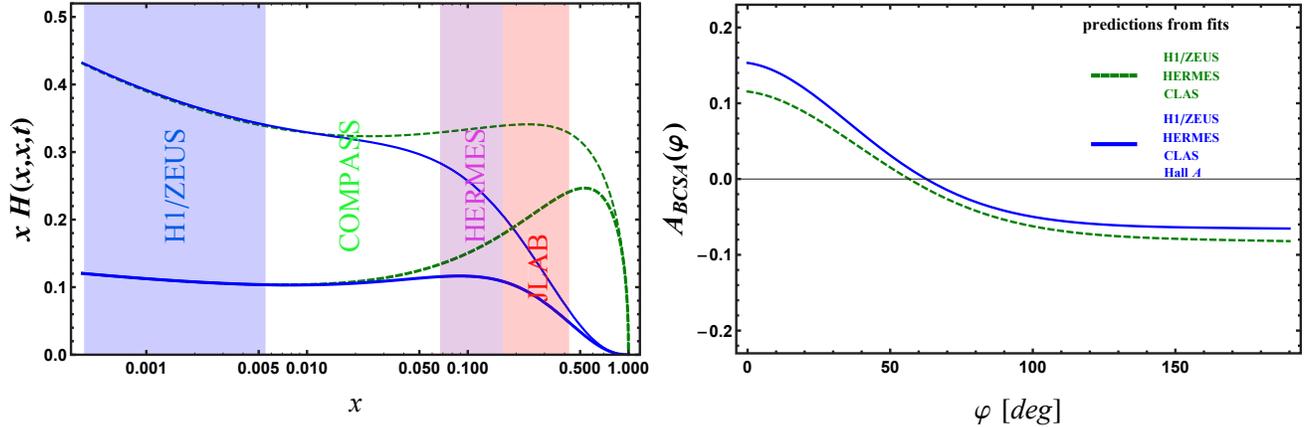}}
\caption{\small
(a) Global GPD $H$ fits (\ref{FitA}) and (\ref{FitB})  at  $t=-0.3\,\GeV^2$ (thick)
and $t=0$ (thin) are displayed as dashed and solid lines, respectively.
(b) Prediction of the BCSA asymmetry (\ref{Def-ChaSpiAsy}) for COMPASS kinematics
($E_{\mu} =160\, \GeV$,  ${\cal Q}^2=2\, \GeV^2$, $t= -0.2\, \GeV^2$) versus $\varphi = \pi -\phi$.
}
\label{fig:Glo}
\end{figure}
The resulting GPDs on the cross-over line are shown in the left
panel of  Fig.~(\ref{fig:Glo}). It is obvious that the two
different fits provide results which are rather different in the
large-$x$ region, but approach each other at smaller $x$.
Essentially, large difference for JLAB kinematics simply indicates
that BSA measurements can be described with two qualitatively
different GPD model scenarios, where the dashed curve represents
the common GPD models with a, let us say, moderate DVCS amplitude,
while the solid curve belongs to GPD models that provide an
enhanced DVCS amplitude, here induced by $\widetilde H$. Whether
this is a realistic scenario is a open problem, which should be
addressed in future studies.

We can now employ our model fits to deliver a prediction for the
COMPASS experiment. In such a fixed target experiment
one would scatter positively or negatively charged muons with
helicity $-1/2$ and $+1/2$  on a proton target.
The preferred observable is the  beam charge-spin asymmetry (BCSA),
\begin{eqnarray}
\label{Def-ChaSpiAsy}
A_{\rm BCSA}(\phi)=
\frac{d\sigma^{\downarrow+}\sigma- d\sigma^{\uparrow-}
}{
d^{\downarrow+}\sigma+ d^{\uparrow-}\sigma}\,,
\end{eqnarray}
which is essentially related to the real part of the interference
term. It is therefore rather sensitive to  details of the spectral
function $\Im{m}\mathcal{H}$. We recall that the sign of the BCA
and, thus, also of the real part of the CFF $\cal H$ changes
somewhere between HERMES and H1/ZEUS kinematics. Knowing the
position of this zero in dependence on the kinematical variables
would be crucial for pinning down GPDs. For the mean values at
COMPASS, we expect a sizeable asymmetry, the sign of which is
fixed by valence-like quarks (or ``Reggeon exchange''), where the
subtraction constant (fixed pole) plays a role, too. We display
our predictions resulting from our two fits for the BCSA
(\ref{Def-ChaSpiAsy}) in the right panel of Fig.~\ref{fig:Glo}
versus the azimuthal angle $\varphi=\pi - \phi$ (i.e., within a
so-called Trento convention). Therefore, the COMPASS kinematics $x
\sim  [10^{-2},10^{-1}]$ is certainly well suited to explore, in
the LO DVCS interpretation, the transition area between sea and
valence-like quarks domination. In the NLO interpretation this
translates into the hope of exploring the interplay between quarks
and gluons.

\section{Summary and conclusions}

In this paper we demonstrated that for the description of
small-$\Bx$ DVCS data the choice of representation, used to set up
GPD models, does not matter too much. We employed different
``languages'' and found that, although explicit transformation
formulae are only partly available, one can easily translate main
results. The problem remains always the same, namely, to find a
flexible parameterization of the skewness function and the
skewness ratio. As a side remark, we explained that the small-$x$
claim, stating that the GPD is rigidly tied to the corresponding
PDF by the conformal skewness ratio, is based on unjustified
mathematical assumptions. Moreover, we demonstrated in detail that
this claim is ruled out at LO. We also revealed that common gluon
GPD models (either RDDA, $t$-decorated PDF, leading PW or
minimalist ``dual'' model) possess the conformal ratio. We showed
then that this is the very feature which is responsible for the
failure of previous attempts to describe DVCS at LO, even if the
quark GPD model is flexible. The ``workaround recipe'' to take
NLO gluon PDFs in an LO description fails for our PDF
parameterizations in DVCS. It also creates an obvious inconsistency
because the same amplitude is described differently in DVCS than
in its DIS limit.

We introduced two flexible GPD parameterizations with respect to
the skewness ratio, namely, one model was set up by adding a
next-leading SO(3) PW and the other one by a model-dependent
resummation of SO(3) PWs. Thereby, we relied on the simplest
assumptions, e.g., (almost) decoupled skewness and $t$-dependence,
and dressed the gluon GPD with the $t$-dependence from the
$J/\Psi$ production analysis. However, it turned out that these
two models are rather similar and possess a rigid parametrization
of the skewness function. Nevertheless, within these
Regge-inspired models we could describe small-$\Bx$ DVCS cross
section data from the H1/ZEUS Collaborations at LO and beyond.
Thereby, the gluonic LO GPD on its cross-over line becomes
negative at lower values of ${\cal Q}^2$ and, moreover, we had a
rather large reparameterization of the skewness effect at NLO for
both the quark and gluon GPDs. Both findings are not expected and
we consider them as model artifacts. So we  conclude  that
the effective nl-PW, or the minimal ``dual'', or any similar model
is not flexible enough to control the evolution over a large lever
arm in ${\cal Q}^2$, and so also the control of the skewness ratio
at larger values of ${\cal Q}^2$ is lost. Also, the fact that at LO
the skewness and $t$-dependence are still correlated in our model
fits reveals that an effective nl-SO(3) PW parameterization
still suffers from rigidity.

We have demonstrated that at NLO the abovementioned small-$x$
claim within the HERA-II DVCS data also does not necessarily hold
true. Within a dipole $t$-ansatz, however, we have confirmed our
previous results that claim is to a great extent valid in the
\cs~scheme, at NLO and NNLO.  Thereby, we observed very small
NNLO corrections, namely, below $1\%$.  This might indicate that
the `holographic' principle, which ties GPD and PDF, arises from a
broken SO(2,1) symmetry. We stress that we do not consider such
`holographic' principle reliable enough to employ it in a fitting
procedure, e.g., to reveal the $t$-dependence of GPDs.

We utilized our flexible models to deliver the transverse
distribution of quarks and gluons and found that they are quite
robust with respect to the change of perturbative order and scheme
conventions. We only observed a slight difference between LO and
beyond LO results, which is attributed to the different fitting
strategies and the remaining rigidity of our models. The results
moderately differ from our previous findings, where we employed
the l-PW model (conforming to the small-$x$ claim). In contrast to
those findings, we observed now that within our flexible GPD
models there is no significant difference between quark and gluon
transverse width. The perturbative prediction that a partonic
shrinkage effect should be washed out at larger scale ${\cal Q}^2
\sim 10\,\GeV^2$ is in agreement with experimental findings,
however, its existence at the lower scale cannot be excluded. We
employed a three-step fitting procedure to pin down the
$t$-dependence of the GPD. Within an exponential $t$-dependence
GPD model there is no shrinkage effect and the partonic transverse
width is about $\sim 0.6-0.65$ fm. This value is compatible with
the one extracted from the $J/\Psi$ photoproduction and with the
radius of the proton disc.  A dipole $t$-dependence ansatz is
accompanied with a small shrinkage effect and the partonic
transverse width is now $0.75-0.8$ fm. The quoted difference of
widths arises essentially from the different extrapolation of the
measured $t$-interval to $t=0$ and essentially provides an
uncertainty in the long range tail of the profile function in
impact space.

Based on the DVCS cross section fits, we refined a previous model
prediction for the beam charge asymmetry. Our prediction is
compatible with preliminary measurements from the H1
Collaboration, which supports our Regge-inspired GPD modelling.
Unfortunately, both theoretical uncertainties and experimental
errors do not allow us to access the chromomagnetic ``pomeron'' and
so neither the anomalous gravitomagnetic moment (or angular
momentum) of sea quarks. An immense reduction of both sources
of error is needed
before one might be able to address these questions. An improvement
of the experimental result might lead to some clarification of the presence of
a third azimuthal angular harmonic, induced by gluon transversity.

Based on the double distribution representation, we have also
built a simple model for the GPD at the cross-over line, where we
implemented the small-$x$ behavior, extracted from H1/ZEUS data.
Using this model and relying  on the scaling hypothesis, we
presented a first ``dispersion relation'' fit to observables for fixed
target DVCS experiments on unpolarized proton target. Assuming
some unexpected properties of $\widetilde H$, we were able to
describe all available small $-t$ data in a first global fit. The
parameters we found match our generic expectations coming from
Regge behavior, large-$x$ counting, and quark soliton model
estimates. Utilizing this GPD as an input, we predict beam
charge-spin asymmetry as measurable at COMPASS. Thereby, it
becomes obvious that COMPASS is needed to reveal the GPD $H$
further. This experiment might also be important for pinning down
the transverse distribution of partons in the transition region
between ``pomeron'' dominance and ``Reggeon'' behavior.

Let us finally emphasize that one can have different partonic
interpretations of our findings. Namely, one can either state that
gluons are very important in DVCS and should be perturbatively
resolved or one sticks to the quark picture (DVCS scheme) in which
gluons only enter the evolution equation. This is a matter of
taste that so far simply provides two qualitatively different GPD
parameterizations and also points in two different directions for
future DVCS studies. We emphasize that in any case an LO description is a
convenient starting point for the phenomenological description of the
growing amount of experimental data. It offers the possibility to
pin down GPD models within rather straightforward fitting
strategies. Such models are phenomenologically valuable for making
contact with dynamical models and lattice simulations, since in
the latter case matching of lattice and perturbative
renormalization is for GPD moments only done at LO.

\subsection*{Acknowledgements}

\noindent For discussions on DVCS measurements we are indebted to
our experimental colleagues C.~M.~Camacho, L.~Favart, F.~X.~Girod,
N.~d'Hose, R.~Kaiser, W.-D.~Nowak, L.~Schoeffel, and D.~Zeiler.
For discussions on GPD representations we would like to thank D.~Diakonov,
M.~Diehl, M.~V.~Polyakov and K.~M.~Semenov-Tian-Shansky. We are grateful to
V.~Guzey and T.~Teckentrup for resolving the incompatibility of
numerical findings, while for an
exchange of opinions on the small-$x$ claim we would like to thank
A.~Martin, C.~Nockles, M.~Ryskin, A.~Shuvaev, and T.~Teubner.
K.K.~is grateful to the Institut f\"{u}r Theoretische Physik II at
Ruhr-Universit\"{a}t Bochum and D.M.~to the Department of Physics
of the Faculty of Science at the University of Zagreb for a warm
hospitality. This work was supported by the Croatian Ministry of
Science, Education and Sport, contract no. 119-0982930-1016, and
by the German Research Foundation contract DFG 436 KRO 113/11/0-1.

{\ }

\noindent
{\bf Note added:} After our manuscript was finalized,
we have noticed the new DVCS measurements from the ZEUS Collaboration \cite{Cheetal08}.
We have convinced ourselves that these data will only mildly
influence our fits and, thus, we do not update here our results.
In other words, we might view our findings as predictions for the
new ZEUS data. After the experimental DVCS analysis of HERA II run
will be finished, we plan to include them in a forthcoming update.


\begin{thebibliography}{100}

\bibitem{Nadetal08}
P.~M. Nadolsky {\em et~al.},
\newblock Phys. Rev. {\bf D78}, 013004 (2008), [0802.0007].

\bibitem{MarStiThoWat09}
A.~D. Martin, W.~J. Stirling, R.~S. Thorne and G.~Watt,
\newblock {Parton distributions for the LHC},
\newblock 0901.0002, 2009.

\bibitem{AbrFraStr95}
H.~Abramowicz, L.~Frankfurt and M.~Strikman,
\newblock Surveys High Energ. Phys. {\bf 11}, 51 (1997), [hep-ph/9503437].

\bibitem{AbrCal98}
H.~Abramowicz and A.~Caldwell,
\newblock Rev. Mod. Phys. {\bf 71}, 1275 (1999), [hep-ex/9903037].

\bibitem{KleYos08}
M.~Klein and R.~Yoshida,
\newblock Prog. Part. Nucl. Phys. {\bf 61}, 343 (2008), [0805.3334].

\bibitem{Adletal01}
H1, C.~Adloff {\em et~al.},
\newblock Phys. Lett. {\bf B517}, 47 (2001), [hep-ex/0107005].

\bibitem{Chekanov:2003ya}
ZEUS, S.~Chekanov {\em et~al.},
\newblock Phys. Lett. {\bf B573}, 46 (2003), [hep-ex/0305028].

\bibitem{Aktas:2005ty}
H1, A.~Aktas {\em et~al.},
\newblock Eur. Phys. J. {\bf C44}, 1 (2005), [hep-ex/0505061].

\bibitem{Aaretal07}
H1, F.~D. Aaron {\em et~al.},
\newblock Phys. Lett. {\bf B659}, 796 (2008), [0709.4114].

\bibitem{Sch07}
L.~Schoeffel,
\newblock {Deeply Virtual Compton Scattering at HERA II},
\newblock 0705.2925, 2007.

\bibitem{Breetal97}
ZEUS, J.~Breitweg {\em et~al.},
\newblock Eur. Phys. J. {\bf C2}, 247 (1998), [hep-ex/9712020].

\bibitem{Breetal98}
ZEUS, J.~Breitweg {\em et~al.},
\newblock Eur. Phys. J. {\bf C6}, 603 (1999), [hep-ex/9808020].

\bibitem{Breetal99}
ZEUS, J.~Breitweg {\em et~al.},
\newblock Eur. Phys. J. {\bf C12}, 393 (2000), [hep-ex/9908026].

\bibitem{Adletal99}
H1, C.~Adloff {\em et~al.},
\newblock Eur. Phys. J. {\bf C13}, 371 (2000), [hep-ex/9902019].

\bibitem{Adletal02}
H1, C.~Adloff {\em et~al.},
\newblock Phys. Lett. {\bf B539}, 25 (2002), [hep-ex/0203022].

\bibitem{Cheetal07}
ZEUS, S.~Chekanov {\em et~al.},
\newblock PMC Phys. {\bf A1}, 6 (2007), [0708.1478].

\bibitem{Bre99a}
ZEUS, J.~Breitweg {\em et~al.},
\newblock Eur. Phys. J. {\bf C14}, 213 (2000), [hep-ex/9910038].

\bibitem{Adletal00a}
H1, C.~Adloff {\em et~al.},
\newblock Phys. Lett. {\bf B483}, 360 (2000), [hep-ex/0005010].

\bibitem{Cheetal05}
ZEUS, S.~Chekanov {\em et~al.},
\newblock Nucl. Phys. {\bf B718}, 3 (2005), [hep-ex/0504010].

\bibitem{Deretal96}
ZEUS, M.~Derrick {\em et~al.},
\newblock Z. Phys. {\bf C73}, 73 (1996), [hep-ex/9608010].

\bibitem{Breetal00}
ZEUS, J.~Breitweg {\em et~al.},
\newblock Phys. Lett. {\bf B487}, 273 (2000), [hep-ex/0006013].

\bibitem{Breetal98a}
ZEUS, J.~Breitweg {\em et~al.},
\newblock Phys. Lett. {\bf B437}, 432 (1998), [hep-ex/9807020].

\bibitem{Adletal00}
H1, C.~Adloff {\em et~al.},
\newblock Phys. Lett. {\bf B483}, 23 (2000), [hep-ex/0003020].

\bibitem{Cheetal02}
ZEUS, S.~Chekanov {\em et~al.},
\newblock Eur. Phys. J. {\bf C24}, 345 (2002), [hep-ex/0201043].

\bibitem{Cheetal04}
ZEUS, S.~Chekanov {\em et~al.},
\newblock Nucl. Phys. {\bf B695}, 3 (2004), [hep-ex/0404008].

\bibitem{Aktetal05}
H1, A.~Aktas {\em et~al.},
\newblock Eur. Phys. J. {\bf C46}, 585 (2006), [hep-ex/0510016].

\bibitem{FraStrWei05}
L.~Frankfurt, M.~Strikman and C.~Weiss,
\newblock Ann. Rev. Nucl. Part. Sci. {\bf 55}, 403 (2005), [hep-ph/0507286].

\bibitem{BooCheRoySch09}
M.~Boonekamp, F.~Chevallier, C.~Royon and L.~Schoeffel,
\newblock Acta Phys. Polon. {\bf B40}, 2239 (2009), [0902.1678].

\bibitem{KhoMarRys00}
V.~A. Khoze, A.~D. Martin and M.~G. Ryskin,
\newblock Eur. Phys. J. {\bf C14}, 525 (2000), [hep-ph/0002072].

\bibitem{DeRKhoMarOraRys02}
A.~De~Roeck, V.~A. Khoze, A.~D. Martin, R.~Orava and M.~G. Ryskin,
\newblock Eur. Phys. J. {\bf C25}, 391 (2002), [hep-ph/0207042].

\bibitem{DonLan86}
A.~Donnachie and P.~V. Landshoff,
\newblock Phys. Lett. {\bf B185}, 403 (1987).

\bibitem{Muea94}
A.~H. Mueller,
\newblock Nucl. Phys. {\bf B415}, 373 (1994).

\bibitem{MuePat94}
A.~H. Mueller and B.~Patel,
\newblock Nucl. Phys. {\bf B425}, 471 (1994), [hep-ph/9403256].

\bibitem{BalLip78}
I.~I. Balitsky and L.~N. Lipatov,
\newblock Sov. J. Nucl. Phys. {\bf 28}, 822 (1978).

\bibitem{KurLipFad77}
E.~A. Kuraev, L.~N. Lipatov and V.~S. Fadin,
\newblock Sov. Phys. JETP {\bf 45}, 199 (1977).

\bibitem{ColFraStr96}
J.~Collins, L.~Frankfurt and M.~Strikman,
\newblock Phys. Rev. {\bf D56}, 2982 (1997), [hep-ph/9611433].

\bibitem{MueRobGeyDitHor94}
D.~M{\"u}ller, D.~Robaschik, B.~Geyer, F.-M. Dittes and J.~Ho\v{r}ej{\v s}i,
\newblock Fortschr. Phys. {\bf 42}, 101 (1994), [hep-ph/9812448].

\bibitem{Rad96}
A.~V. Radyushkin,
\newblock Phys. Lett. {\bf B380}, 417 (1996), [hep-ph/9604317].

\bibitem{Ji96a}
X.~Ji,
\newblock Phys. Rev. {\bf D55}, 7114 (1997), [hep-ph/9609381].

\bibitem{MueSch05}
D.~M{\"u}ller and A.~Sch{\"a}fer,
\newblock Nucl. Phys. {\bf B739}, 1 (2006), [hep-ph/0509204].

\bibitem{Die03a}
M.~Diehl,
\newblock Phys. Rept. {\bf 388}, 41 (2003), [hep-ph/0307382].

\bibitem{BelRad05}
A.~V. Belitsky and A.~V. Radyushkin,
\newblock Phys. Rept. {\bf 418}, 1 (2005), [hep-ph/0504030].

\bibitem{BelJiFen03}
A.~V. Belitsky, X.~Ji and F.~Yuan,
\newblock Phys. Rev. {\bf D69}, 074014 (2004), [hep-ph/0307383].

\bibitem{FraKoeStr95}
L.~Frankfurt, W.~Koepf and M.~Strikman,
\newblock Phys. Rev. {\bf D54}, 3194 (1996), [hep-ph/9509311].

\bibitem{GolKro05}
S.~V. Goloskokov and P.~Kroll,
\newblock Eur. Phys. J. {\bf C42}, 281 (2005), [hep-ph/0501242].

\bibitem{GolKro07}
S.~V. Goloskokov and P.~Kroll,
\newblock Eur. Phys. J. {\bf C53}, 367 (2008), [0708.3569].

\bibitem{FraFreStr98}
L.~L. Frankfurt, A.~Freund and M.~Strikman,
\newblock Phys. Rev. {\bf D 58}, 114001 (1998), [hep-ph/9710356],
\newblock erratum D 59 (1999) 119901E.

\bibitem{BalKuc00}
I.~I. Balitsky and E.~Kuchina,
\newblock Phys. Rev. {\bf D 62}, 074004 (2000), [hep-ph/0002195].

\bibitem{CapFazFioJenPac06}
M.~Capua, S.~Fazio, R.~Fiore, L.~Jenkovszky and F.~Paccanoni,
\newblock Phys. Lett. {\bf B645}, 161 (2007), [hep-ph/0605319].

\bibitem{FazJen08}
S.~Fazio and L.~Jenkovszky,
\newblock {Exclusive diffraction and Pomeron trajectory in ep collisions},
\newblock 0811.1018, 2008.

\bibitem{DonDos00}
A.~Donnachie and H.~G. Dosch,
\newblock Phys. Lett. {\bf B 502}, 74 (2001), [hep-ph/0010227].

\bibitem{McDSanSha02}
M.~McDermott, R.~Sandapen and G.~Shaw,
\newblock Eur. Phys. J. {\bf C22}, 655 (2002), [hep-ph/0107224].

\bibitem{FavMac04}
L.~Favart and M.~V.~T. Machado,
\newblock Eur. Phys. J. {\bf C34}, 429 (2004), [hep-ph/0402018].

\bibitem{Mac07}
M.~V.~T. Machado,
\newblock Braz. J. Phys. {\bf 37}, 555 (2007).

\bibitem{KopSchSid08}
B.~Z. Kopeliovich, I.~Schmidt and M.~Siddikov,
\newblock Phys. Rev. {\bf D79}, 034019 (2009), [0812.3992].

\bibitem{BelMueKir01}
A.~V. Belitsky, D.~M{\"u}ller and A.~Kirchner,
\newblock Nucl. Phys. {\bf B629}, 323 (2002), [hep-ph/0112108].

\bibitem{GuzTec06}
V.~Guzey and T.~Teckentrup,
\newblock Phys. Rev. {\bf D74}, 054027 (2006), [hep-ph/0607099].

\bibitem{GuzTec08}
V.~Guzey and T.~Teckentrup,
\newblock Phys. Rev. {\bf D79}, 017501 (2009), [0810.3899].

\bibitem{FreMcD01b}
A.~Freund and M.~McDermott,
\newblock Phys. Rev. {\bf D65}, 074008 (2002), [hep-ph/0106319].

\bibitem{FreMcD01a}
A.~Freund and M.~F. McDermott,
\newblock Phys. Rev. {\bf D65}, 091901 (2002), [hep-ph/0106124].

\bibitem{FreMcDStr02}
A.~Freund, M.~McDermott and M.~Strikman,
\newblock Phys. Rev. {\bf D67}, 036001 (2003), [hep-ph/0208160].

\bibitem{KumMuePas07}
K.~Kumeri{\v c}ki, D.~M{\"u}ller and K.~Passek-Kumeri{\v c}ki,
\newblock Nucl. Phys. B {\bf 794}, 244 (2008), [hep-ph/0703179].

\bibitem{Mue06}
D.~M{\"u}ller,
\newblock {Pomeron} dominance in deeply virtual {Compton} scattering and the
  femto holographic image of the proton,
\newblock hep-ph/0605013, 2006.

\bibitem{KumMuePas08}
K.~Kumeri{\v c}ki, D.~M{\"u}ller and K.~Passek-Kumeri{\v c}ki,
\newblock Eur. Phys. J. {\bf C58}, 193 (2008), [0805.0152].

\bibitem{KivMan01}
N.~Kivel and L.~Mankiewicz,
\newblock Eur. Phys. J. {\bf C21}, 621 (2001), [hep-ph/0106329].

\bibitem{Ter05}
O.~V. Teryaev,
\newblock Analytic properties of hard exclusive amplitudes,
\newblock hep-ph/0510031, 2005.

\bibitem{FraFreGuzStr97}
L.~Frankfurt, A.~Freund, V.~Guzey and M.~Strikman,
\newblock Phys. Lett. {\bf B418}, 345 (1998), [hep-ph/9703449],
\newblock Erratum-ibid. B429 (1998) 414.

\bibitem{Che97}
Z.~Chen,
\newblock Nucl. Phys. {\bf B525}, 369 (1998), [hep-ph/9705279].

\bibitem{DieIva07}
M.~Diehl and D.~Y. Ivanov,
\newblock Eur. Phys. J. {\bf C52}, 919 (2007), [0707.0351].

\bibitem{AniTer07}
I.~V. Anikin and O.~V. Teryaev,
\newblock Phys. Rev. {\bf D76}, 056007 (2007), [0704.2185].

\bibitem{PolWei99}
M.~V. Polyakov and C.~Weiss,
\newblock Phys. Rev. {\bf D60}, 114017 (1999), [hep-ph/9902451].

\bibitem{Ter01}
O.~V. Teryaev,
\newblock Phys. Lett. {\bf B510}, 125 (2001), [hep-ph/0102303].

\bibitem{Dia02}
D.~Diakonov,
\newblock Prog. Part. Nucl. Phys. {\bf 51}, 173 (2003), [hep-ph/0212026].

\bibitem{BelFreMue99}
A.~V. Belitsky, A.~Freund and D.~M{\"u}ller,
\newblock Nucl. Phys. {\bf B574}, 347 (2000), [hep-ph/9912379].

\bibitem{Fre99}
A.~Freund,
\newblock Phys. Lett. {\bf B472}, 412 (2000), [hep-ph/9903488].

\bibitem{Rad97}
A.~V. Radyushkin,
\newblock Phys. Rev. {\bf D56}, 5524 (1997), [hep-ph/9704207].

\bibitem{BelGeyMueSch97}
A.~V. Belitsky, B.~Geyer, D.~M{\"u}ller and A.~Sch{\"a}fer,
\newblock Phys. Lett. {\bf B421}, 312 (1998), [hep-ph/9710427].

\bibitem{Shu99}
A.~G. Shuvaev,
\newblock Phys. Rev. {\bf D60}, 116005 (1999), [hep-ph/9902318].

\bibitem{PolShu02}
M.~V. Polyakov and A.~G. Shuvaev,
\newblock {On} 'dual' parametrizations of generalized parton distributions,
\newblock hep-ph/0207153, 2002.

\bibitem{KirManSch05a}
M.~Kirch, A.~Manashov and A.~Sch{\"a}fer,
\newblock Phys. Rev. {\bf D72}, 114006 (2005), [hep-ph/0509330].

\bibitem{BroLep79}
G.~P. Lepage and S.~J. Brodsky,
\newblock Phys. Lett. {\bf B87}, 359 (1979).

\bibitem{EfrRad80}
A.~Efremov and A.~Radyushkin,
\newblock Phys. Lett. {\bf B94}, 245 (1980).

\bibitem{EfrRad80a}
A.~Efremov and A.~Radyushkin,
\newblock Theor. Math. Phys. {\bf 42}, 97 (1980).

\bibitem{BroLep80}
G.~Lepage and S.~Brodsky,
\newblock Phys. Rev. {\bf D22}, 2157 (1980).

\bibitem{LanPol70}
P.~V. Landshoff, J.~C. Polkinghorne and R.~D. Short,
\newblock Nucl. Phys. {\bf B28}, 225 (1971).

\bibitem{BroCloGun73}
S.~J. Brodsky, F.~E. Close and J.~F. Gunion,
\newblock Phys. Rev. {\bf D8}, 3678 (1973).

\bibitem{BelMueKirSch00}
A.~V. Belitsky, D.~M{\"u}ller, A.~Kirchner and A.Sch{\"a}fer,
\newblock Phys. Rev. {\bf D64}, 116002 (2001), [hep-ph/0011314].

\bibitem{HwaMue07}
D.~S. Hwang and D.~M{\"u}ller,
\newblock Phys. Lett. {\bf B660}, 350 (2008), [0710.1567].

\bibitem{Rad98a}
A.~Radyushkin,
\newblock Phys. Rev. {\bf D59}, 014030 (1999), [hep-ph/9805342].

\bibitem{Rad98}
A.~V. Radyushkin,
\newblock Phys. Lett. {\bf B449}, 81 (1999), [hep-ph/9810466].


\bibitem{Nor00}
J.~D. Noritzsch,
\newblock Phys. Rev. {\bf D62}, 054015 (2000), [hep-ph/0004012].

\bibitem{ManKirSch05}
A.~Manashov, M.~Kirch and A.~Sch{\"a}fer,
\newblock Phys. Rev. Lett. {\bf 95}, 012002 (2005), [hep-ph/0503109].

\bibitem{Pol07a}
M.~V. Polyakov,
\newblock {Educing} {GPDs} from amplitudes of hard exclusive processes,
\newblock 0711.1820, 2007.

\bibitem{ShuBieMarRys99}
A.~G. Shuvaev, K.~J. Golec-Biernat, A.~D. Martin and M.~G. Ryskin,
\newblock Phys. Rev. {\bf D60}, 014015 (1999), [hep-ph/9902410].

\bibitem{MusRad99}
I.~V. Musatov and A.~V. Radyushkin,
\newblock  Phys. Rev. {\bf D61}, 074207 (2000), [hep-ph/9905376].

\bibitem{MarNocRysShuTeu08}
A.~D. Martin, C.~Nockles, M.~G. Ryskin, A.~G. Shuvaev and T.~Teubner,
\newblock {The power of the Shuvaev transform},
\newblock 0812.3558, 2008.

\bibitem{Khu63}
N.~N. Khuri,
\newblock Phys. Rev. {\bf 132}, 914 (1963).

\bibitem{Ven68}
G.~Veneziano,
\newblock Nuovo. Cim. {\bf A57}, 190 (1968).

\bibitem{Sem08}
K.~M. Semenov-Tian-Shansky,
\newblock Eur. Phys. J. {\bf A36}, 303 (2008), [0803.2218].

\bibitem{PolSem08}
M.~V. Polyakov and K.~M. Semenov-Tian-Shansky,
\newblock Eur. Phys. J. {\bf A40}, 181 (2009), [0811.2901].

\bibitem{DieKug07a}
M.~Diehl and W.~Kugler,
\newblock Phys. Lett. {\bf B660}, 202 (2008), [0711.2184 [hep-ph]].

\bibitem{DieFelJakKro98}
M.~Diehl, T.~Feldmann, R.~Jakob and P.~Kroll,
\newblock Eur. Phys. J. {\bf C8}, 409 (1999), [hep-ph/9811253].

\bibitem{BroDieHwa00}
S.~J. Brodsky, M.~Diehl and D.~S. Hwang,
\newblock Nucl. Phys. {\bf B596}, 99 (2001), [hep-ph/0009254].

\bibitem{DieFelJakKro00}
M.~Diehl, T.~Feldmann, R.~Jakob and P.~Kroll,
\newblock Nucl. Phys. {\bf B596}, 33 (2001), [hep-ph/0009255],
\newblock Erratum-ibid. B605 (2001) 647.

\bibitem{MukMusPauRad02}
A.~Mukherjee, I.~V. Musatov, H.~C. Pauli and A.~V. Radyushkin,
\newblock Phys. Rev. {\bf D67}, 073014 (2003), [hep-ph/0205315].

\bibitem{Pob02}
P.~V. Pobylitsa,
\newblock Phys. Rev. {\bf D66}, 094002 (2002), [hep-ph/0204337].

\bibitem{Pob02a}
P.~V. Pobylitsa,
\newblock Phys. Rev. {\bf D70}, 034004 (2004), [hep-ph/0211160].

\bibitem{Pob02b}
P.~V.~Pobylitsa,
Phys.\ Rev.\  {\bf D67}, 094012 (2003), [hep-ph/0210238].


\bibitem{Pol98}
M.~V. Polyakov,
\newblock Nucl. Phys. {\bf B555}, 231 (1999), [hep-ph/9809483].

\bibitem{JiLeb00}
X.-D. Ji and R.~F. Lebed,
\newblock Phys. Rev. {\bf D63}, 076005 (2001), [hep-ph/0012160].

\bibitem{KobOku62}
I.~Y. Kobzarev and L.~B. Okun,
\newblock Zh. Eksp. Teor. Fiz. {\bf 43}, 1904 (1962),
\newblock [Sov. Phys. JETP 16, 1343 (1963)].

\bibitem{Ter99}
O.~V. Teryaev,
\newblock {Spin structure of nucleon and equivalence principle},
\newblock hep-ph/9904376, 1999.

\bibitem{Ter06}
O.~V. Teryaev,
\newblock AIP Conf. Proc. {\bf 915}, 260 (2007), [hep-ph/0612205].

\bibitem{BurMilNow08}
M.~Burkardt, A.~Miller and W.~D. Nowak,
\newblock Rept. Prog. Phys. {\bf 73}, 016201 (2010) [0812.2208].

\bibitem{Ji96}
X.~Ji,
\newblock Phys. Rev. Lett. {\bf 78}, 610 (1997), [hep-ph/9603249].

\bibitem{KumMuePas08a}
D.~M{\"u}ller, K.~Kumeri{\v c}ki and K.~Passek-Kumeri{\v c}ki,
\newblock {GPD sum rules: a tool to reveal the quark angular momentum},
\newblock 0807.0170, 2008.

\bibitem{Broetal07}
QCDSF-UKQCD, D.~Brommel {\em et~al.},
\newblock PoS {\bf LAT2007}, 158 (2007), [0710.1534].

\bibitem{Hagetal07}
LHPC, P.~H{\"a}gler {\em et~al.},
\newblock Phys. Rev. {\bf D77}, 094502 (2008), [0705.4295].

\bibitem{Sar82}
M.~Sarmadi,
\newblock PhD thesis, Pittsburgh Univ., 1982.

\bibitem{DitRad81}
F.-M. Dittes and A.~Radyushkin,
\newblock Sov. J. Nucl. Phys. {\bf 34}, 293 (1981).

\bibitem{MikRad85}
S.~Mikhailov and A.~Radyushkin,
\newblock Nucl. Phys. {\bf B254}, 89 (1985).

\bibitem{MikVla08}
S.~V. Mikhailov and A.~A. Vladimirov,
\newblock Phys. Lett. {\bf B671}, 111 (2009), [0810.1647].

\bibitem{MelMuePas02}
B.~Meli{\' c}, D.~M{\"u}ller and K.~Passek-Kumeri{\v c}ki,
\newblock Phys. Rev. {\bf D68}, 014013 (2003), [hep-ph/0212346].

\bibitem{BraKorMue03}
V.~M. Braun, G.~P. Korchemsky and D.~M{\"u}ller,
\newblock Prog. Part. Nucl. Phys. {\bf 51}, 311 (2003), [hep-ph/0306057].

\bibitem{AbrSte}
M.~Abramowitz and I.~Stegun,
{\em Handbook of Mathematical Functions}, Dover Publications,  1970, New York.


\bibitem{JamRos75}
F.~James and M.~Roos,
\newblock Comput. Phys. Commun. {\bf 10}, 343 (1975).

\bibitem{Nor03}
J.~D. Noritzsch,
\newblock Phys. Rev. {\bf D69}, 094016 (2004), [hep-ph/0312137].

\bibitem{CheKuhSte00}
K.~G. Chetyrkin, J.~H. Kuhn and M.~Steinhauser,
\newblock Comput. Phys. Commun. {\bf 133}, 43 (2000), [hep-ph/0004189].

\bibitem{Aidetal96}
H1, S.~Aid {\em et~al.},
\newblock Nucl. Phys. {\bf B470}, 3 (1996), [hep-ex/9603004].

\bibitem{Ale02}
S.~Alekhin,
\newblock Phys. Rev. {\bf D68}, 014002 (2003), [hep-ph/0211096].

\bibitem{MarStiTho06}
A.~D. Martin, W.~J. Stirling and R.~S. Thorne,
\newblock Phys. Lett. {\bf B636}, 259 (2006), [hep-ph/0603143].

\bibitem{Pumetal02}
J.~Pumplin {\em et~al.},
\newblock JHEP {\bf 07}, 012 (2002), [hep-ph/0201195].

\bibitem{IvaSchSzyKra04}
D.~Y. Ivanov, A.~Schafer, L.~Szymanowski and G.~Krasnikov,
\newblock Eur. Phys. J. {\bf C34}, 297 (2004), [hep-ph/0401131].

\bibitem{MarRysTeu99}
A.~D. Martin, M.~G. Ryskin and T.~Teubner,
\newblock Phys. Rev. {\bf D62}, 014022 (2000), [hep-ph/9912551].

\bibitem{MarNocRysTeu07}
A.~D. Martin, C.~Nockles, M.~G. Ryskin and T.~Teubner,
\newblock Phys. Lett. {\bf B662}, 252 (2008), [0709.4406].

\bibitem{Bur02}
M.~Burkardt,
\newblock Int. J. Mod. Phys. {\bf A18}, 173 (2003), [hep-ph/0207047].

\bibitem{RPP08}
Particle Data Group, C.~Amsler {\em et al.},
\newblock Phys. Lett. {\bf B667}, 1 (2008).


\bibitem{StrWei03}
M.~Strikman and C.~Weiss,
\newblock Phys. Rev. {\bf D69}, 054012 (2004), [hep-ph/0308191].

\bibitem{Gui08a}
M.~Guidal,
\newblock Eur. Phys. J. {\bf A37}, 319 (2008), [0807.2355].

\bibitem{Yua03}
F.~Yuan,
\newblock Phys. Rev. {\bf D69}, 051501 (2004), [hep-ph/0311288].

\bibitem{BroCloGun71}
S.~J. Brodsky, F.~E. Close and J.~F. Gunion,
\newblock Phys. Rev. {\bf D5}, 1384 (1972).

\bibitem{BroCloGuo72}
S.~J. Brodsky, F.~E. Close and J.~F. Gunion,
\newblock Phys. Rev. {\bf D6}, 177 (1972).

\bibitem{BelMue08}
A.~V. Belitsky and D.~M{\"u}ller,
\newblock Phys. Rev. {\bf D79}, 014017 (2009), [0809.2890].

\bibitem{Airetal06}
HERMES, A.~Airapetian {\em et al.}
\newblock Phys. Rev. {\bf D75}, 011103 (2007), [hep-ex/0605108].

\bibitem{Airetal08}
HERMES, A.~Airapetian {\em et~al.},
\newblock JHEP {\bf 06}, 066 (2008), [0802.2499].

\bibitem{Giretal07}
CLAS, F.~X. Girod {\em et~al.},
\newblock Phys. Rev. Lett. {\bf 100}, 162002 (2008), [0711.4805].

\bibitem{Cametal06}
Jefferson Lab Hall A and Hall A DVCS, C.~M. Camacho {\em et~al.},
\newblock Phys. Rev. Lett. {\bf 97}, 262002 (2006), [nucl-ex/0607029].

\bibitem{Ell07}
F.~Ellinghaus,
\newblock {DVCS at HERMES: Recent Results},
\newblock 0710.5768, 2007.

\bibitem{PenPolGoe99}
M. Penttinen, M. V. Polyakov, and K. Goeke,
\newblock Phys. Rev. {\bf D62}, 014024 (2000), [hep-ph/9909489].


\bibitem{GoePolVan01}
K.~Goeke, M.~V. Polyakov and M.~Vanderhaeghen,
\newblock Prog. Part. Nucl. Phys. {\bf 47}, 401 (2001), [hep-ph/0106012].

\bibitem{SchBofRad02}
P.~Schweitzer, S.~Boffi and M.~Radici,
\newblock Phys. Rev. {\bf D66}, 114004 (2002), [hep-ph/0207230].

\bibitem{Wak07}
M.~Wakamatsu,
\newblock Phys. Lett. {\bf B648}, 181 (2007), [hep-ph/0701057].

\bibitem{Goeetal07}
K.~Goeke {\em et~al.},
\newblock Phys. Rev. {\bf D75}, 094021 (2007), [hep-ph/0702030].

\bibitem{PolVan08}
M.~V. Polyakov and M.~Vanderhaeghen,
\newblock {Taming Deeply Virtual Compton Scattering},
\newblock 0803.1271, 2008.

\bibitem{Gui08}
M.~Guidal,
\newblock Nucl. Phys. Proc. Suppl. {\bf 184}, 234 (2008), [0803.1592].



\bibitem{Cheetal08}
ZEUS, S.~Chekanov {\em et~al.},
\newblock JHEP {\bf 0905}, 108 (2009), [0812.2517]

\end{thebibliography}

\end{document}